\documentclass[11pt]{article}

\usepackage{fullpage}
\usepackage{amsmath}
\usepackage{amsfonts}
\usepackage{amssymb}
\usepackage{amsthm}
\usepackage{graphicx}
\usepackage{rotate,rotating}
\usepackage{setspace}
\usepackage{comment}
\usepackage{geometry}
\usepackage{enumerate}
\usepackage{epstopdf}
\usepackage{enumitem}
\usepackage{pdfpages,xcolor}
\usepackage{graphicx}
\usepackage{subcaption}
\usepackage{caption}
\definecolor{navyblue}{rgb}{0,0,0.8} %
\usepackage[colorlinks, linkcolor=navyblue, citecolor=navyblue, urlcolor=navyblue, pdfborder={0 0 1}, citebordercolor={1 1 1}]{hyperref}

\usepackage{afterpage} %

\usepackage{booktabs}
\usepackage{array}
\usepackage{ulem}
\usepackage{tikz}
\usepackage{natbib}

\usepackage{ragged2e} %

\usepackage{longtable}
\usepackage{multirow}
\usepackage{booktabs}
\usepackage{array}
\usepackage{csvsimple}
\usepackage{graphicx}
\usepackage{pdflscape}

\usepackage{xcolor}
\usepackage{soul}

\usepackage{listings}
\usepackage{floatrow}
\usepackage{xcolor} %

\usepackage{placeins} %
\usepackage{appendix}

\allowdisplaybreaks
\graphicspath{{imgs/}}
\makeatletter
\def\input@path{{tables/}}
\makeatother

\lstdefinestyle{promptstyle}{
    basicstyle=\ttfamily\scriptsize,  %
    backgroundcolor=\color{gray!10}, %
    breaklines=true,             %
    numbers=none,                %
    numberstyle=\tiny\color{gray}, %
    keywordstyle=\color{blue}\bfseries, %
    stringstyle=\color{orange},  %
    commentstyle=\color{green!50!black}\itshape, %
    tabsize=4,                   %
    showstringspaces=false       %
    lineskip=-10.5pt                  %
}
\newcommand{\displayprompt}[1]{
    \begin{singlespace}
    \lstinputlisting[style=promptstyle]{#1}
    \end{singlespace}
}

\begin{document}
\doublespacing

\newcommand{\Es}[2]{\mathbb{E}_{#1}\left[{#2}\right]}
\newcommand{\Ec}[2]{\mathbb{E}\left[{#1}\middle|{#2}\right]}
\newcommand{\Ets}[1]{\mathbb{E}\left[{#1}\middle|\fFTS\right]}
\newcommand{\E}[1]{\mathbb{E}\left[#1\right]}
\newcommand{\Ej}{\mathbb{E}}  %

\newcommand{\Var}[1]{\mathbb{V}\mathrm{ar}\left[#1\right]}
\newcommand{\Varj}{\mathbb{V}\mathrm{ar}}
\newcommand{\Varc}[2]{\mathbb{V}\mathrm{ar}\left[{#1}\middle|{#2}\right]}
\newcommand{\Covs}[2]{\mathbb{C}\mathrm{ov}_{#1}\left[{#2}\right]}
\newcommand{\Cov}[1]{\mathbb{C}\mathrm{ov}\left[{#1}\right]}

\newcommand{\T}{^{\top}}
\newcommand{\inv}{^{-1}}

\newcommand{\IS}{\mathbb{S}}

\newcommand{\Paren}[1]{\left({#1}\right)}
\newcommand{\Brack}[1]{\left[{#1}\right]}
\newcommand{\Curli}[1]{\left\{{#1}\right\}}
\newcommand{\Norm} [1]{\left \lVert{#1}\right \rVert}
\newcommand{\Abs}  [1]{\left \lvert{#1}\right \rvert}
\newcommand{\Eval} [2]{\left.{#1}\right\rvert_{#2}}

\newcommand{\lamtld}{\widetilde{\lambda}}
\newcommand{\Vtld}{\widetilde{V}}
\newcommand{\vtld}{\widetilde{v}}
\newcommand{\xtld}{\widetilde{x}}
\newcommand{\ztld}{\widetilde{z}}

\newcommand{\vhat}{\widehat{v}}

\newcommand{\Xl }{\mathcal{X}}

\newcommand{\xzer}{x^0}
\newcommand{\xstar}{x^*}
\newcommand{\xx}{x}

\newcommand{\lamzer}{\lambda^0}  %
\newcommand{\lamhat}{\widehat{\lambda}}
\newcommand{\sit}{_{i, t}}
\newcommand{\defeq}{:=}

\newcommand{\showauthors}{true} %

\newcommand{\astfootnote}[1]{%
\let\oldthefootnote=\thefootnote%
\setcounter{footnote}{0}%
\renewcommand{\thefootnote}{\fnsymbol{footnote}}%
\footnote{#1}%
\let\thefootnote=\oldthefootnote%
}
\begin{center}
{\LARGE
    Explaining Realized Returns: Roll's $R^2$ 40 Years Later%
    \astfootnote{\ifthenelse{\equal{\showauthors}{true}}{%
        We thank
        Federico Bandi,
        John Campbell,
        Richard Roll,
        Dacheng Xiu,
        and Kangying Zhou,
        as well as participants at the
        EFMA ``AI and Asset Management'' Special Session
        for valuable comments and suggestions.
        We thank Xinran Feng for his excellent research assistance.
        AQR Capital Management is a global investment management firm, which may or may not apply similar investment techniques or methods of analysis as described herein. The views expressed here are those of the authors and not necessarily those of AQR.
        Send correspondence to Bryan Kelly, bryan.kelly@yale.edu. 
    }~
    }
}
\\[30pt]

\ifthenelse{\equal{\showauthors}{true}}{
\begin{tabular}[t]{ccc}
Bryan Kelly & Yinan Su & Chao Zhang
\\
{\small \parbox{1.8in}{\centering \singlespacing Yale University \\ AQR Capital Management \\ NBER}} & 
{\small \parbox{1.8in}{\centering \singlespacing Johns Hopkins}} & 
{\small \parbox{1.8in}{\centering \singlespacing HKUST (GZ)}}	
\end{tabular}
}~
\\[30pt]
November 16, 2026
\\[30pt]

\end{center}
\setcounter{footnote}{0}%

\begin{abstract}
Can we explain asset price movements even with hindsight? Reinterpreting \citeauthor{roll1988r2}'s \citeyearpar{roll1988r2} AFA Presidential Address, ``$R^2$,'' as a modern empirical exercise, we show that conventional macroeconomic, cash-flow, expectations, and flow variables explain less than 10\% (often with negative $R^2$) of monthly U.S. equity return variation out of sample and lack daily observations. We use generative AI to extract reasons stated in Wall Street Journal market summaries while filtering out price-related language. These reasons explain around 30--40\% of daily and monthly return variation out of sample. Their content spans interpretable themes, including corporate news, macroeconomic conditions, monetary policy, sentiment, and geopolitics.

\end{abstract}

{
\scriptsize
\vspace{30pt}

\vspace{-8pt}
}

\newpage
\doublespacing

\section{Introduction}

\vspace{10pt}
\begin{center}
\begin{minipage}{0.95\linewidth}
{\it 
    Many financial economists seem to believe that, with hindsight, they
    could explain most asset-price movements with authenticated information. \dots ~
    By observing and measuring these influences ex post, one could presumably ``explain'' stock price movements with an astronomical degree of accuracy. The $R^2$ should be close to $1.0$.
}
    \vspace*{-10pt}
    \begin{flushright}
        $R^2$, Richard Roll, American Finance Association Presidential Address, 1988%
    \end{flushright}
\end{minipage}
\end{center}
\vspace*{20pt}

Prices move every day, and every day market participants, journalists, and researchers attach reasons to those movements.
The explanations are often concrete: earnings news, inflation data, monetary policy, geopolitical risk, investor sentiment, or changes in expected growth.
Yet the discipline has much less evidence on a basic quantitative question: how much of realized market variation can these reasons actually explain?
Roll's \citeyearpar{roll1988r2} AFA Presidential Address made this question precise.
If financial prices are driven by real-world events and researchers can observe the relevant drivers after the fact, then an ex-post explanation of price movements might appear almost mechanical.
The relevant $R^2$ should be close to one.
Roll showed instead that available explanations delivered disappointing explanatory power.
The result exposed a measurement gap at the center of empirical finance: markets are interpreted in narrative terms, but the narratives have been difficult to turn into disciplined, high-$R^2$ explanations.

This paper revisits Roll's problem in this spirit, but it does not update Roll's original empirical analysis directly.
We reinterpret his question as a modern measurement problem: how much of realized U.S.\ stock market fluctuation can be explained, out of sample, by the real-world reasons stated in financial news?
This is not a return forecasting exercise.
The reasons we study are observed after the market movement occurs, and the object is not whether one could have predicted the return in advance.
Nor is the paper an attempt to identify the causal effect of one particular shock.
The object is variation explained: after the market movement has occurred and after the financial press has described what moved the market, can researchers recover a stable mapping from those stated reasons to returns?\footnote{As \citet{cochrane2025causation} emphasizes, causation does not imply variation. This paper is about measuring variation explained, not about identifying causal effects.}

The question is: do we understand why prices move?
Financial prices aggregate fundamental information, but they may also reflect sentiment, institutional liquidity conditions, and interactions among these factors.
The empirical question is whether, and to what extent, these factors can be observed independently of the asset prices being explained, and whether those observations have a stable explanatory relationship with realized returns.
We evaluate this relationship statistically without committing to a particular structural model, while leaving room for structural models to interpret it and identify causal mechanisms.

Our premise is that financial news is a natural empirical record of the reasons for price movements.
News articles do not contain all information seen by investors, and stated reasons are not guaranteed to be structural causes.
But they are the public language through which market movements are interpreted, and they provide a concrete object for measuring how far such interpretations can go.

\begin{figure}[!t]
	\centering
	\caption{Out-of-sample explanatory power of market returns by news and non-news explanations
    	\label{fig:comp_non_news}}
	\includegraphics[width=\textwidth]{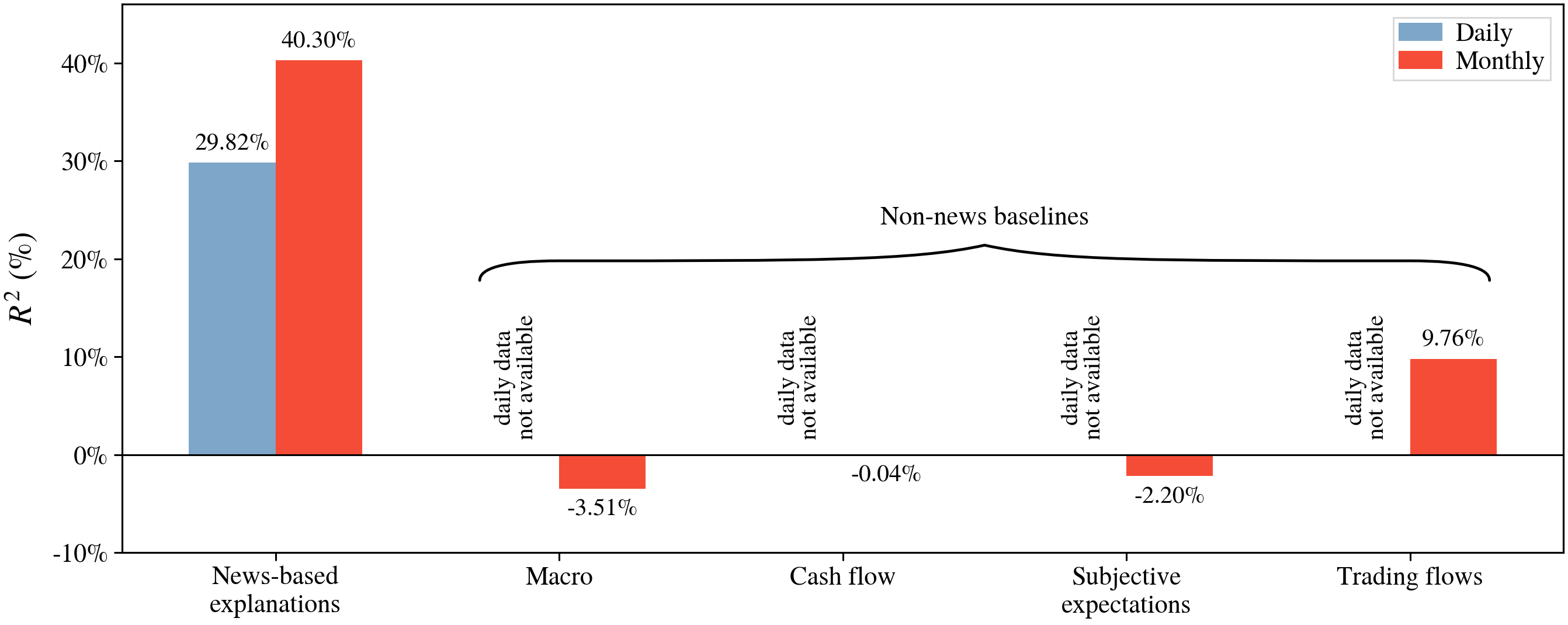}
    \footnotesize\justifying
    \noindent Note: The figure reports out-of-sample $R^2$ values for explaining realized aggregate U.S.\ stock market returns. Non-news benchmarks use structured macroeconomic, cash-flow, expectations, sentiment, and flow variables.
    ``Daily data not available'' indicates that these variables are not observed at the daily frequency.
    News-based reasons use newspaper-stated explanation texts extracted from the Wall Street Journal and filtered to remove direct price and return language. The mapping from reason text to returns is estimated using an embedding regression approach. See benchmark results in Section \ref{subsec:nonnews} and news-based results in Section \ref{subsec:embedding_regression}.
    \vspace*{-10pt}
\end{figure}

We first show that Roll's puzzle remains alive in our modern empirical setting: conventional quantitative variables have little reliable explanatory power for realized market returns.
We focus on explaining daily and monthly returns of the aggregate U.S.\ stock market.
The aggregate market is the natural starting point for the core $R^2$ problem because it is the most visible and widely discussed asset-pricing object, but the framework can potentially be extended to sectors, individual stocks, and other asset classes.
We evaluate explanations by out-of-sample $R^2$, which asks whether the mapping from candidate explanations to returns is stable over time and guards against overfitting in settings with many potential variables.
We construct conventional baselines with quantitative measures from macroeconomic statistics, cash-flow variables, investor sentiment and subjective expectations, and trading flows.
At the monthly frequency, these structured variables have little explanatory power.
Most groups have negative out-of-sample $R^2$, and trading flows deliver roughly 10\%.
At the daily frequency, these variables are not available because the underlying data arrive monthly or less frequently.
These results identify an empirical failure at a foundational level of finance. (See Figure \ref{fig:comp_non_news})

News-based reasons change this picture.
We construct the explanations from a single article per trading day in the Wall Street Journal's ``Today's Markets'' column.
These articles provide broad coverage of daily market developments and typically state the reasons associated with those movements.
Using only a single article is intentionally conservative.
It gives a transparent proof of concept, but it likely understates how much of market variation could be explained by a broader corpus of market commentary, including full news coverage or the market-commentary documents produced by investment-bank research departments.
We use generative AI to extract portions of text that the article itself presents as explanations for market movements.
We then search for and remove language that directly describes the price movement being explained.
This masking step is central.
It prevents the explanatory variable from mechanically containing the return.
After this filtering step, the retained explanations are verifiably non-tautological, containing no restatements of the price changes they seek to explain.

The resulting reasons dataset records, for each trading day, the stated events and narratives that financial media cite as drivers of market fluctuations.
The reasons cover corporate news, macroeconomic developments, monetary policy, investor sentiment, geopolitical developments, and other recurrent themes.
We organize them into a hierarchical taxonomy using AI-assisted summarization with human validation.
This taxonomy is not merely descriptive.
It helps turn repeated market narratives into an ontology of economic drivers, allowing us to ask which kinds of events repeatedly appear as explanations.
Evaluated separately, every reason category delivers positive out-of-sample explanatory power. The leading categories---corporate news, macroeconomic activity, and investor sentiment—each achieve an \(R^2\) between 6\% and 9\%. The taxonomy also reveals economically interpretable temporal structure: corporate-news reasons follow the quarterly earnings cycle; monetary-policy reasons recur at the FOMC frequency and track shifts between easing and tightening; and geopolitical and pandemic narratives rise sharply around the corresponding events. These patterns and the underlying reason text are discussed further below.

We map reason text into returns in two ways.
The first is embedding regression.
Each day's reason text is represented by a high-dimensional LLM embedding, and a regularized linear model maps embeddings into realized returns.
The OOS evaluation shows that the relation between the semantic content of news explanations and realized returns is stable over time and achieves $R^2$ values of 30--40\% at both daily and monthly frequencies.
The second method uses zero-shot inference.
We directly prompt a generative AI model to read the masked reason text and state the return implication.
The exercise is ``zero shot'' because the mapping is not estimated on our reason-return sample; it instead relies on the LLM's knowledge to infer the implication of the stated reasons from each day's news.
The intuition is that if the relation between events and market movements is part of broadly shared market knowledge, a general-purpose language model may recover that relation without sample-specific estimation.
This method also explains daily market returns with an out-of-sample $R^2$ of roughly 25\%. We also show that the zero-shot inference method fails when aggregated to the monthly frequency in its raw form, but can achieve a monthly out-of-sample $R^2$ of nearly 50\% with a simple adjustment that filters out the temporal persistence in the raw zero-shot predictions.

Lastly, we address the concern that the reasons reported in market commentary are merely ex-post attributions constructed after journalists observe the realized return. 
We present several pieces of evidence that are difficult to reconcile with a pure ex-post attribution account. 
First, we identify textual evidence that some of the extracted reasons are anticipated by market participants ex ante.
For example, when progress in U.S.--China trade negotiations was cited as the reason for the market rally on October 11, 2019, we find that earlier coverage had already reported that investors were awaiting trade-negotiation news and expected it to move the market.
We systematically match the reasons to earlier news coverage and find that a nontrivial number of reasons across different types are anticipated in a similar way.
This suggests that the market, in many cases, may have already understood the potential sources of uncertainty that are relevant for the market return before the actual directional realization of the uncertainty.
Moreover, we show that the anticipated reasons identified by our conservative matching method have stronger explanatory power.
Second, we show that the occurrence of the reasons aligns with the timing of 12 types of scheduled events that we collected, such as FOMC meetings and major macroeconomic data releases, which are widely regarded as relevant drivers of market fluctuations.
We also show that on monetary-policy event days, reason-implied returns align with high-frequency market responses measured in narrow windows around FOMC communications, which the literature uses as a proxy for the market's reaction to monetary-policy news.

{\bf Literature.} 
Our paper is a reinterpretation of \citeauthor{roll1988r2}'s \citeyearpar{roll1988r2} original question as a modern empirical task.\footnote{The empirical part of \citet{roll1988r2} explains stock returns using systematic and industry factors and compares factor-model explanatory power and return distributions on dates with and without firm-specific public news. This is a different empirical exercise from ours.}
In addition to his AFA Presidential Address, \citet{roll1984orange} uses orange-juice futures as a particularly transparent setting for linking prices to fundamentals. Yet he finds that the most obvious observable fundamental---weather---explains only a small share of price volatility. In ``What Moves Stock Prices?,'' \citet{cutler1989what} similarly focus on explaining aggregate market returns. They use a broad set of macroeconomic and financial variables, including leads and lags of these variables, to explain monthly market returns.\footnote{\citet{cutler1989what} evaluate explanatory power using in-sample $R^2$ adjusted for the number of regressors and conclude that it is difficult to explain more than one-third of return variance using these variables. Adjusted $R^2$ addresses the mechanical increase in in-sample fit from adding regressors, but it is not an out-of-sample evaluation and is therefore not directly comparable to our results.} These papers pioneered the study of explanations for realized returns and demonstrated the empirical difficulty of the task. An important connection to our paper is their recognition of financial news as a potential source of explanations for price movements.\footnote{\citet{roll1984orange} manually reviews the \textit{Wall Street Journal}, \citet{roll1988r2} uses the Dow Jones newswire (the ``Broad Tape'') to identify whether a firm-specific news event occurred,
and \citet{cutler1989what} examine the \textit{New York Times}.} Constrained by the technology available at the time, these studies use only event incidence or categories, which provide little improvement in explanatory power.

A related literature connects news to realized price movements. We shift the empirical object from the occurrence or classification of news events to the content of stated explanations and its explanatory power. \citet{chan2003stock} compares subsequent returns following news-associated and no-news price movements, finding post-news drift and no-news reversal. \citet{boudoukh2019information} improve the identification of firm-level news days by distinguishing economically relevant news, revealing a stronger relationship between public news and contemporaneous return volatility. Both \citet{jeon2022news} and \citet{aitsahalia2024many} link firm-level news to the incidence and magnitude of stock-return jumps. \citet{aitsahalia2024many} study the occurrence and magnitude of intraday jumps, irrespective of direction, and their association with the arrival of public news, finding that around 80\% of intraday jumps are not matched to public news within their event windows, while approximately 96\% of overnight jumps are matched to news. \citet{baker2021triggers} are closely related to our work because they use newspaper-stated explanations of aggregate-market jumps across 19 countries. Their analysis, however, focuses on manually classified explanations for extreme market movements and asks {\it what types} of events trigger major market jumps. We focus on the question of {\it how much} market variation, extreme or otherwise, can be explained; with our reason taxonomy, we can also assess how much {\it each type} of reason explains.\footnote{More broadly, many other papers use newspaper and newswire text to study asset-pricing questions ranging from information transmission and liquidity to return predictability, volatility, and common risk factors. See, for example, \citet{mitchell1994impact}, \citet{tetlock2007giving}, \citet{tetlock2008more}, \citet{dougal2012journalists}, \citet{garcia2013sentiment}, \citet{tetlock2014information}, \citet{manela2017news}, \citet{calomiris2019news}, \citet{bybee2023narrative}, and \citet{bybee2024business}.}

AI is critical to our empirical design. It allows us to process the contextual content of financial news at scale and thereby revisit a classical question in finance. This use of AI connects our paper to a growing literature that uses LLMs to extract information from financial text, including \citet{chen2022expected}, \citet{lopez2023can}, \citet{jha2024chatgpt}, \citet{hansen2024chatgpt}, \citet{didisheim2026inefficient}, and \citet{kelly2026scaling}.\footnote{The broader and earlier literature on textual analysis in economics and finance is surveyed by \citet{gentzkow2019text} and \citet{loughran2020textual}.}
Our objective, however, is not to explore the capabilities of LLMs but to use them as tools to address a classical empirical question in finance. 

\citet{koijen2026assessing} assess the benefits of agentic AI systems in finance in a task that involves explaining firm-level returns around earnings announcements. The paper is closely related to ours in that the task is also to explain realized returns contemporaneously, rather than to predict future returns ex ante. However, our goals and approaches differ in several respects. We explain the time series of aggregate market returns, whereas they explain firm-level returns around earnings announcements. We use financial news text, whereas they use earnings-call transcripts and earnings surprises. More fundamentally, these choices reflect their focus on using a live, real-time task to evaluate the benefits of agentic AI. By contrast, our goal is to understand how much realized market variation can be explained by observable information even in hindsight. We measure the explanatory power over a historical sample through leakage-controlled out-of-sample evaluation.

\section{Explaining realized returns: framework and non-news baselines}

We begin by specifying the empirical object for evaluating explanations of market fluctuations, which is a modern reinterpretation of Roll's $R^2$.
An explanation of a market movement must identify observations that are neither about nor derived from asset prices and must map those observations to returns in a stable way.
These requirements allow conventional structured variables, news-based reasons, and other candidate explanations to be evaluated in a common framework.
With this framework in place, we evaluate standard non-news explanations--macroeconomic statistics, cash flows, trading flows, subjective expectations, and combinations of these variables--and show that they leave most market variation unexplained.\footnote{\citet{kitchens2024news} follow Roll's original empirical setting, using an indicator for whether firm-specific news appeared to compare factor-model $R^2$ on news and no-news days. They conclude that selective disclosure resolves Roll's $R^2$ puzzle in that setting.}

\subsection{Criteria for explanations of market fluctuations}

We impose two criteria on candidate explanations of market fluctuations.
Explanations or reasons must be a set of observations 
\begin{itemize}
    \item[] Criterion 1. neither about nor derived from asset prices,
    \item[] Criterion 2. consistently mapped to asset price fluctuations.
\end{itemize}
The first criterion requires explanatory information to describe real-world events rather than concurrent asset prices.
It rules out explanations that use transformations of the same price fluctuations they seek to explain.
This distinction matters because several influential empirical exercises generate very high model fits while answering a different question from ours.
For example, the Campbell-Shiller decomposition of returns into cash-flow news and discount-rate news delivers an identity with $R^2 = 100\%$ up to the linear approximation.\footnote{\citet{martin2026option} make the same point: ``But the literature that follows Campbell (1991) does not address---and does not seek to address---Roll's puzzle, for the simple reason that realized returns are part of the state vector.'' In their own analysis, they explain realized returns on individual stocks using analysts' cash-flow forecast errors and revisions, along with option-implied expected returns and shocks to those expectations.}
As another example, a calibration exercise in \citet{gabaix2021search} attributes approximately 90\% of stock-return variance to flow shocks.
These exercises are useful and revealing in their respective settings.
However, their variance attributions do not establish explanatory power under our criteria.
Appendix \ref{appsubsec:variance_attribution} discusses the differences in detail.

The second criterion requires a stable mapping from explanatory observations to market fluctuations.
If cash-flow shocks are sufficient explanations, then comparable cash-flow growth shocks should be associated with comparable market reactions over time.
If market reactions differ systematically, then cash-flow shocks alone are not sufficient explanations; other observations are needed to determine the market's response.
Having a consistent mapping is necessary, though not sufficient, for showing that one has found the true reasons for market fluctuations.
It does not by itself establish whether the mapping from reasons to returns conforms to a particular economic model or channel.
We adopt a statistical approach to establishing the mapping to evaluate the explanatory power of candidate reasons, while recognizing that structural economic modeling remains necessary for discipline, interpretation, and identification of specific causal channels.

The criteria are deliberately permissive about researcher timing.
The explanations should be observable to market participants at the time of price formation, but researchers may recover those observations and estimate the mapping after the fact, i.e., ``with hindsight'' in Roll's original language. Explaining realized returns is therefore fundamentally different from predicting future returns: prediction requires the predictor information and the future return  to respect chronological order to prevent look-ahead bias.

\subsection{Estimating the mapping and evaluating the explanatory power}
\label{subsec:estimation_evaluation}

The empirical implementation of these criteria is a regression of returns on candidate explanations:
\begin{align}
    r_{t} = f(x_{t}; \theta) + \epsilon_{t},
\end{align}
where $r_{t}$ is the return at time $t$, $x_{t}$ is the vector of candidate explanations observed at time $t$, $f(\cdot; \theta)$ is the mapping from explanations to return with parameters $\theta$ subject to estimation, and $\epsilon_{t}$ is the unexplained residual.

\newcommand{\rhat}{\widehat{r}}
\newcommand{\ehat}{\widehat{\epsilon}}

The $R^2$ from this regression measures the explanatory power of the candidate explanations, defined as 
$R^2 = 1 - \Paren{\sum_{t\in\text{eval}} (r_{t} - \widehat{r}_{t})^2} /\Paren{\sum_{t\in\text{eval}} \Paren{r_{t} - \bar{r}}^2}
$, where 
$\widehat{r}_{t}
= f(x_{t}; \widehat{\theta})
$ is the fitted value and $\bar{r} = \frac{1}{|\text{eval}|}\sum_{t\in\text{eval}} r_{t}$ is the evaluation sample mean.\footnote{The denominator uses deviations from the evaluation-sample mean. This convention has little effect at the daily frequency, where the mean return is small, but it sets a slightly more demanding benchmark at the monthly frequency than a denominator that is not demeaned.}
	    
Criteria 1 and 2 together define a concurrent regression problem.
Criterion 1 requires the regressors to be free of price leakage: $x_t$ should not directly encode the return being explained.
Criterion 2 requires the regression to produce high explanatory power out of sample, indicating that the mapping from candidate reasons to returns is stable enough to account for realized market fluctuations.

The detailed implementation of this regression follows a standard machine-learning approach that facilitates comparison across candidate explanations.
We run concurrent time-series regressions of S\&P 500 returns ($r_{t}$) on high-dimensional observations of candidate reasons ($x_{t}$).
The representation $x_t$ can be high-dimensional and may include lagged values.
For the news-based model, the unit of observation is daily or monthly; for the non-news baseline comparison, the highest feasible frequency is monthly.

For baseline results, we estimate the regression with ridge-regularized linear models.
Specifically, the parameters solve the following optimization problem in the training set:
$
    \min_{\beta} \sum_{t \in \text{train}} \Paren{r_{t} - \beta \T x_{t}}^2 + \lambda \|\beta\|^2_2
$.
We also report specifications with lasso regularization, where the penalty term is $\lambda \|\beta\|_1$.

We evaluate the explanatory power of the estimated relationship with out-of-sample $R^2$ using a five-fold cross-validation split.
This is a strict out-of-sample evaluation.
Specifically, we split the sample period into five folds.
We estimate the mapping, including all hyperparameter tuning, on four folds and then apply the fitted model $f(\cdot, \widehat{\theta})$ to the left-out fold to produce out-of-sample fitted values $\widehat{r}_{t}$.
Repeating this procedure five times gives an out-of-sample fitted value for every observation in the full sample.
The reported $R^2$ is computed from these cross-validated fitted values across the full sample.
We calculate $R^2$ as $R^2 = 1 - \Paren{\sum_{t\in\text{test}} (r_{t} - \widehat{r}_{t})^2} /\Paren{\sum_{t\in\text{test}} \Paren{r_{t} - \bar{r}}^2}$, where $\widehat{r}_{t}$ is the fitted value and $\bar{r} = \sum_{t\in\text{test}} r_{t}/ |\text{test}|$ is the full-sample mean.
The full time series sample period is 2002--2024, aligned with our news-based reasons dataset.

Cross-validation is suitable here because the exercise explains realized returns with hindsight rather than predicting future returns before information arrives.
The goal of the cross-validation is to evaluate the stability of the mapping and to guard against overfitting in high-dimensional settings.
Expanding- or rolling-window evaluations commonly used in prediction are not used here because the estimation sample need not precede the evaluation sample in calendar time.

\subsection{Non-news explanations have weak explanatory power}
\label{subsec:nonnews}

We evaluate four existing sets of explanations that are commonly viewed as candidate drivers of market fluctuations: macroeconomic statistics, cash-flow variables, subjective expectations, and trading flows.
We show their explanatory power is weak or nonexistent when evaluated at the monthly frequency, while higher frequency observations are not available.
The implication is that explaining market fluctuations remains a difficult task, and that news-based reasons provide a materially more informative source of explanatory variation.

Table \ref{tab:non_news_summary} summarizes the baseline non-news explanations used for comparison.
We construct these time series from well-known literatures, data sources, and established procedures.
Construction details are provided in Appendix \ref{appsubsec:non_news}.
Two points are worth emphasizing.
First, many series are originally observed quarterly, including GDP and firm earnings.
The monthly data used here are obtained through interpolation or repetition, either by the data providers or by us following the same frequency-transformation procedure.\footnote{For macro and cash-flow variables, the data providers (FRED-MD and Shiller, respectively) have transformed the data to monthly frequency through interpolation or repetition. For subjective expectations and trading flows, we apply the same linear transformation procedure used by Shiller to convert quarterly data to monthly data. See details in Appendix~\ref{appsubsec:non_news}.}
This low frequency is an inherent limitation of structured data for explaining market returns.
Second, to satisfy criterion 1, we exclude financial-market variables such as stock indexes and exchange rates from FRED-MD.
Similarly, we exclude price-earnings ratios from Shiller's cash-flow data and retain only dividends and earnings.

\begin{table}[!t]
    \caption{Non-news explanations of market fluctuations}
    \label{tab:non_news_summary}
    \centering
    \begin{tabular}{>{\raggedright\arraybackslash}p{2.8cm} 
                    >{\raggedright\arraybackslash}p{12cm} 
                    >{\raggedright\arraybackslash}p{0.5cm}}
        \toprule
        Explanation & Content, main reference, and source & dim.\ \\
        \midrule
        Macro & Macroeconomic statistics: output, labor, etc., FRED-MD dataset excluding financial market indexes \citep{mccracken2016fred} & 99 \\[20pt]
        Cash flow & Dividends and earnings, aggregated at the S\&P 500 level, from \citeauthor{shiller2015irrational}'s \citeyearpar{shiller2015irrational} webpage & 4 \\[20pt]
        Subjective expectations & Analyst forecasts of dividend and earnings, aggregated at the S\&P 500 level \citep{delao2021subjective} & 4 \\[20pt]
        Trading flows & Net equity flows of various investor sectors, such as banks, pension funds, and broker dealers, constructed following \citet{gabaix2021search} & 12 \\%
        \bottomrule
    \end{tabular}
    \footnotesize\justifying\noindent
    Note: The table summarizes the structured non-news explanation groups used as benchmarks for aggregate U.S.\ stock market returns. The last column reports the number of variables in each group.
\end{table}

All non-news explanations considered in this paper are observed at monthly or lower frequency, whereas news can be used at the daily frequency.
Even at the monthly frequency, non-news explanations produce $R^2$ values far below those of news-based reasons, and most are negative.
This shows that conventional structured data explain little realized market variation (see Figure \ref{fig:comp_non_news}).

We also examine alternative specifications, including lagged variables to capture dynamic relationships and ensembles of the four non-news explanations formed either by stacking variables or averaging fitted values from individual models.
None of these variations materially improves explanatory power.

Among the four benchmarks, trading flows have the highest explanatory power, with a monthly $R^2$ of roughly 10\%.
This suggests that flows explain price fluctuations to some extent, consistent with mechanisms in which flows have price impact or correlate with investor sentiment and other fundamental drivers of market movements \citep{shleifer1990noise, gabaix2021search}.\footnote{At the same time, flows may respond to price movements, for example through performance-chasing behavior by some investor sectors \citep{lou2012flow, ben2024discontinued}.
This raises a possible violation of criterion 1, especially because flow data are observed at low frequency and part of the measured flow may occur after the price movement it is intended to explain.
Such reverse causality could affect the measured $R^2$.
Addressing flow endogeneity and providing a definitive estimate of its effect on the measured $R^2$ are beyond our objective.
The literature on identifying the price-impact coefficient, or elasticity, has developed methods to address flow endogeneity; see, for example, \citet{gabaix2024granular, gabaix2021search, haddad2025casual}.}

The other three explanations (macroeconomic statistics, cash flow, and subjective expectations) do not have robust explanatory power for market fluctuations.
These findings may simply reflect that the measured variables have little connection to market movements.
Alternatively, macroeconomic and cash-flow developments may indeed drive returns, but the market may respond when investors learn about those developments, before they appear in the observed data and with delays that vary across events.\footnote{\citet{chahrour2026exchange} document a similar timing pattern in foreign exchange markets, where real exchange rates move on news about future U.S.\ productivity before the associated macroeconomic changes show up in the statistics. They infer the news shocks with a VAR estimated on macroeconomic and financial data, rather than from news text.}
Even when the timing aligns, the market response to a given change in these variables may be complex and time-varying and depend on economic and financial conditions that our specifications do not capture.
Regardless of which interpretation applies, the main conclusion remains unchanged: standard structured variables provide only weak explanations for realized market variation, even under the favorable hindsight design.

\section{Extracting reasons from news}

This section describes how we distill the many events described in financial news into a structured set of textual reasons that can be used to explain market fluctuations.
The premise is that financial news media identify and summarize the events that matter for markets on a given day.
We focus on a single Wall Street Journal article per trading day that provides broad coverage of market movements and discusses the reasons for those movements, a source whose format and quality are relatively consistent over the sample period.

We extract the specific statements that, according to the article, are associated with or explain the day's market movements.
Generative AI reads each article, identifies the events described in it, and locates the exact excerpts that the article presents as explanations for market movements.
We then validate the extracted reason text to ensure that it does not leak descriptions of financial-market movements themselves, as required by criterion 1.
The result is a daily dataset of newspaper-stated reason excerpts that can be merged with realized market outcomes.

\subsection{``Today's Markets'' articles from the Wall Street Journal}

We use a single article per trading day from the Wall Street Journal (WSJ) ``Today's Markets'' column, which provides broad coverage of major market movements.
The ``Today's Markets'' column, sometimes also labeled as ``Monday's Markets'', ``Tuesday's Markets'', ..., or ``U.S.\ Markets'' appears in the WSJ on most trading days. 
Table \ref{tab:article_examples} shows examples of the articles in our sample.
These articles have a consistent format and style over the sample period.
They typically include a summary of broad equity-index movements and, in many cases, also report price changes in bond, commodity, currency, or international markets.
As illustrated in the examples, many articles state the DJIA or S\&P 500 closes higher or lower by how many points or percentage, or describe the market movements in more loose terms such as ``Global stocks dropped Tuesday'' or ``U.S. stocks rallied to new all-time highs''.

Most importantly, the articles often organize market movements around real-world events and explicitly link price changes to those events.
For example, the articles state that ``Stocks headed higher Friday {\it after} the November jobs report eased fears of an economic slowdown'', or that ``Global stocks dropped Tuesday {\it as} trade tensions between the U.S. and China intensified''.
In these cases, words such as ``{\it after}'' and ``{\it as}'' explicitly connect market movements to real-world developments.
In other cases, the connection is expressed through more nuanced language and must be inferred from context.
These articles therefore provide a natural corpus for tracing the stated reasons for market movements.

These articles usually do not provide deep analysis of a single event, unlike many other WSJ articles and columns.
Instead, they summarize the market day and attribute movements to a set of salient developments.
Their formulaic language helps the extraction task.
At the same time, expanding to a richer news corpus could improve explanatory power and extend the framework beyond broad market-index movements.

We index each article by the trading day covered by its reporting content so that it can be merged with market data.
The articles are published in the print edition on the morning after the trading day.
For example, an article recapping Monday's market movements is published in the Tuesday print edition, though the same article is often available online shortly after the Monday close.
We construct the trading-day index from publication metadata and verify it by comparing article content with merged price and return data.

We compile a dataset of a single article per trading day for the sample period from 2002 to 2024. Nearly all trading days in the sample have a corresponding article.

\newcommand{\hlreasonf}[1]{{\sethlcolor{cyan!25}\hl{#1}}}
\newcommand{\hlreason}[1]{{\sethlcolor{cyan!50}\hl{#1}}}

\begin{table}[!t]
\centering

\begin{minipage}[t]{0.32\linewidth}
\centering
{\normalsize\bfseries 
    Stocks Rise After Jobs Report
\par}
\vspace{0.25em}
{\scriptsize Worth Civils, Dec 08, 2006\par}
\vspace{0.6em}
\raggedright\justifying %
\setlength{\parindent}{0pt}
\setlength{\parskip}{3pt}
\scriptsize

    Stocks headed higher Friday after \hlreason{the November jobs report eased fears of an economic slowdown without giving the Federal Reserve much of a reason to raise interest rates any further}. But at the same time, the Fed is unlikely to cut rates any time soon.

    As traders weighed these outcomes, stocks fluctuated. Up some 50 points earlier in the day, the Dow Jones Industrial Average closed 29.08, or 0.2\%, higher at 12307.49, led by Citigroup, C which added 2.3\% amid \hlreason{rumors of a management shake-up and sale of some of its businesses}. But weighing on the blue-chip average was 3M, MMM which fell 1.3\% after \hlreason{Prudential downgraded the stock to 'neutral' from 'overweight' on valuation concerns}. 
    ...

\end{minipage}
\hfill
\begin{minipage}[t]{0.32\linewidth}
\centering
{\normalsize\bfseries 
    Stocks Set Record Highs After Fed Statement
\par}
\vspace{0.25em}
{\scriptsize Alexandra Scaggs, Sep 18, 2013\par}
\vspace{0.6em}
\raggedright\justifying 
\setlength{\parindent}{0pt}
\setlength{\parskip}{3pt}
\scriptsize

    U.S. stocks rallied to new all-time highs as \hlreason{the Federal Reserve surprised markets by announcing that its stimulus measures would stay in place}.

    Treasury prices also rose, with the benchmark 10-year yield settling at 2.706\%, its lowest since Aug.\ 12. Prices of mortgage-backed securities leapt even more. Oil futures pushed higher, and the dollar slumped against most major currencies.

    \hlreason{The Federal Open Market Committee said in its post-meeting policy-setting statement that it would continue to buy \$85 billion in bonds each month}. \hlreason{Investors expected the central bank to pare its bond-buying programs by between \$10 billion and \$15 billion at this meeting}.
    ...
\end{minipage}
\hfill
\begin{minipage}[t]{0.32\linewidth}
\centering
{\normalsize\bfseries 
    Escalating Trade Threats Slam Global Shares
\par}
\vspace{0.25em}
{\scriptsize David Hodari, Allison Prang, \\Jun 19, 2018\par}
\vspace{0.6em}
\raggedright\justifying %
\setlength{\parindent}{0pt}
\setlength{\parskip}{3pt}
\scriptsize
    Global stocks dropped Tuesday as \hlreason{trade tensions between the U.S. and China intensified}, sparking selling from Shanghai to New York.  

    The declines came after \hlreason{President Donald Trump raised the stakes in Washington's trade conflict with China, asking his administration on Monday to draw up a fresh list of Chinese goods worth \$200 billion on which to impose tariffs}. 

    That move marked the latest in a series of back-and-forth measures between the countries that left investors around the globe worried the conflict would escalate into an all-out trade war. 
    ...

\end{minipage}

\caption{Examples of news articles with extracted reasons highlighted}
\label{tab:article_examples}
\vspace{4pt}
\begin{minipage}{0.98\textwidth}
\footnotesize
Note: The table reports excerpts from Wall Street Journal market-recap articles. Highlighted text indicates passages extracted by the AI pipeline as stated reasons for market movements. Direct descriptions of the market movement itself are excluded from the reason text.
\end{minipage}
\end{table}

\subsection{Extracting reason text from news articles with AI}

We use generative AI to extract from each article the specific text that explains the day's market movements while excluding language that directly describes the movements themselves.
The design follows criterion 1: the explanatory variable should contain the stated reason, not the return or a verbal description of the return.

We design a two-step prompting pipeline for extraction and validation.\footnote{The full system prompts are reported in Appendix \ref{appsubsec:prompts}.}
The first step prompts the LLM to read one article at a time and return excerpts that are reasons for financial asset price movements.
Specifically, the prompt instructs the LLM to identify pairs of excerpts: one stating the reason and the other stating the corresponding price movement.
The key instruction is to output a JSON list with separate \texttt{reason} and \texttt{price\_movement} fields, and to ``only use direct quotes from the news article.''
Although our object of interest is the reason text, this paired output schema guides the LLM to identify the reason-price association structure in the article.
It also helps separate reason language from market-movement language, preventing descriptions of price changes from being mixed into the reason extractions.
We explicitly instruct the LLM to extract exact quotes from the article so that it does not inject its own interpretation or reasoning.

The second step guards against leakage from extracted text that still contains information about market movements themselves.
Such leakage would violate criterion 1 and mechanically inflate explanatory power.
We supply the reason text from the first step to the LLM one excerpt at a time and prompt it to flag any excerpt that contains information about financial-market prices.
The prompt instructs the LLM to search for descriptions of market indices, sectors, individual stocks, and other assets such as bonds, commodities, and currencies.
It also instructs the LLM to flag subtle price-movement language such as ``pulled back'', ``rebounded'', or ``buoyed''.
Flagged excerpts are removed from the final reason-text dataset.
Excerpts that pass this validation step contain stated non-price reasons, not price movements or verbal transformations of returns.

Table \ref{tab:article_examples} shows examples of the extracted reason text (highlighted in blue) in the original article.
These examples illustrate that the extracted reasons describe real-world events, such as economic data releases and policy or trade developments, and are linked to market movements by the article.
They also show that the extraction process excludes direct descriptions of market movements themselves, such as ``Stocks headed higher Friday'' or ``sparking selling from Shanghai to New York''.
Section \ref{sec:reason_content} below provides a detailed analysis of the content of the extracted reason text, including more examples and a taxonomy of the reasons.

This extraction is a plain reading comprehension task for the LLM,
rather than an exercise of financial reasoning or inference.
Generative AI makes it feasible to process heterogeneous articles in which market movements and stated reasons are intertwined in unstructured language, and to do so consistently across the full sample.
At this stage, the LLM is not treated as an independent source of information and is not asked to infer how markets work.
The financial media state the perceived relationship between market movements and their reasons; the LLM extracts that stated relationship in a directly verifiable form.
Human experts can simply perform the same task for a small number of articles to verify the extraction, but not at the required scale.
\footnote{Our use of LLMs at this stage is therefore distinct from work that evaluates whether LLMs can predict financial market movements better than human experts, as in \citet{chen2025chatgpt, lopez2023can}.}

\section{The content of extracted reasons}
\label{sec:reason_content}

\subsection{Basic properties of the extracted reasons}

\begin{figure}[!t]
  \centering

  \begin{subfigure}[t]{0.5\linewidth}
    \centering
    \caption{Snippet length}
    \includegraphics[width=\linewidth]{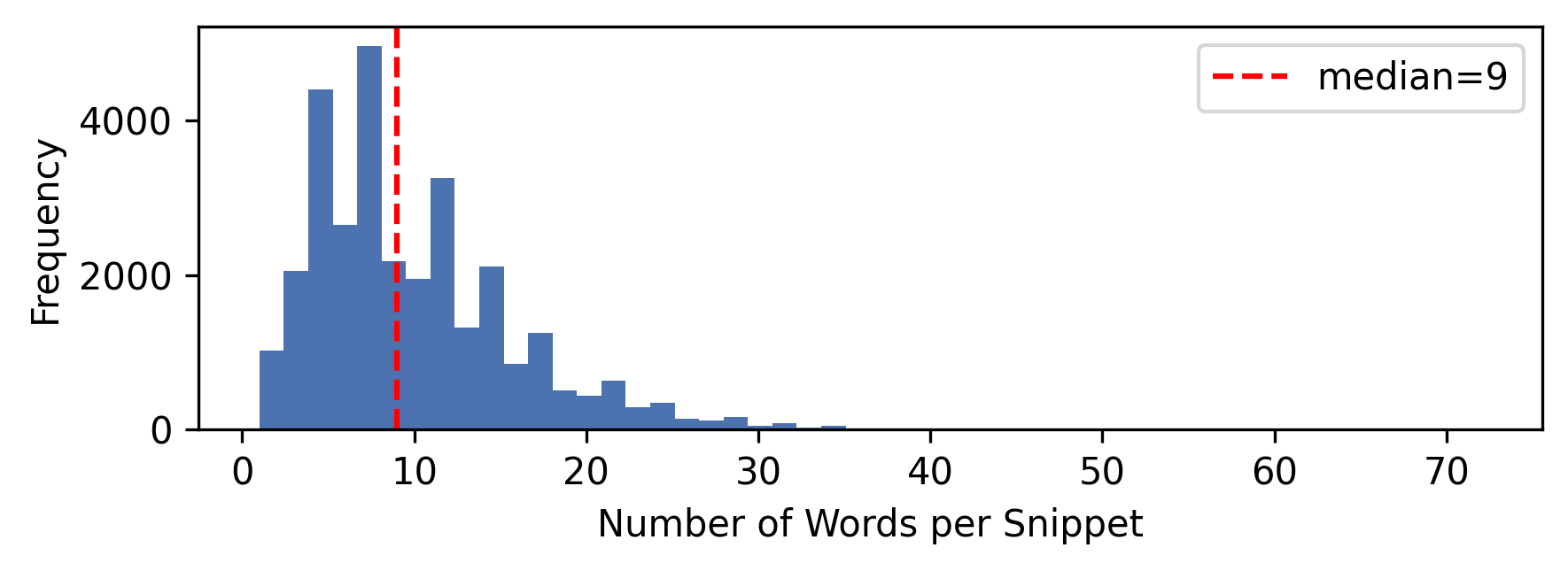}
    \label{fig:dist_snippet_length}
  \end{subfigure}\hfill
  \begin{subfigure}[t]{0.5\linewidth}
    \centering
    \caption{Snippets per day}
    \includegraphics[width=\linewidth]{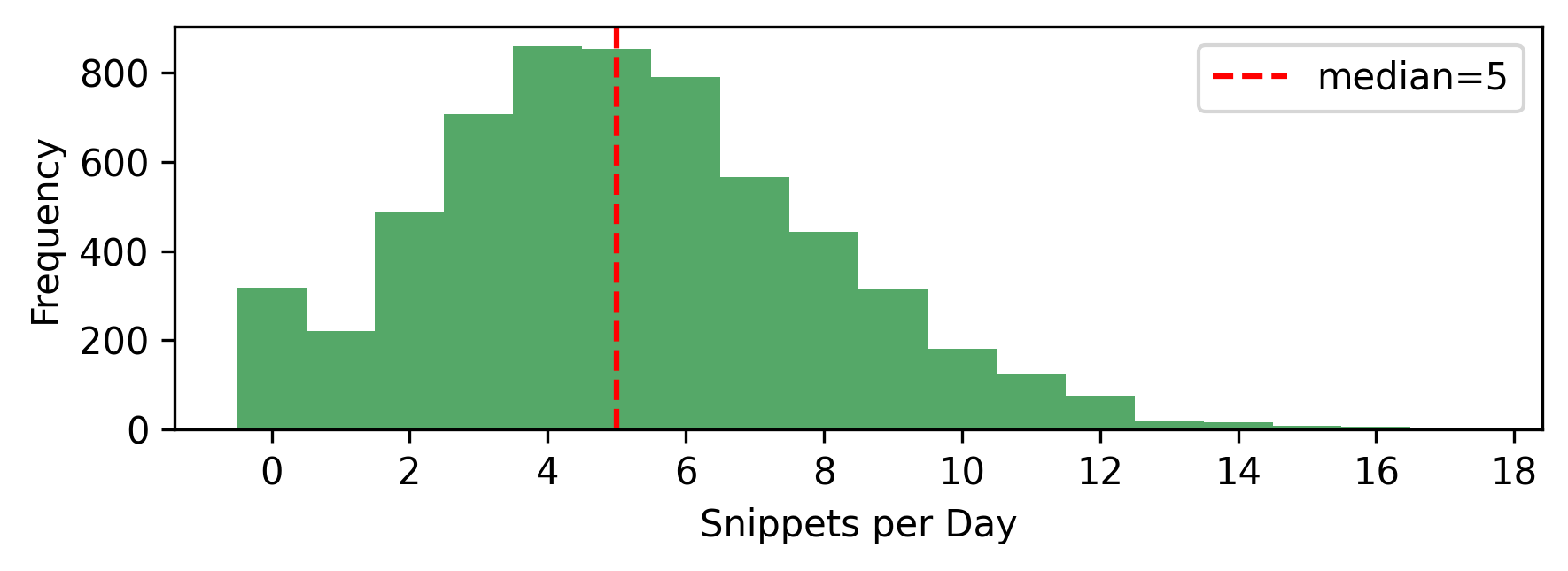}
    \label{fig:dist_snippets_per_day}
  \end{subfigure}

  \begin{subfigure}[t]{0.5\linewidth}
    \centering
    \caption{Composite length}
    \includegraphics[width=\linewidth]{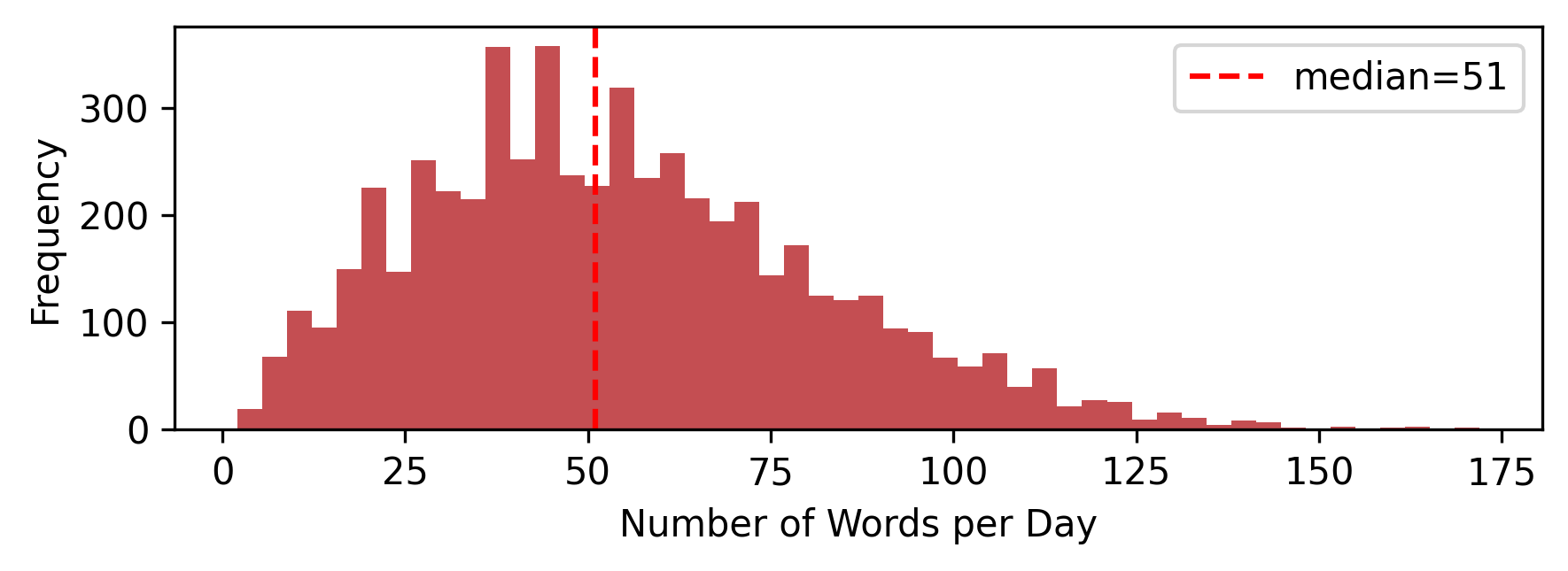}
    \label{fig:dist_composite_length}
  \end{subfigure}\hfill
  \begin{subfigure}[t]{0.5\linewidth}
    \centering
    \caption{Days with no reason snippets}
    \includegraphics[width=\linewidth]{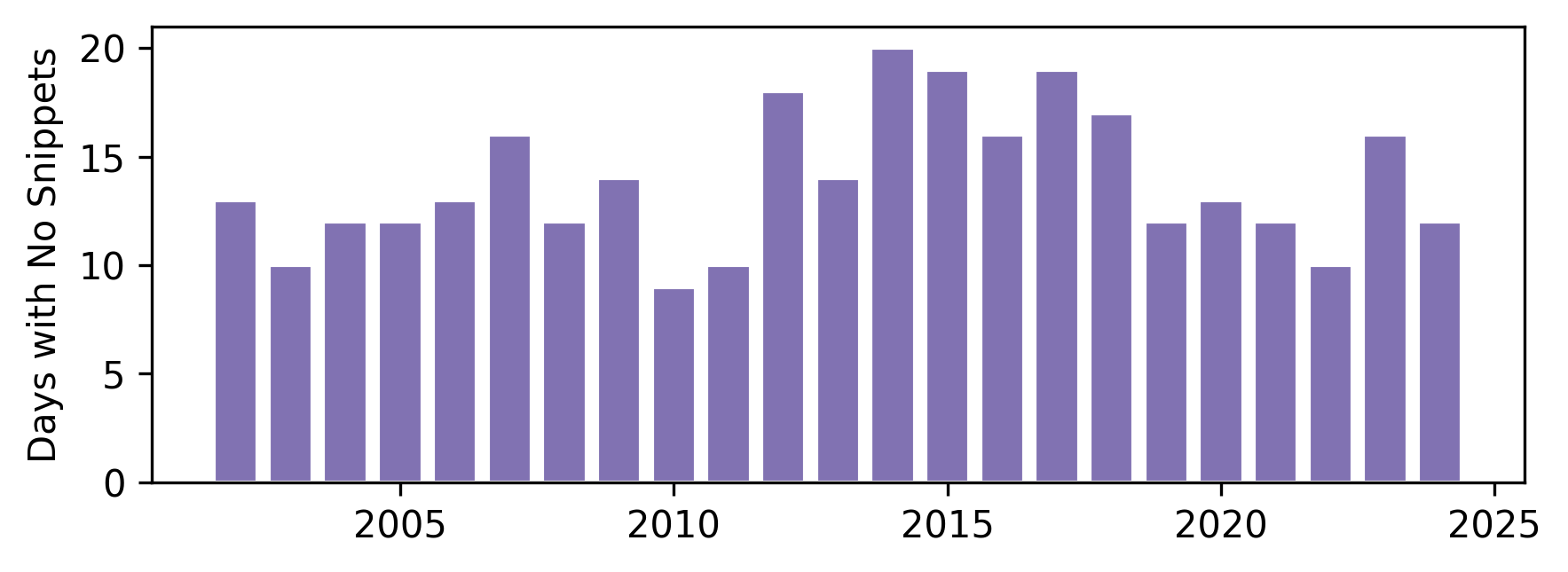}
    \label{fig:dist_no_snippet_days}
    
  \end{subfigure}
    \par
    \justifying  
    \footnotesize
    \noindent
    Note: The figure summarizes the distribution of extracted reason text in the Wall Street Journal sample. Panel A reports the number of words per reason snippet. Panel B reports the number of snippets extracted per trading day. Panel C reports the number of words in each daily reason composite, defined as the concatenation of all snippets on the day. Panel D reports days with no extracted reason snippets after the leakage screen. On days with no reason snippets, the implied return is set to zero for both embedding regression and zero-shot inference.
  \caption{Basic properties of the extracted reasons}
  \label{fig:overall}
\end{figure}

The extraction procedure produces {\it reason snippets}: excerpts from the original article that directly explain the market movements of the day.
A trading day can have multiple snippets, or none if the article contains no qualifying reason after the leakage screen.
For return-reason regressions, we concatenate all snippets from a day into a single {\it reason composite}.
Figure \ref{fig:overall} reports the basic distributional properties of snippets and composites, including snippet length, the number of snippets per day, composite length, and the number of days with no extracted reason.

\subsection{Textual reason taxonomy method}
\label{sec:clustering_method}

The extracted reason text covers a wide range of topics and events that influence financial markets.
We organize the reasons into a hierarchical taxonomy: reason snippets are assigned to clusters and sub-clusters, which allows us to describe the content of the extracted reasons and later evaluate the explanatory power of different reason types.

We form the hierarchical structure and assign reason text to it using generative AI with human-in-the-loop support.
The procedure has three steps:
1) summarize candidate cluster names from sampled reason snippets,
2) assign each reason snippet to a cluster, and
3) validate the cluster names and assignments with human feedback.\footnote{The procedure follows the iterative prompt design in the narrative formation process of \citet{zhou2025professional}, but we adapt it by
a) summarizing from random batches rather than sampling by time periods (month, year, and five-year),
b) involving more human validation, and 
c) forming the hierarchical structure top down (first meta-clusters then sub-clusters) rather than forming narratives and then grouping them.}
The first step generates candidate cluster names by prompting the LLM to summarize common themes from 12 sample batches of 2000 %
reason snippets to overcome context-window limitations that prevent processing all snippets at once.
We draw shuffled subsets of reason snippets and prompt the LLM to propose at most 10 cluster names by identifying recurring themes in each batch.
We then pool the proposed cluster names, prompt the LLM to deduplicate and merge similar themes, and ask it to return the most frequent and meaningful cluster names.
The proposed names are reviewed by humans, yielding a final list of 11 meta-cluster names.
The second step loops over all snippets and prompts the LLM to assign each snippet to one cluster from the list, or to mark it as ``unassigned'' if it does not fit.
The third step reviews the cluster names and assignment results.
If the unassigned group is too large, we rerun the first step on that group to generate candidate names; if assignments are unsatisfactory, we adjust cluster names to better fit the snippets.
Once the cluster list is satisfactory, we rerun the assignment step for all snippets.

After forming meta-clusters, we repeat the same procedure within each meta-cluster to form sub-clusters.
Human guidance is also used in sub-clustering to ensure coherence and balanced cluster sizes.
For smaller meta-clusters with fewer snippets, no sub-clusters are formed because the meta-cluster is already sufficiently specific for displaying the content of extracted reasons.

\begin{figure}[!t]
    \centering
    \caption{Validating reason clustering via t-SNE visualization}
    \vspace*{-10pt}
    \label{fig:cluster_tsne}
    \includegraphics[width=0.7\textwidth]{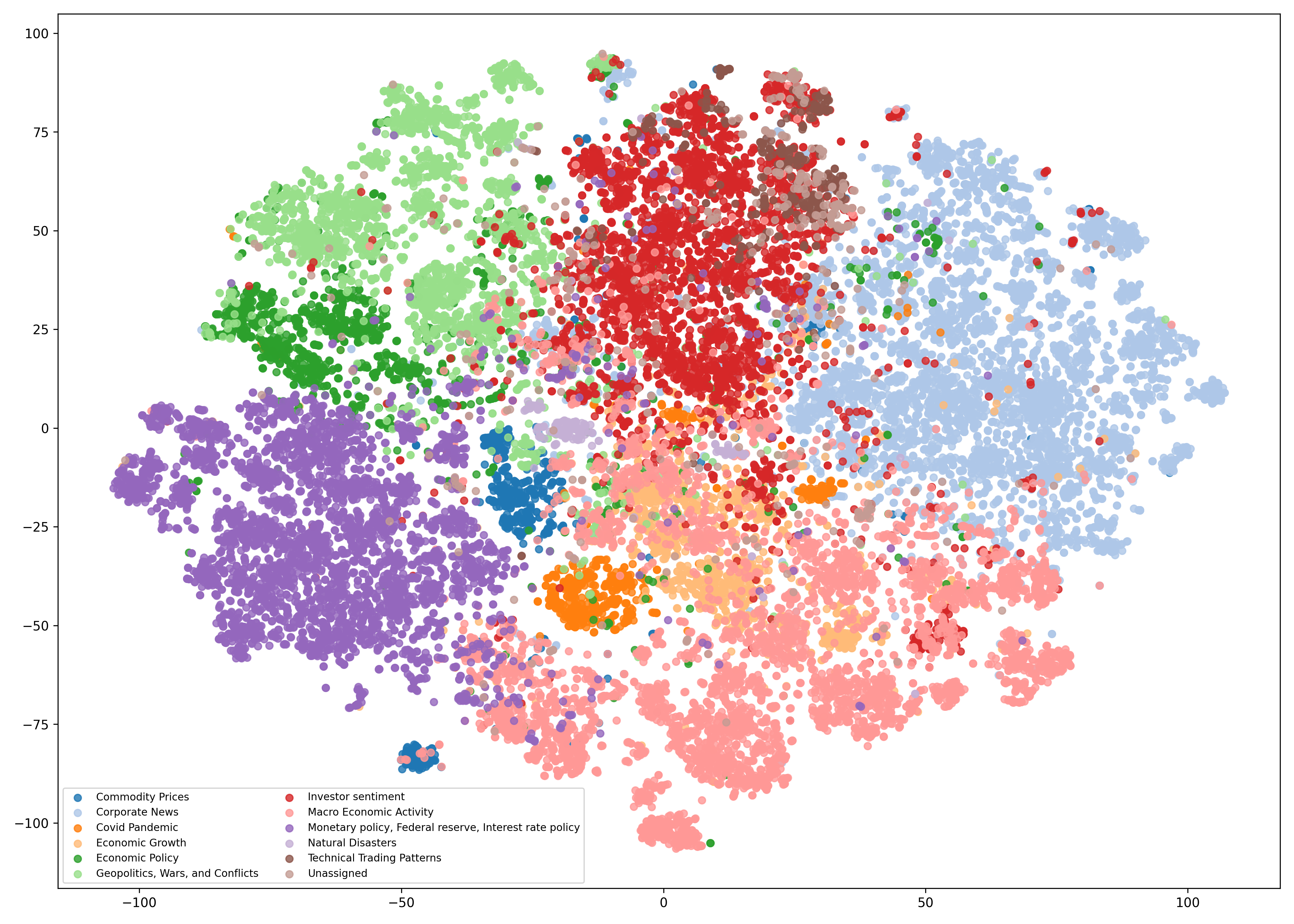}
    \par\footnotesize\justifying
    \noindent
    Note: Each point represents an extracted reason snippet from the Wall Street Journal sample. Snippets are represented by 3,072-dimensional language-model embeddings and projected into two dimensions using t-SNE. Colors indicate the meta-cluster assignments produced by the AI-assisted clustering procedure described in Section \ref{sec:clustering_method}.
\end{figure}

We validate the clustering procedure with t-SNE visualization.
T-SNE ($t$-distributed Stochastic Neighbor Embedding) is a standard dimensionality reduction technique for visualizing high-dimensional data such as textual embeddings \citep{maaten2008visualizing}.
We apply t-SNE to map each $3,072$-dimensional reason embedding to a two-dimensional point so that semantically similar snippets are geometrically close in the visualization.
Figure \ref{fig:cluster_tsne} shows that the clusters (assigned with human-in-the-loop prompting) are well separated in the embedding space, and that reasons in the same cluster indeed tend to cluster together geometrically.
This provides a visual check that the taxonomy captures meaningful structure in the reason text.
The figure also reveals economically natural imperfections: most ``investor sentiment'' reasons cluster together, but some blend into ``macroeconomic activities'' and ``corporate news,'' reflecting the fact that sentiment language often appears in the context of macroeconomic or corporate developments.\footnote{We also validate sub-clusters with the same t-SNE visualization applied within each meta-cluster. The results are similar and available upon request.}

\subsection{Taxonomy results: the interpretable content of extracted textual reasons}

\begin{figure}[!t]
    {
        \raggedright
        \caption{Reason taxonomy and time-series frequency of meta-clusters}
        \label{fig:cluster_treeplot}
        \vspace*{-10pt}
        \includegraphics[
            width=0.7\textwidth,
            trim={0 0 0 4cm}, 
            clip %
        ]{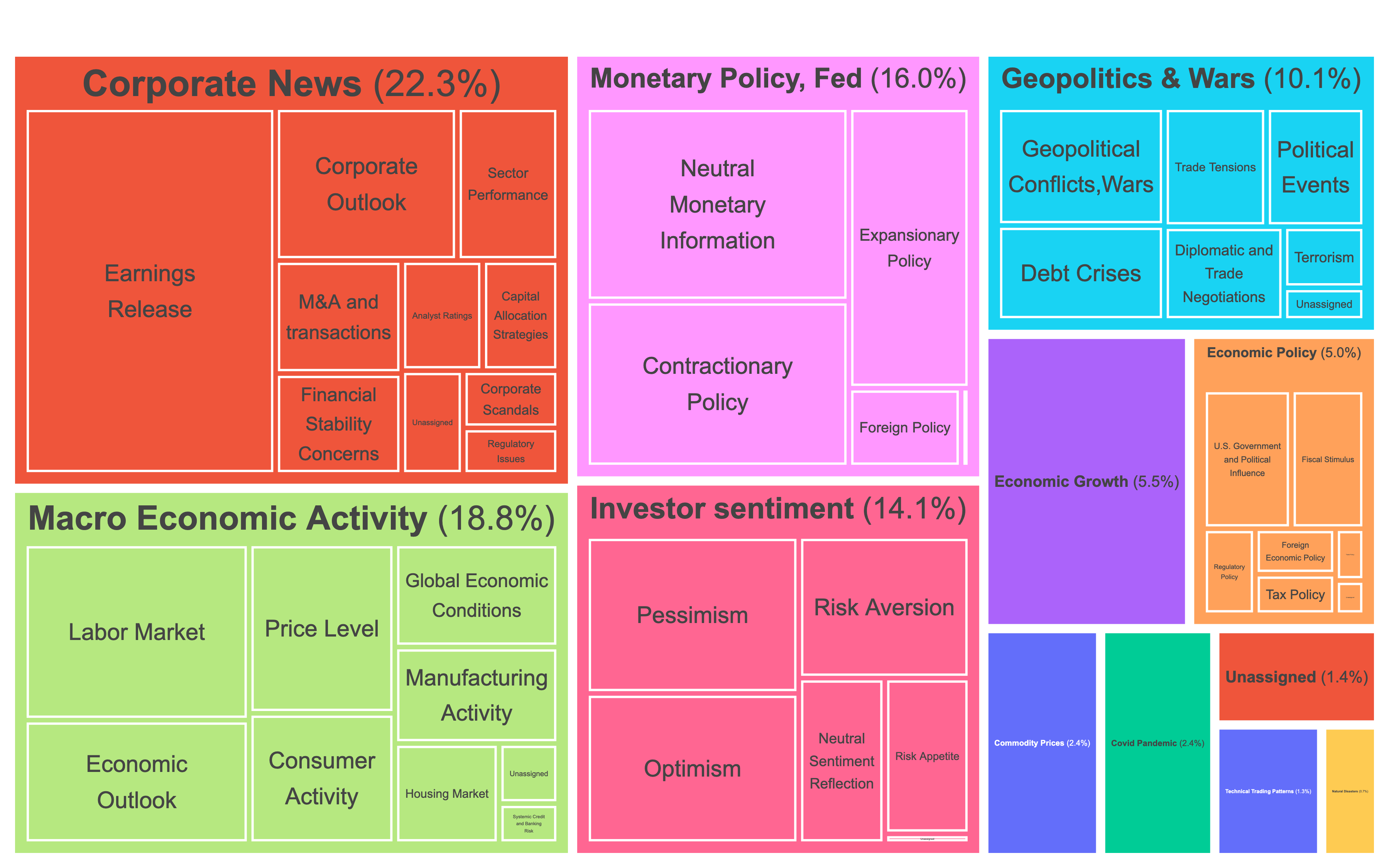}%
        \vspace*{-10pt}
        \includegraphics[width=1\textwidth]{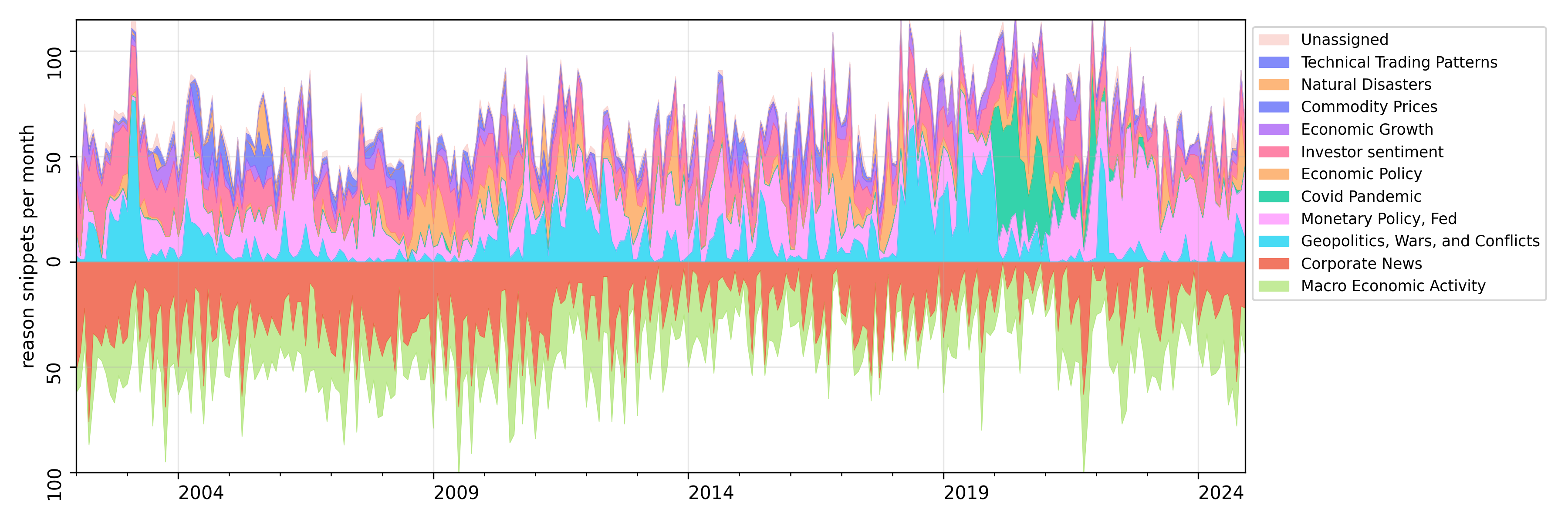}
        \vspace*{-10pt}
    }
    \par
    \justifying
    
    \footnotesize
    \noindent
    Note: Panel A reports the hierarchical taxonomy of extracted reason snippets. Block sizes are proportional to the number of snippets assigned to each cluster or sub-cluster. Panel B reports monthly snippet counts for each meta-cluster over the sample period. Colors are matched across panels.
\end{figure}

The clustering procedure organizes the extracted reason snippets into a hierarchical structure consisting of 12 meta-clusters, most of which are further divided into sub-clusters.
Each reason snippet is assigned to a single sub-cluster, which in turn belongs to one meta-cluster.
Figure \ref{fig:cluster_treeplot} displays the hierarchical structure with the name and relative size of each cluster and sub-cluster. 
This structure makes the content of the extracted reasons easier to inspect.
Table \ref{tab:reason_cluster_examples} reports an extensive list of original reason snippets organized by cluster and sub-cluster, allowing readers to see the concrete language behind the taxonomy.

The bottom panel of Figure \ref{fig:cluster_treeplot} shows the time-series evolution of each meta-cluster, measured by the number of reason snippets assigned to the cluster each month.
The figure reinforces that news-based reasons are not a static dictionary of labels.
Corporate news is a persistent component of market explanations and displays the regular quarterly rhythm associated with earnings cycles.
Geopolitical reasons arrive in waves, including terrorism and Middle East conflicts in the early 2000s, the European debt crisis, the first Trump administration's trade disputes, the Russia-Ukraine war, and the Israel-Gaza conflict.
COVID-related reasons appear sharply around the pandemic and then fade.
Monetary-policy reasons become especially prominent after the pandemic, when inflation and central-bank tightening dominate market interpretation.

The remainder of this section walks through the major clusters, emphasizing their connection to real-world events and their usefulness for explaining financial-market dynamics.

\newcommand{\firstcolformat}[1]{%
  \multirow[t]{3}{*}{\parbox[t]{1\linewidth}{#1}}%
}
\newcommand{\twocoltitle}[1]{%
          \multicolumn{2}{c}{\normalsize{#1}} \\[4pt]}

\begin{table}[!t]
	\centering
	\scriptsize\singlespacing
	\renewcommand{\baselinestretch}{0.89}\selectfont
	\setlength{\extrarowheight}{2pt} %
	\caption{Part A: Example reasons and clusters (four largest meta-clusters)}
	\label{tab:reason_cluster_examples}
	\vspace{-6pt}
	\begin{tabular}{p{0.20\linewidth}p{0.75\linewidth}}
		\toprule
		\twocoltitle{Meta-cluster (percentage)}
		Sub-cluster (percentage) & Example reason snippets \\
		\midrule
		\twocoltitle{Corporate news (22\%)}
		
\firstcolformat{Earnings Release (47\%)} & Corporate earnings so far have come in above expectations, with 87\% of companies posting  stronger-than-expected profits and more than two thirds beating revenue expectations\\[4pt]

\firstcolformat{Corporate Outlook (14\%)} & Wal-Mart Stores said that pre-Christmas sales hit expectations\\
& Apple, which unveiled its latest iPhones last week\\[4pt]

\firstcolformat{M\&A and transactions (7\%)} & The frenzy surrounding LinkedIn’s initial public offering on Thursday\\
& auto-parts maker ArvinMeritor is making an unsolicited bid for Dana\\[4pt]

\firstcolformat{Analyst Ratings (5\%)} & Merrill Lynch analyst Joseph Osha issued negative comments on the chip maker and cut his first-quarter and 2006 earnings expectations\\

\ignorespacesafterend
		\\
		\twocoltitle{Macroeconomic news (19\%)}
		
\firstcolformat{Labor Market (25\%)} & the number of Americans applying for unemployment benefits fell to 787,000 last week, according to data from the Labor Department\\[4pt]

\firstcolformat{Economic Outlook (17\%)} & another weak round of U.S. economic data
\hspace{10pt} improving economic data\\[4pt]

\firstcolformat{Consumer Activity (12\%)} & a weak consumer-confidence reading 
\hspace{10pt} strong retail-sales figures\\[4pt]

\firstcolformat{Global Economy (11\%)} & a report that China’s oil imports fell in January and February\\[4pt]

\firstcolformat{Price Level (15\%)} & Consumer prices rose in April at a slower pace than the previous month, but still faster than economists had expected\\[4pt]

\firstcolformat{Manufacturing Activity (10\%)} & manufacturing activity came in weaker than expected\\[4pt]

\firstcolformat{Housing Market (7\%)} & Home resales fell 4.3\% in August from July’s pace, the National Association of Realtors said\\[4pt]

\firstcolformat{Systemic Credit and Banking Risk (1\%)} & a report came out showing increased delinquency rates on subprime mortgages worries about subprime lending\\[4pt]
\ignorespacesafterend
		\\
		\twocoltitle{Monetary policy, Federal reserve, Interest rate policy (16\%)}
		
\firstcolformat{Neutral Monetary Information (37\%)} & Wall Street began to handicap monetary policy under Bernanke\\
& Mr. Poole said in an afternoon speech that interest rates seem ‘about right’ to deal with the current levels of inflation in the economy\\[4pt]

\firstcolformat{Contractionary Policy (32\%)} & San Francisco Fed President John Williams said Friday the central bank should raise interest rates twice this year\\
& fears that it will raise rates once the economy strengthens\\
& suggested the Federal Reserve would remain on a slow path of interest-rate increases\\[4pt]

\firstcolformat{Expansionary Policy (25\%)} & bets that Fed stimulus will power stocks higher\\
& investors weighed the likelihood that the Fed will continue to emphasize low interest rates coming out of its policy-setting meeting Wednesday\\
& reinvigorating hopes the Fed will cut interest rates in the coming months\\%

\ignorespacesafterend
		\\
		\twocoltitle{Investor Sentiment (14\%)}
		
\firstcolformat{Optimism (27\%)} & investors regained confidence after skirting several dire scenarios\\
& investors showed renewed interest in megacap technology companies\\
[4pt]

\firstcolformat{Pessimism (28\%)} & The poor showing surprised investors\\
& The warnings dashed some of the stock market’s highflying expectations\\[4pt]

\firstcolformat{Risk Aversion (21\%)} & some players apparently forced out of unprofitable positions to raise cash\\
& investors struck a cautious stance on the economy \hspace{10pt} renewed investor caution\\[4pt]

\firstcolformat{Risk Appetite (11\%)} & investors’ risk appetite grew\hspace{10pt} the so-called risk trade was back in vogue\\[4pt]

\firstcolformat{Neutral Sentiment (12\%)} & in a quiet, preholiday trading session \hspace{10pt} her comments did not include any surprises\\

\ignorespacesafterend
		\\[-8pt]
		\bottomrule
	\end{tabular}
\end{table}

\begin{table}[!t]
	\centering
	\scriptsize\singlespacing
	\renewcommand{\baselinestretch}{0.9}\selectfont
	\setlength{\extrarowheight}{2pt} %
	\addtocounter{table}{-1}
	\caption{Part B: Example reasons and clusters (5th to 11th meta-clusters)}
	\vspace{-6pt}
	\begin{tabular}{p{0.20\linewidth}p{0.75\linewidth}}
		\toprule
		\twocoltitle{Geopolitics, Wars, and Conflicts (10\%)}
		
\firstcolformat{Conflicts and Wars (25\%)} & reports of supply disruptions from Iraq\hspace{10pt} elevated tensions between the West and Russia\\[4pt]

\firstcolformat{Debt Crises (20\%)} & amid worries that debt-heavy governments like Italy’s could weigh on the banking sector\\[4pt]

\firstcolformat{Trade Tensions (16\%)} %
& President Trump said China wants to strike a trade deal, signaling a potential de-escalation in trade tensions between the world’s two largest economies\\

\firstcolformat{Diplomatic and Trade Negotiations (14\%)} & Some investors expect an agreement will lift the outlook for the global economy\\
& Investors welcomed news that senior U.S. and Chinese officials were committed to carrying out the phase-one trade accord signed in January\\[4pt]
\ignorespacesafterend
		\\
		\twocoltitle{Economic Policy (5\%)}
		
\firstcolformat{Fiscal Stimulus (28\%)} & White House officials and House Speaker Nancy Pelosi opened the door to passing a coronavirus relief package\\[4pt]

\firstcolformat{U.S. Government and Political Influence (34\%)} & reports of progress over the debt ceiling
\hspace{10pt}
a skeptical reception on Capitol Hill for the Treasury Department’s proposed rescue of Wall Street\\

\firstcolformat{Regulatory Policy (12\%)} & concern surrounding the White House’s plans to overhaul financial regulation\\[4pt]

\ignorespacesafterend
		\\
		\twocoltitle{Five smallest clusters without sub-clusters}
		
\firstcolformat{Economic Growth (5\%)} & once again dimming hopes for a speedy recovery in the credit markets\\
& cheered investors hoping for a robust economic recovery\\[4pt]

\firstcolformat{Covid Pandemic (2\%)} & promising vaccine developments\\
& States from Florida to California to Arizona to Texas are recording increases in coronavirus cases after lifting restrictions put in place to limit the virus’s spread\\[4pt]

\firstcolformat{Commodity Prices (2\%)} & Some investors and analysts said that while falling oil prices are likely to continue to hurt energy companies\\[4pt]

\firstcolformat{Technical Trading Patterns (1\%)} & pre-placed orders are triggered via trading firms’ computer systems\\
& several large block trades totaling nearly 40 million shares hit the tape around 1 p.m.\\[4pt]

\firstcolformat{Natural Disasters (1\%)} & investors weighed the fallout from Katrina\\

\ignorespacesafterend
		\\[-8pt]
		\bottomrule
		\multicolumn{2}{@{}p{\linewidth}@{}}{\footnotesize Note: Example reason snippets for the clusters in the news-reason taxonomy.} \\
	\end{tabular}
\end{table}

\begin{figure}[!t]
    \centering
    \caption{Seasonality in corporate-news reasons}%
    \label{fig:corp_season}
    \includegraphics[width=.8\textwidth]{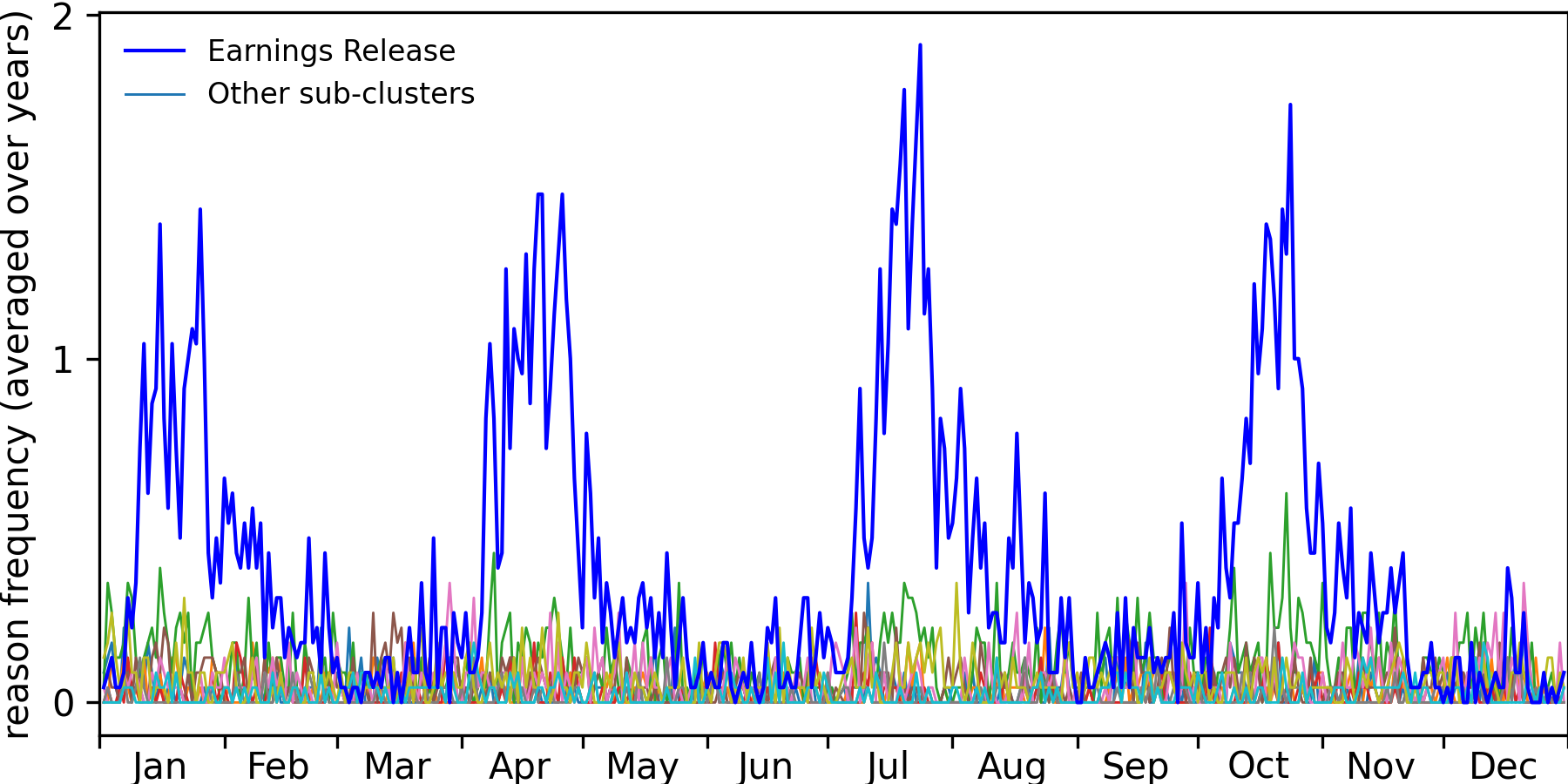} %
    \flushleft\footnotesize
    Note: The figure reports the number of corporate-news reason snippets extracted on each calendar day of the year, averaged across years in the sample. Each curve represents one sub-cluster within the corporate-news meta-cluster.
\end{figure}

The largest meta-cluster is ``corporate news'', which contains sub-clusters such as earnings releases, corporate outlook, M\&A, and transactions.
Earnings-related news accounts for nearly half of corporate-news snippets and closely tracks the quarterly earnings-release cycle, with peaks in January, April, July, and October.
Figure \ref{fig:corp_season} displays this seasonal pattern, while other corporate-news sub-clusters have no comparable time-of-year variation.
The alignment with known disclosure schedules, including earnings releases and conference calls, provides an external validation that the extracted reason text is grounded in real-world information events.\footnote{Section \ref{subsec:spectral} uses spectral analysis to formally examine seasonality and cycles across clusters.}

Most corporate-news reasons pertain to one or a few specific companies, such as {\it Apple} unveiling new iPhone models or {\it Walmart} reporting holiday-season sales that hit expectations (see excerpts in Table \ref{tab:reason_cluster_examples}).
The rest describe corporations in aggregate, as in the excerpt ``Corporate earnings so far have come in above expectations''. 
Although our goal is to explain aggregate market movements, it is unsurprising that many firm-level news events are relevant.
Academic explanations include firm granularity in the aggregate market \citep{gabaix2024granular}, the propagation of firm-level shocks through input-output networks \citep{acemoglu2012network,herskovic2020firm}, and investor attention to prominent firms \citep{da2011search}.
Many reasons in the corporate cluster align with these mechanisms.

\begin{figure}[!t]
    \centering
    \caption{Most frequently mentioned companies in reason snippets over time}%
    \label{fig:corp_names}
    \vspace*{-10pt}
    \includegraphics[width=.8\textwidth]{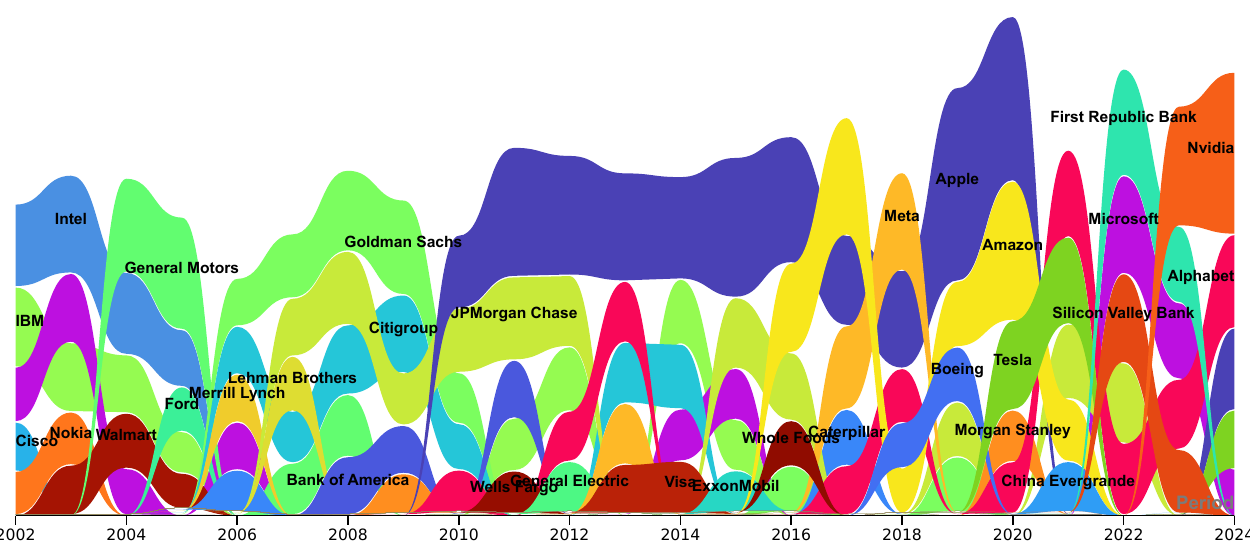} 
    \flushleft\justifying\footnotesize
    Note: The figure reports the five companies most frequently mentioned in reason snippets within rolling two-year windows.
    The horizontal axis marks the start year of each window. The final window contains only 2024 observations because the sample ends that year.
    Company names are extracted using prompting-based named entity recognition (NER), with entity normalization to merge name variants.
    Band thickness is proportional to each company's share of all company mentions within the window, and bands are ordered vertically by frequency rank. Company names are labeled near the peak of each band.
\end{figure}

Figure \ref{fig:corp_names} shows the most frequently mentioned companies in the reason snippets over time.
We extract company names from the reason snippets using prompting-based named entity recognition (NER) with entity normalization to merge name variants.
The visualization shows the changing composition of firm-level news that is relevant for aggregate market movements.
Technology companies appear frequently, but their identities change over time.
The first top-five list of the sample includes legacy technology firms
{\it Intel, IBM, Cisco, Nokia} 
in the early 2000s, which give way to {\it Nvidia, Alphabet, Apple, Tesla} in 2024; {\it Microsoft} appears in the top five at both the beginning and end of the sample.
Financial companies such as {\it Goldman Sachs, JPMorgan, Lehman Brothers} top the list around the 2008 financial crisis, while {\it First Republic Bank} and {\it Silicon Valley Bank} appear briefly around their failures in 2023.
Blue-chip companies such as {\it GM, Ford, Caterpillar, Walmart, Boeing, ExxonMobil} also make the list at various times, reflecting their importance in the economy and market.

The second largest meta-cluster is ``macroeconomic news.''
Most reasons in this cluster are journalistic reports centered on macroeconomic data releases, with phrasing such as 
``{\it ... consumer confidence reading}'', 
``{\it ... retail sales figures}''
``{\it report that China's oil imports ...}'', 
``{\it ... according to data from the Labor Department}'', 
shown in the examples in Table \ref{tab:reason_cluster_examples}.
The list of sub-clusters mirrors the typical categories of macro statistics: job market, consumer activity, price levels, manufacturing, and housing.
The list bears a striking resemblance to the structuring of macroeconomic statistics adopted by datasets such as FRED-MD \citep{mccracken2016fred}, which are widely used in macroeconomics and finance research.

Despite the similarity in categories, news-based macroeconomic reasons contain information that is difficult to recover from structured macroeconomic statistics alone.
Journalistic writing often provides context and interpretation of raw data, most importantly by comparing releases with contemporaneous market expectations.
This expectation component is evident in phrases such as ``{\it ... but still faster than economists had expected}'' and ``{\it ... came in weaker than expected}.''\footnote{A more detailed comparison between news-based macroeconomic expectations and quantitative expectation data is a useful direction for future work.}
Macroeconomic readings in news text also arrive in market time, whereas structured datasets such as FRED-MD compile macro data at regular intervals typically aligned with month- or quarter-end.
These features make macroeconomic news text a more useful source of explanatory information than structured macro statistics alone.

\begin{figure}[!t]
    \centering
    \caption{Expansionary and contractionary monetary-policy reasons with interest rates}
    \label{fig:monetary}
    \includegraphics[width=0.8\textwidth]{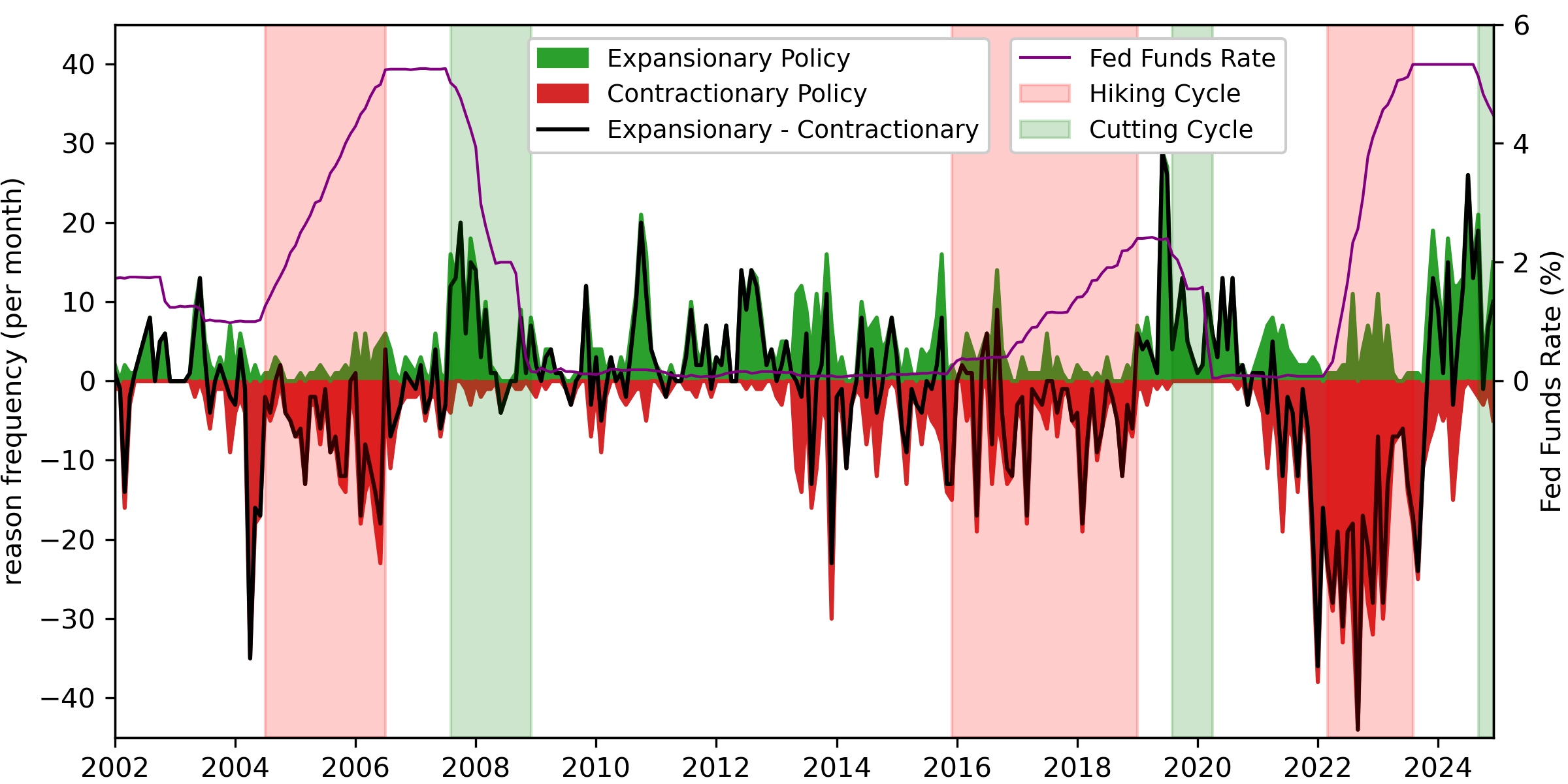}

    \noindent\justifying\footnotesize
    Note: The left axis reports monthly reason frequencies (counts per month) for two sub-clusters of the Monetary Policy meta-cluster: Expansionary Policy (green area), Contractionary Policy (red area, plotted downward), and their difference (black line). The right axis reports the federal funds target rate (purple line). Red and green bands denote rate-hiking and rate-cutting periods, respectively.
\end{figure}

The third largest meta-cluster is ``Monetary policy, Federal reserve, interest rate policy.''
Figure \ref{fig:monetary} displays the time series evolution of the two major sub-clusters within this meta-cluster: ``Expansionary Policy'' and ``Contractionary Policy,'' along with the Federal Funds target rate for reference.

The figure shows patterns that align with known monetary-policy dynamics.
The two monetary-policy sub-clusters, expansionary and contractionary, move roughly in opposite directions over time (corr = $-0.12$).
The overall level of net monetary-policy reasons is highly correlated with actual federal-funds-rate dynamics.
Defining a hike-cut index with values $-1$, 0, and $+1$ for rate-hike, neutral, and rate-cut days, respectively, the correlation between this index and net monetary-policy reason frequency is 0.48.

This coarse analysis shows broad alignment between the monetary-policy sub-clusters and interest-rate dynamics, but the reason text contains richer information than the rate movement itself.
The reason cluster variations display higher-frequency variation and often appear to lead the onset of rate-hike or rate-cut regimes.
This is evident in snippets that discuss expectations, with phrases such as ``{\it ... reinvigorating hopes the Fed will cut interest rates ...}'' and ``{\it ... investors weighed the likelihood that the Fed will ...}'' (from Table \ref{tab:reason_cluster_examples}).
These excerpts contain forward-looking judgments about monetary policy from the perspective of investors.
    
The monetary-policy reasons also capture policy narratives beyond contemporaneous interest-rate changes.
During the zero-lower-bound period from 2009 to 2015, expansionary monetary-policy reasons rose around the three rounds of quantitative easing, while contractionary-policy reasons spiked around events such as the 2013 taper tantrum.

\begin{figure}[!t]
    \centering
    \caption{Evolving content of geopolitics reasons}
    \label{fig:geopolitics_evolve}
	\begin{subfigure}{0.48\textwidth}
        \caption{Sub-cluster frequencies over time}%
        \includegraphics[width=\textwidth]{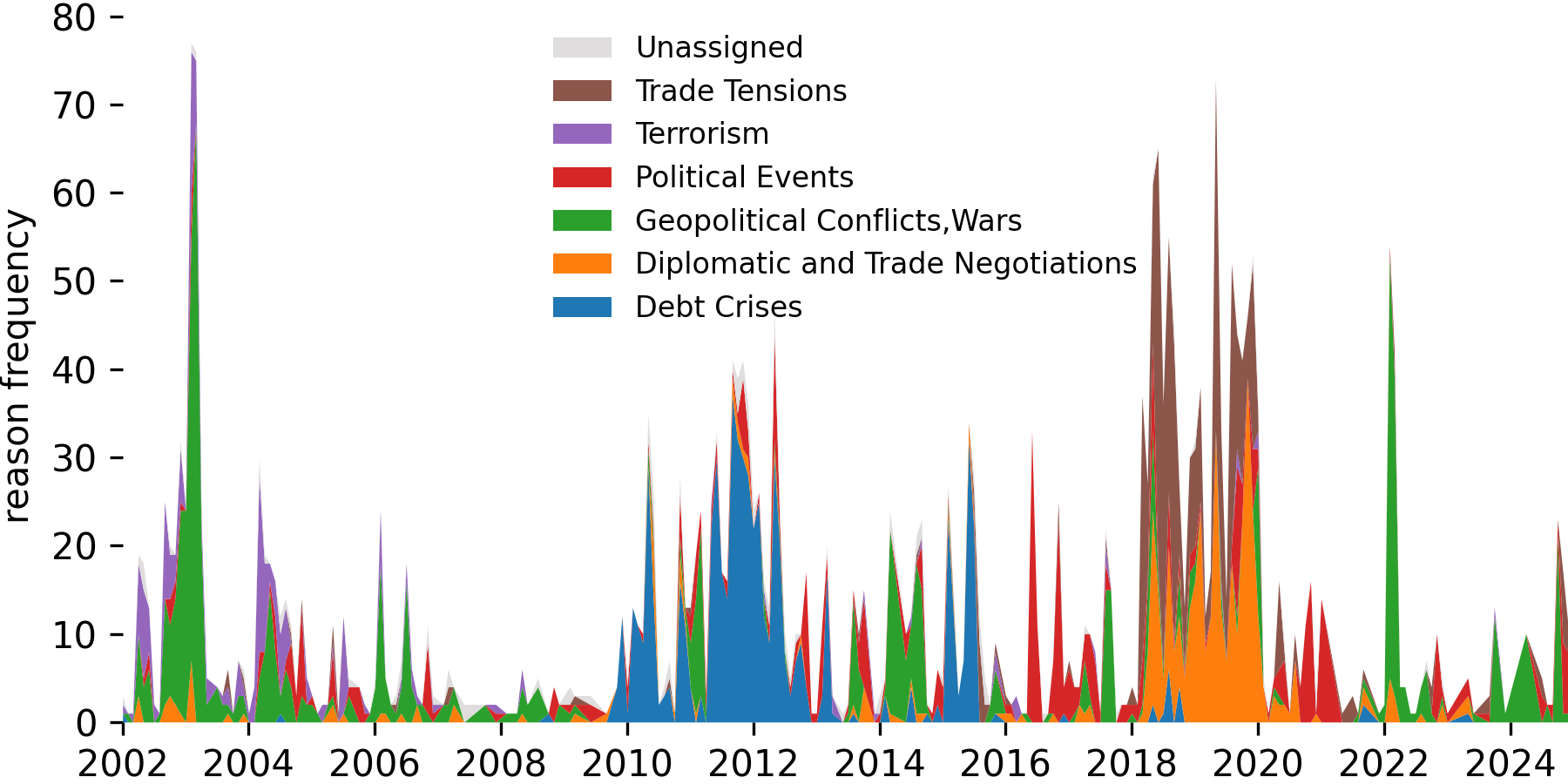}%
    \end{subfigure}%
    \hfill
    \begin{subfigure}{0.48\textwidth}
        \caption{Word stream: keywords per year}%
        \includegraphics[width=\textwidth]{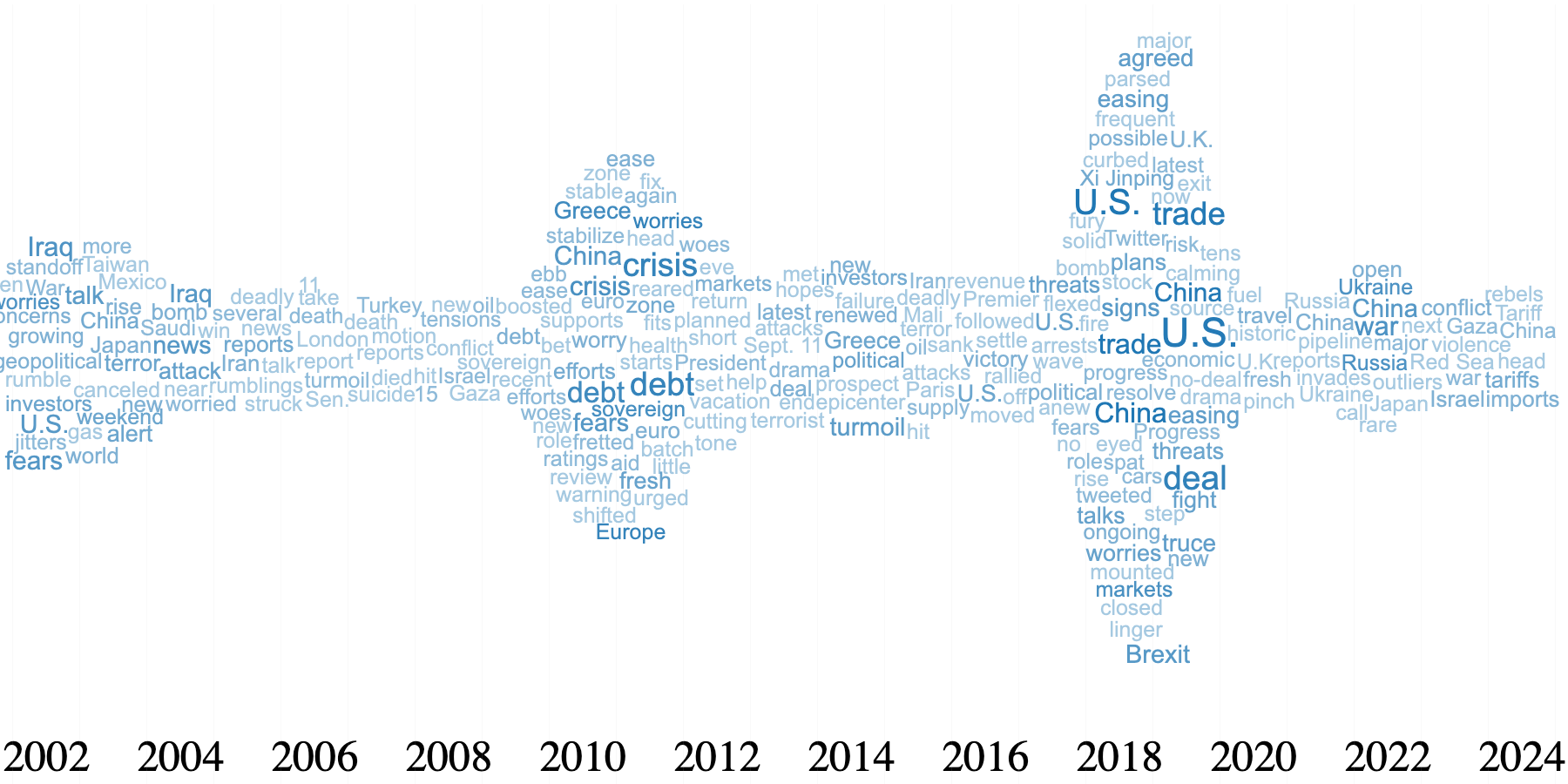}%
    \end{subfigure}%
    \par\footnotesize\justifying\noindent
    Note: Panel A reports monthly reason frequencies for sub-clusters within the geopolitics meta-cluster. Panel B reports a word-stream visualization of frequent keywords in geopolitics reason snippets by year.
\end{figure}

Figure \ref{fig:geopolitics_evolve} displays the evolving content of geopolitics-related reasons over time.
The intensity of geopolitical explanations varies substantially, with distinct waves and changing content.
Conflicts and wars dominate the early 2000s, especially around terrorism and Middle East events; European debt-crisis narratives become prominent around 2010--2012; trade and diplomatic negotiations rise sharply during 2018--2020; Brexit appears around 2018--2019; the Russia-Ukraine war spikes in 2022; and Gaza-Israel news appears in 2023 after the October 7 attacks.
Even within a single broad meta-cluster, the relevant market-moving geopolitical narrative is therefore far from static.

\subsection{The seasonality and periodicity of different reason types}
\label{subsec:spectral}

Different types of extracted reasons exhibit distinct seasonal and periodic patterns.
We use spectral analysis to uncover these patterns in the time series of reason counts for each meta-cluster.
The frequency domain of news-reason time series reveals distinctions that are difficult to see from time-domain plots alone.

\begin{figure}[!t]
    \centering
    \caption{Periodic cycles of different reason types via spectral analysis}%
    \label{fig:fft}
    \includegraphics[width=0.8\textwidth]{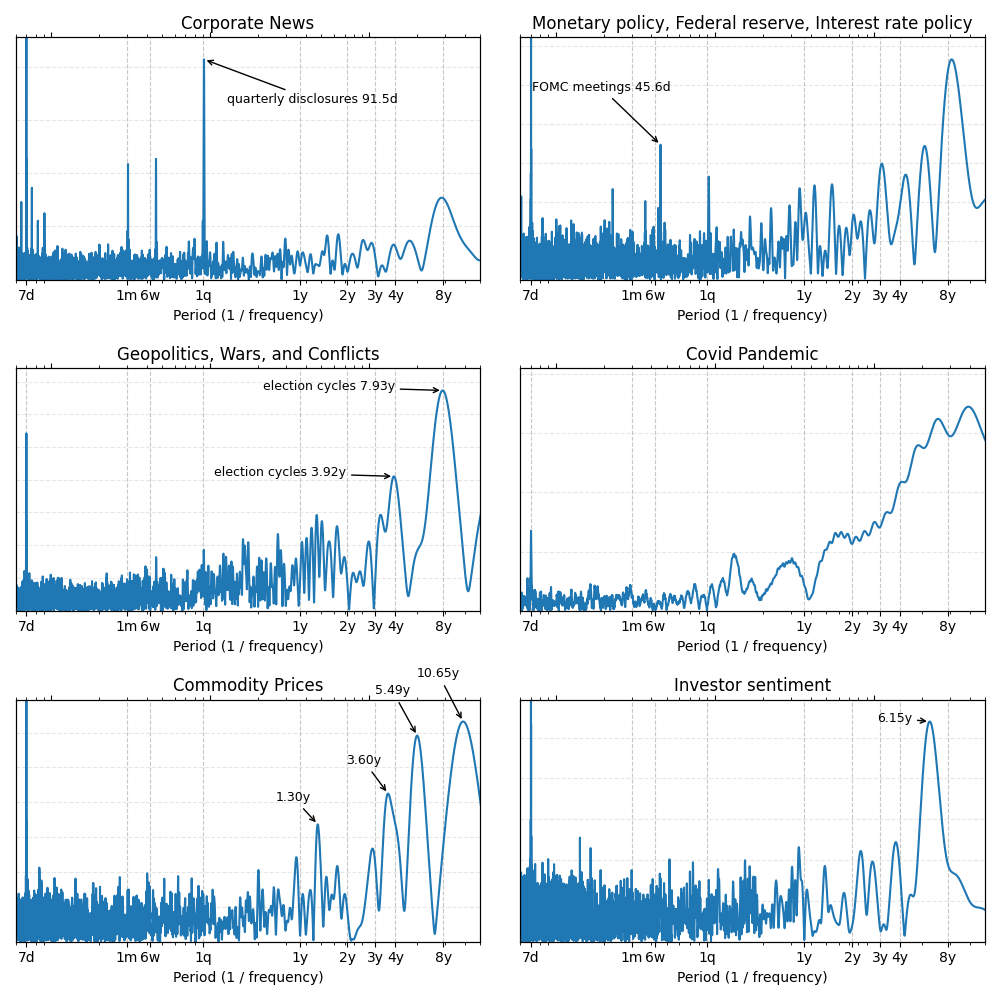}

    \noindent\justifying\footnotesize
    Note: The figure reports amplitude spectra for daily reason-count time series for six meta-clusters, computed using the Discrete Fourier Transform described in Footnote \ref{fn:dft}. The bottom horizontal axis reports period or cycle length, with selected labels at 7 days (7d), 30 days (1m), 42 days (6w), 90 days (1q), and yearly multiples. The top horizontal axis reports the corresponding frequency in log scale. The vertical axis reports amplitude, and selected peaks are labeled by period.
\end{figure}

Figure \ref{fig:fft} displays the amplitude spectrum obtained via Discrete Fourier Transform (DFT) for the daily time series of reason counts for six selected meta-clusters.
Intuitively, DFT amplitude measures the ``correlation'' or ``alignment'' between the input time series and sinusoids of a given frequency.%
\footnote{The formula is the standard DFT:
$\text{amplitude}(f) = \left| \sum_{t=1}^{T} x_t \, e^{-i 2\pi f t} \right|$, where $x_t$ is the input time series---the demeaned daily reason count for a meta-cluster---and $f$ is the frequency. \label{fn:dft}
Relatedly, \citet{bandi2025conditional} use Haar wavelets (rather than sinusoids) to analyze the scale-specific cycles of conditional risk measures.}
Corporate-news reasons exhibit strong quarterly cycles, with a peak at 91.5 days corresponding to the quarterly earnings-reporting cycle, which is also visible in Figure \ref{fig:corp_season}.
Monetary-policy reasons show a dominant cycle at 45.6 days, reflecting the cadence of Federal Open Market Committee (FOMC) meetings, which occur every 6 to 7 weeks.
In contrast, the other meta-clusters mainly exhibit low-frequency cycles with no discernible periods under one year.
Geopolitical reasons fluctuate with political-election cycles of approximately four years and, with even greater amplitude, eight years.
Commodity-news reasons show periodicity with labeled peaks at 1.30, 3.60, 5.49, and 10.65 years.
These cycles do not coincide with artificial calendar intervals, suggesting that they may reflect intrinsic market dynamics such as inventory horizons, investment cycles, or business cycles.
Investor-sentiment reasons exhibit a strong 6.15-year cycle, which may be related to longer-run sentiment swings \citep{akerlof2010animal}.
Finally, the COVID pandemic, as a one-off event, shows a cycle length approaching infinity.

\section{Explanatory power of news-based reasons}

\subsection{LLM embedding regressions and their explanatory power}
\label{subsec:embedding_regression}

Our baseline method for estimating the reason-return relationship is LLM embedding regression.
The method first converts daily reason text into high-dimensional numerical representations, or embeddings, and then estimates a regularized linear regression of market returns on these embeddings.

In particular, we use OpenAI's \texttt{text-embedding-3-large} model to convert each day's reason-composite text into a 3,072-dimensional embedding.
As a smaller-scale comparison, we also use 768-dimensional BERT embeddings.
For monthly exercises that compare with non-news baselines, we aggregate daily reasons to the monthly frequency by summing daily embeddings within each month.\footnote{Alternatively, we also estimate the regression at the daily frequency and then aggregate the fitted values to monthly. That result is discussed further in Appendix \ref{appsec:zero_shot_har}. Here, we aggregate the embeddings first to be consistent with the non-news baselines, where the predictors are only available at the monthly frequency.}
Given reason embeddings $\{x_{t}\}$ at either the daily or monthly frequency, we conduct high-dimensional ridge and lasso regressions to explain market returns.\footnote{%
On days when no reason snippets are extracted, the embedding is a zero vector, and the implied market movement is zero: $\widehat{r}_t = 0$. This is aligned with the idea that the market price does not move when there is no news.}
For each $t$, the model produces an out-of-sample fitted value $\widehat{r}_t = f(x_t; \widehat{\theta})$, where the mapping $f(~\cdot~; \widehat{\theta})$ is estimated from a cross-validation sample that excludes day $t$.
Explanatory power is evaluated with out-of-sample $R^2$ by comparing $\widehat{r}_t$ with the realized return $r_t$ across all days or months in the sample.
The regression and cross-validation procedure is the same as for the non-news baselines; the only difference is the input data $x_t$.
See Section \ref{subsec:estimation_evaluation} for more details.

In this procedure, the LLM converts unstructured reason text into structured numerical data for statistical analysis.
Embedding models are designed to preserve semantic and contextual information in text, allowing statistical models to use the economic content of language.
This is why we prefer embeddings to topic models for the baseline analysis.
Our objects are short reason snippets, and much of their market-relevant information is in the specific event, direction, and context rather than in a broad topic label.
LLM embeddings provide high-fidelity numerical representations of this content, whereas a topic model would intentionally discard much of that granularity.
We use pre-trained embedding models as a bridge from natural language to numerical regressors for explaining market movements \citep{gentzkow2019text,chen2022expected}.

\begin{table}[!t] %
	\caption{Explanatory power of embedding regression models (OOS $R^2$, \%)}
	\label{tab:R2}
    \begin{center}
        \begin{tabular}{llll}
        \toprule
            \multicolumn{4}{l}{A. Daily} \\
            ~ & ~ & Ridge & Lasso \\
            ~ & text-embedding-3-large (ndim = 3,072) & {\bf 29.82} & 28.99 \\
            ~ & BERT (ndim = 768) & 12.60 & 12.08 \\[8pt]
            \multicolumn{4}{l}{B. Monthly} \\
            ~ & ~ & Ridge & Lasso \\
            ~ & text-embedding-3-large (ndim = 3,072) & {\bf 40.30} & 32.54 \\
            ~ & BERT (ndim = 768) & 24.35 & 21.86 \\
        \bottomrule
		\end{tabular}
	\end{center}
	\justifying %
	\footnotesize
    Note: The table reports out-of-sample $R^2$ values in percent from embedding regressions of S\&P 500 returns on embeddings of daily news-reason composites. Daily and monthly specifications are evaluated at the corresponding return frequency. Out-of-sample fitted values are constructed by five-fold cross-validation, with model tuning performed only on training folds. Bold numbers are benchmark specifications also highlighted in Figure \ref{fig:comp_non_news}.
\end{table}

Table \ref{tab:R2} shows that news reasons have high explanatory power for market returns.
Using the benchmark text-embedding-3-large model with ridge regression, daily $R^2$ is around 30\% and monthly $R^2$ is around 40\%.
Monthly explanatory power is far above the non-news baselines in Figure \ref{fig:comp_non_news}; at the daily frequency, there is essentially no comparable non-news baseline.

Ridge consistently outperforms lasso across frequencies and embedding models, suggesting that market-relevant reason information is distributed across many embedding dimensions rather than concentrated in a small sparse subset.\footnote{The comparison of ridge and lasso results is similar to that of \citet{shen2025can} albeit our application is in a higher signal-to-noise environment.}
More flexible methods that capture nonlinear, interactive, or dynamic relationships may further improve explanatory power, but we leave that extension for future research.

The smaller BERT embedding model yields substantially lower $R^2$ values than the larger model, regardless of frequency or estimation method.
This suggests that embedding quality matters for preserving the market-relevant information in reason text.
Differences in training corpus scale, parameter count, and embedding dimension may all contribute to the performance gap between BERT and OpenAI's embedding model.\footnote{BERT gives contextualized embeddings at the token level. We obtain the sentence-level reason embedding by averaging the token-level embeddings, commonly known as the ``mean pooling'' method.
The text-embedding-3-large model is more suitable for our application as it directly provides input-level embeddings with the dedicated embedding head, and is also trained on more relevant tasks such as semantic similarity and retrieval tasks for the whole input text.}

\subsection{Explanatory powers of different types of reason}
\label{subsec:2r2}

We next decompose overall explanatory power by reason type, using the clusters described in Section \ref{sec:reason_content}.
The goal is to identify which aspects of news-based reasons are more important for explaining market movements.

The estimated model provides a mapping from reason text to an implied market movement.
We apply this mapping to the individual reason snippets on each day, which have already been assigned to clusters, to obtain each cluster's implied market movement $\rhat_{c,t}$.
For this exercise, we focus on the daily specification using the benchmark text-embedding-3-large model with ridge regression, which has the highest overall $R^2$ in Table \ref{tab:R2}.

Cluster-level evaluation must account for the fact that reason types emerge and fade over time.
A cluster is active on a day if at least one reason snippet is assigned to that cluster.
When a cluster is inactive, its implied market movement is set to zero: $\rhat_{c,t} = 0$.

Given each cluster's sequence of implied returns, we evaluate how well that cluster explains realized market returns using cluster-level $R^2$ metrics.
These cluster-level evaluations do not re-estimate any models; they use the model already estimated on pooled reason text.

Accordingly, we report two $R^2$ metrics for each cluster.
The {\it overall $R^2$} evaluates a cluster's explanatory power across all days in the sample, setting its implied return to zero when the cluster is inactive.
The {\it active $R^2$} applies the same formula only to days when the cluster appears in the news.
The overall $R^2$ therefore reflects both the explanatory power of the cluster when it is active and the frequency with which it appears.
Appendix \ref{appsubsec:2r2} provides formal definitions and shows that overall $R^2$ is approximately the product of active $R^2$ and the active-day fraction, with equality when the conditional variance of market returns on active days equals the full-sample variance.

\begin{figure}[!t]
    \centering
    \caption{Explanatory power and activity of reason types}
    \label{fig:cluster_radar}
    \raisebox{-0.225\textwidth}{\Large ~~~~~}
    \begin{subfigure}{0.55\textwidth}
        \centering
        \caption{Overall $R^2$}%
        \includegraphics[width=\textwidth]{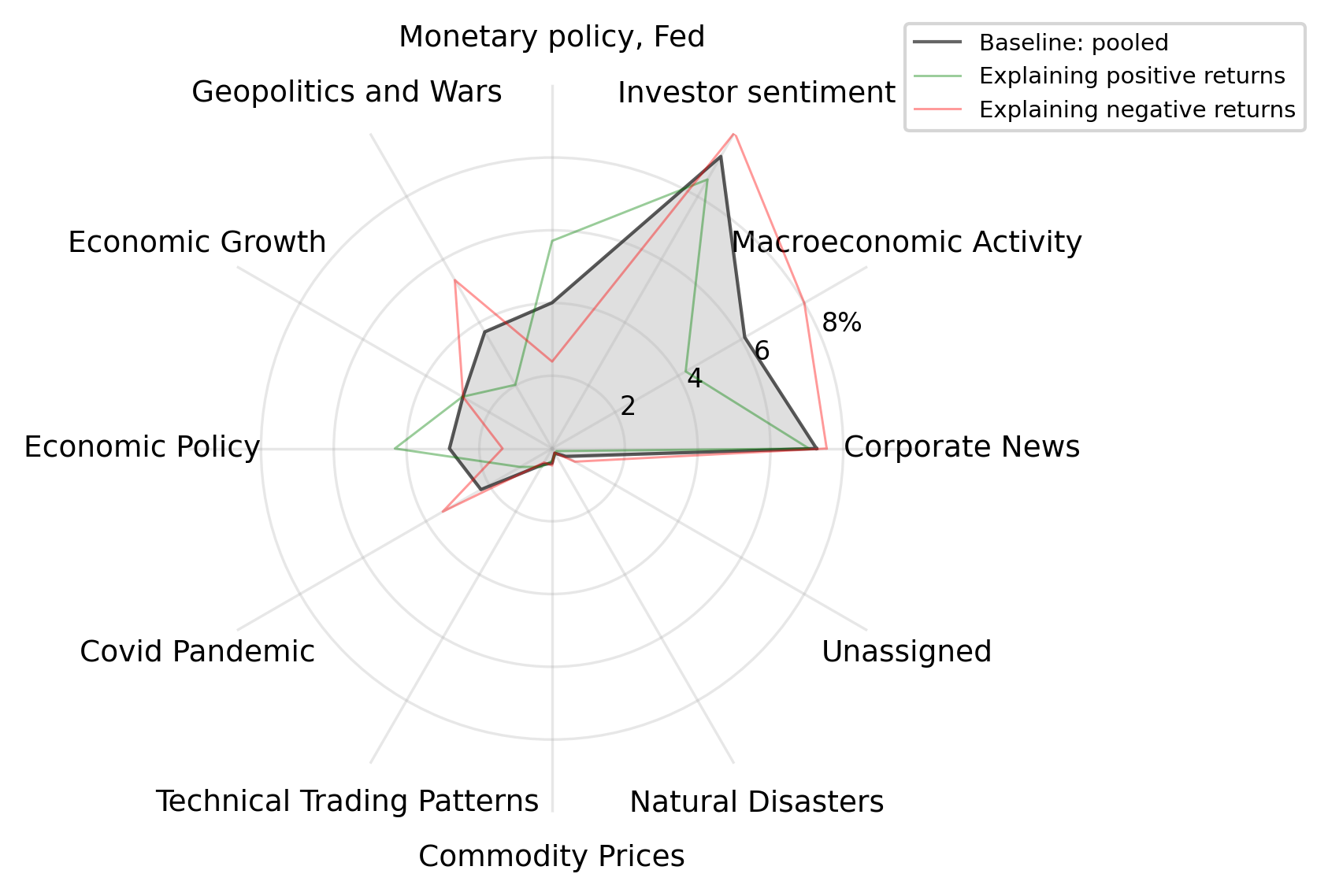}%
    \end{subfigure}%
    \raisebox{-0.225\textwidth}{\Large ~~~$\approx$}
    \\[6pt]
    \begin{subfigure}{0.45\textwidth}
        \caption{Active-day $R^2$}%
        \includegraphics[width=\textwidth]{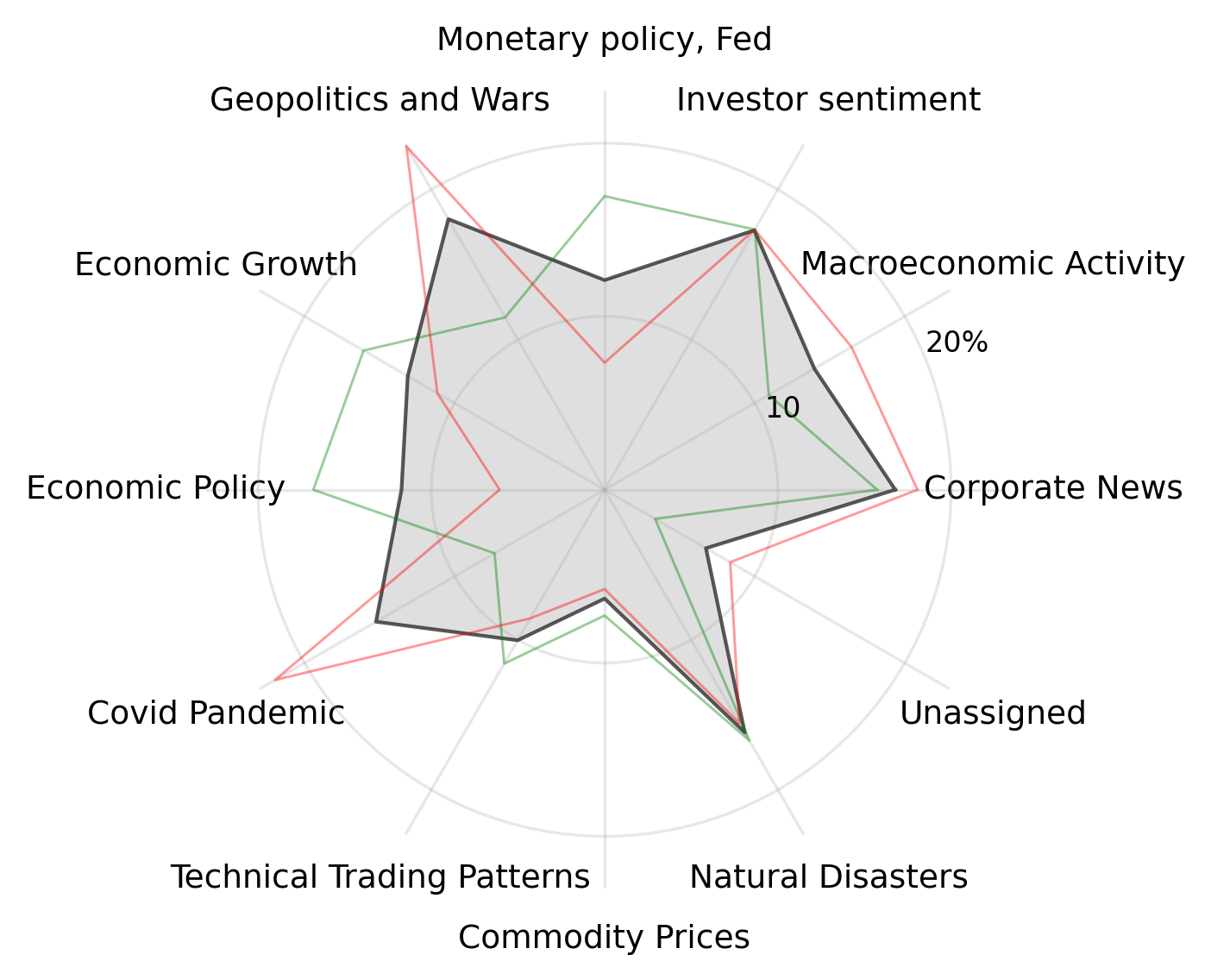}%
    \end{subfigure}%
    \hfill
    \raisebox{-0.225\textwidth}{\Large $\times$}
    \hfill
    \begin{subfigure}{0.45\textwidth}
        \caption{Active-day share}%
        \includegraphics[width=\textwidth]{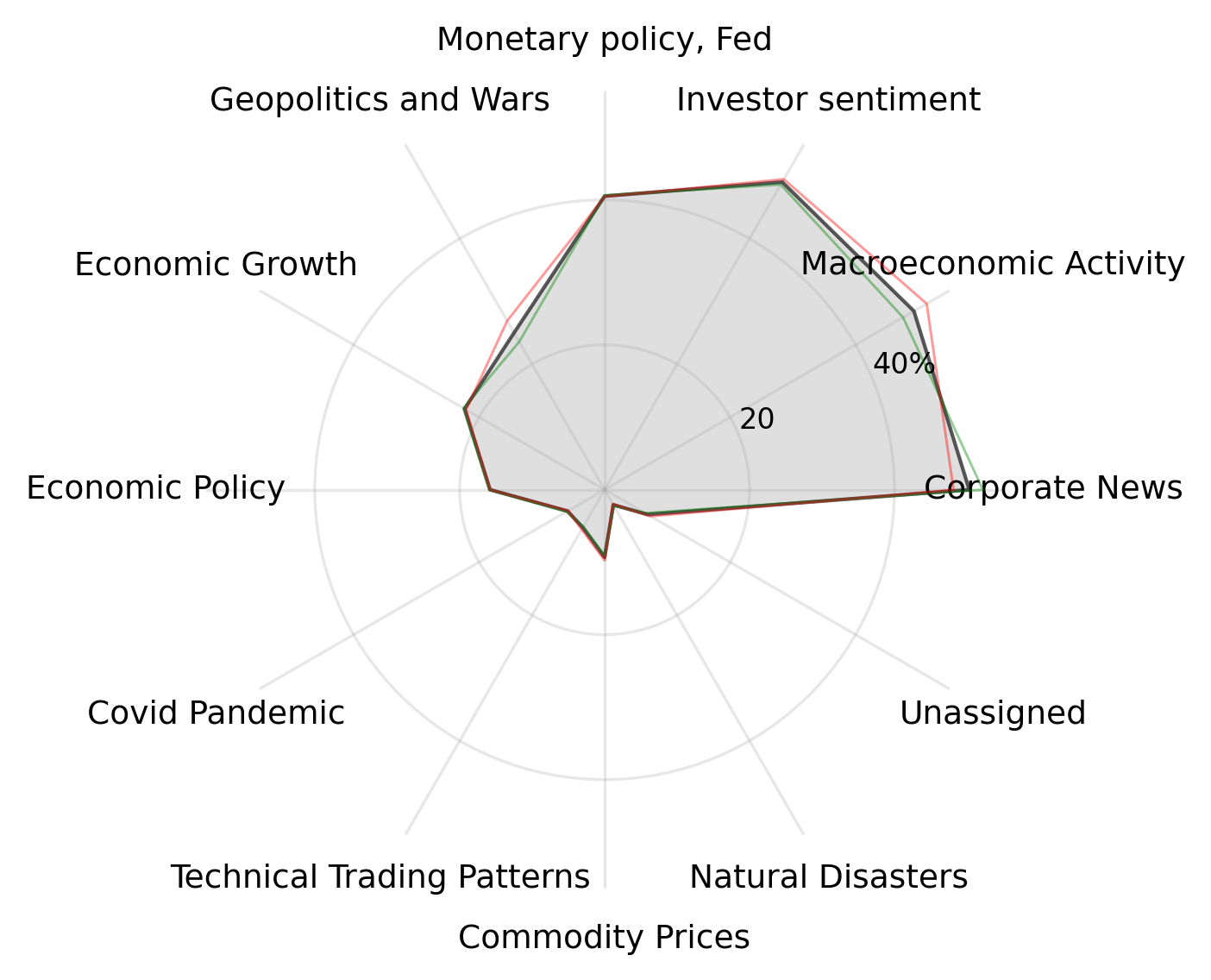}%
    \end{subfigure}%
    ~\\
    \flushleft\justifying\footnotesize
    Note: Panel A reports $R^2_{\text{overall},c}$, the overall explanatory power of reason cluster $c$ for market returns, setting the cluster's implied return to zero when the cluster is inactive. Panel B reports $R^2_{\text{active},c}$, computed only on days when cluster $c$ is active. Panel C reports the fraction of days on which cluster $c$ is active. Exact definitions are given in Appendix \ref{appsubsec:2r2}. The approximate relationship ``Panel A $\approx$ Panel B $\times$ Panel C'' is formalized and numerically validated in Appendix \ref{appsubsec:2r2}. Black lines use the full sample; green and red lines repeat the calculation for positive- and negative-return days, respectively. All $R^2$ values are out-of-sample and use the pooled embedding-regression model in Table \ref{tab:R2} without cluster-specific re-estimation.
\end{figure}

Figure \ref{fig:cluster_radar} reports per-cluster explanatory power.
Panel A shows overall $R^2$, Panel B shows active $R^2$, and Panel C shows the fraction of days each cluster is active, preserving the approximate decomposition ``A $\approx$ B $\times$ C''.
We also evaluate explanatory power separately for positive and negative market-return days, shown by the green and red lines, obtained by applying the same procedure to the corresponding subsamples.

Each reason type individually has positive $R^2$.
This is not mechanical.
Each reason type uses the same model estimated on pooled reason text, without re-estimation, and all $R^2$ values are evaluated out of sample under the cross-validation procedure.
The result indicates that the mapping from reason text to market returns is robust within each reason type.

The clusters with the highest overall $R^2$ are Corporate news, Macroeconomic activity, and Investor sentiment.
Each explains roughly 6--9\% of daily market returns on its own.
These are meaningful levels of explanatory power at the daily frequency.
At the same time, cluster-level $R^2$ values are much lower than the roughly 30\% $R^2$ obtained when all reason types are pooled.
The Monetary policy cluster explains about 4\%; Geopolitics and Wars, Economic growth, Economic policy, and Covid pandemic clusters have low single-digit $R^2$ values; and smaller clusters have $R^2$ values close to zero.

The decomposition shows that cross-cluster differences in overall explanatory power are mostly driven by differences in active-day frequency.
Conditional on being active, most clusters have active $R^2$ values around 10--15\%, with the exceptions of Commodity prices and Unassigned, whose active $R^2$ values are below 10\%.
For example, when Covid pandemic reasons are present, they explain a high level of market returns with active $R^2$ around 15\%.
But the Covid cluster appears on only about 5\% of sample days, which leads to a much lower overall $R^2$.
(The overall $R^2$ is around 2\%, greater than $15\% \times 5\%$; the gap is due to higher variance of market returns conditional on Covid active days. Appendix \ref{appsubsec:2r2} formalizes this relationship.)

This pattern suggests that explanatory power comes from the presence of specific reason snippets.
When a reason snippet is present, it tends to be informative for market returns regardless of reason type; when a category of world events has no market-relevant development, it is typically absent from the news.
The relative prevalence of reason types, and the evolution of news cycles in which types become active or inactive, therefore help answer {\it what} is moving the market at a given time.

Finally, we examine explanatory power separately for positive and negative market returns, shown by the green and red lines.
Overall, most clusters have relatively balanced explanatory power across return signs, with some deviations in active $R^2$.
For example, Geopolitics and Wars and Covid have higher active $R^2$ for negative returns, consistent with the intuition that these categories are often associated with bad news.
This does not mean that these clusters only contain bad news or only explain price declines; positive updates, such as vaccine-development news and conflict de-escalation, are also important in these clusters, as shown in Table \ref{tab:reason_cluster_examples}.
By contrast, Monetary policy and Economic policy lean more toward explaining positive returns.

The positive and negative lines in active-day percentage (Panel C) are nearly identical.
This indicates that the directional relationship between news and market movements comes from the content of the reason text, not simply from the topic label.
Knowing that a reason belongs to a broad theme is not enough to infer the sign of the market movement, which highlights the value of embedding regressions relative to approaches that rely only on topic labels.

\subsection{Zero-shot learning with Generative AI and its explanatory power}

In addition to estimating the reason-return mapping with embedding regressions, we use generative AI for zero-shot inference by directly asking an LLM to infer the market movement implied by a given reason text.\footnote{We used \texttt{gpt-4o-mini-2024-07-18}, with temperature set to 0 and one retained response per trading day (no averaging across generations). Its reported knowledge cutoff is October 2023.}
The rationale is that if a stable mapping from real-world events to market movements exists and is broadly understood by market participants and observers, then that mapping is likely captured in the trained LLM and can be replicated in the inference process.
In that case, an LLM may be able to approximate the reasoning process that maps reason text into implied market movement.

Unlike embedding regression, this method does not statistically estimate the mapping from historical reason-return pairs.
It is therefore ``zero-shot'' in the sense that it uses no task-specific training sample of realized returns.
The model is given the task description and the reason text, but no examples of realized returns from our sample on which to learn coefficients.

In particular, the prompt instructs the LLM to read the day's reason snippets and ``assess their implications for market returns, and predict the potential impact on the overall stock market.''
The prompt requires a structured output: an implied market return in ``percentage decimal'' form and a short textual explanation.\footnote{See the complete prompt instructions reported in Appendix \ref{appsubsec:prompts}.}
We parse the output string to obtain the numerical daily implied return, denoted $\rhat_{t}^\text{zero-shot}$.
This evaluation is out of sample in the sense that there is no task-specific training sample, so cross-validation is neither applicable nor needed.

The daily zero-shot $R^2$ is 25.58\% (see Table \ref{tab:GenAI_R2} Panel A), comparable to the embedding-regression $R^2$ of 29.82\%. 
This result suggests the existence of a stable mapping from news reason text to market returns and that part of this mapping is encoded in the LLM's prior knowledge.

\begin{table}[!t]
\caption{Explanatory power of zero-shot, embedding regression, before and after filtering ($R^2$, \%)}
\label{tab:GenAI_R2}
    \begin{center}
        \begin{tabular}{llll}
        \toprule
            \multicolumn{4}{l}{A. GenAI zero-shot learning and its filtering} \\
            ~ & ~       & raw ($\rhat_{t}^\text{zero-shot}$) & HAR filtered 
            ($\ehat_{t}^{\text{zero-shot}}$)\\
            ~ &  daily  & 25.58 & 31.88 \\
            ~ &  monthly & $-115.03$ & 49.84 \\[8pt]
            \multicolumn{4}{l}{B. Embedding regression (Ridge) and its filtering} \\
            ~ & ~       & raw ($\widehat{r}_{t}^\text{emb-reg}$) & HAR filtered 
            ($\ehat_{t}^\text{emb-reg}$) \\
            ~ &  daily  & {\bf 29.82} & 33.88 \\
            ~ &  monthly& 33.06 & 44.42 \\[8pt]
            \multicolumn{4}{l}{Repeated from Table \ref{tab:R2} for comparison}
            \\
            ~&\multicolumn{3}{l}{monthly Ridge regression of aggregated embeddings: {\bf 40.30}} \\
        \bottomrule
		\end{tabular}
\end{center}
\justifying\footnotesize
    Note: The table reports $R^2$ values in percent for implied return series from generative-AI zero-shot inference and embedding regression, before and after applying HAR filtering to remove persistence bias. The reported series are $\rhat_{t}^\text{zero-shot}$, $\rhat_{t}^\text{emb-reg}$, $\ehat_{t}^{\text{zero-shot}}$, and $\ehat_{t}^\text{emb-reg}$. Monthly $R^2$ values are calculated by summing daily implied-return series to the monthly frequency and applying the same $R^2$ formula to monthly market returns, except for the last row, which reports ridge regression estimated directly at the monthly frequency using monthly aggregated embeddings. Bold numbers are benchmark results from Figure \ref{fig:comp_non_news} and
    Table \ref{tab:R2}. 
\end{table}

Monthly aggregation provides an additional diagnostic of the time-series properties of reason-implied returns.
Simply summing the raw daily zero-shot outputs within each month fails to explain monthly returns, producing an $R^2$ of $-115.03\%$ (Table \ref{tab:GenAI_R2} Panel A second row left).
We show that this failure reflects systematic biases in the reason-implied return series: an unconditional bias toward pessimism and, more importantly, a persistence bias, meaning that news-based reasons are sticky and slow to adjust to new information.
The persistence bias is largely innocuous at the daily frequency, but it accumulates when daily implied returns are summed to the monthly frequency, producing exaggerated implied monthly returns that tend to overshoot realized returns.

A simple HAR filtering procedure removes the mean and predictable persistence from the implied-return series without using realized market returns.
The filtered implied return series not only rescues the raw zero-shot returns but also improves it beyond the embedding-regression benchmark, raising the monthly $R^2$ to $49.84\%$.
The daily $R^2$ is also improved from $25.58\%$ to $31.88\%$ (Table \ref{tab:GenAI_R2} Panel A right column ``HAR-filtered'').
Panel B shows that the same filtering procedure improves embedding-regression results as well, raising daily $R^2$ from $29.82\%$ to $33.88\%$ and monthly $R^2$ from $33.06\%$ to $44.42\%$.
It suggests the persistence bias also exists in the embedding-regression implied-return series.

Appendix \ref{appsec:zero_shot_har} reports the HAR filtering procedure and the underlying bias diagnostics  in detail. We note the filtering procedure is applied to the implied-return series without using realized market returns.
The filtered series therefore still satisfies Criterion 1: it is not derived from asset returns.

\section{Evidence against news-based reasons as ex-post attributions}

\subsection{The ex-post attribution concern and the counterarguments}

The previous sections show that extracted reasons explain a significant share of realized market variation.
A natural concern is that news-based reasons are not true drivers of price movements, but ex-post attributions constructed by journalists after observing realized returns.
In the strongest version of this concern, the attributed reasons are invented or fabricated to fit the return.
A more subtle version is that the reasons are factual events that occurred during the return horizon, but their attribution as causes is not valid.
On any given day, many news events are available--consumer, monetary, industrial, geopolitical, and so on.
Some can rationalize a positive market move, while others can rationalize a negative move.
Journalists may selectively choose the subset of news whose direction matches the realized return and present those events as ``the reasons.''

This section provides evidence against that interpretation.
We do not claim to identify every news-based reason as a genuine causal factor.
Instead, we present evidence that is difficult to reconcile with a pure ex-post attribution story.
We first show that the market anticipates many news-based reasons before they are realized.
Moreover, the reasons identified as anticipated have stronger explanatory power than average reasons.
Second, news-based reasons align with real-world events recorded in structured data, such as monetary-policy announcements and macroeconomic data releases, which are difficult for journalists to manipulate and are widely understood to be market-moving events.

At a broader level, our question is how well we understand why markets fluctuate.
If journalists systematically attribute market movements to unrelated events, then either market participants find such reporting unconvincing and of little value, or the reporting reflects a broader lack of understanding among investors.
Either interpretation would itself be informative about the state of financial explanation.

\subsection{The ex-ante anticipation of news-based reasons }

First, we provide suggestive evidence that realized reasons are anticipated by the market in many cases.
For example, progress in U.S.-China trade negotiations is cited as the reason for a market rally on October 11, 2019.
We can trace this reason to earlier news articles stating that investors were waiting for trade-negotiation news and expected it to be a key market-moving factor (see Table \ref{tab:anticipated_realized_examples}, Row 2).
In such cases, market participants are already waiting for specific events that are expected to move prices, and the news-based reasons we capture reflect the realization of those anticipated events.
The type of uncertainty is anticipated, while the reason text captures the directional resolution of that uncertainty.
When an event appears first as an anticipated reason and later as a realized reason, it is difficult to interpret the later explanation as a purely ex-post attribution.

We use a conservative procedure to identify whether a realized reason was anticipated before its realization.
The procedure is designed to illustrate anticipated reasons cleanly and is likely underpowered, so the results should be interpreted as a lower bound on their prevalence.
We process ``Today's Markets'' articles with AI prompts to identify anticipation text: language stating, or strongly suggesting, that the market is waiting for a future event that may move prices.
For each anticipation, we extract the anticipated event in text form and the implied date window when the event is expected to occur.

We then search within that future window and compare the anticipated event with the extracted reason snippets observed during the window.
A match requires more than sharing a broad topic; the realized reason must describe the realization of the same underlying event.
The final output partitions reason snippets into anticipated and unanticipated groups.
Anticipated reasons are also paired with the anticipation text that describes the market's expectation before realization.

\begin{table}[!t]
    \centering
    \scriptsize\singlespacing
\renewcommand{\baselinestretch}{0.9}\selectfont
    \setlength{\extrarowheight}{2pt}
    \setlength{\parindent}{0pt}
    \caption{Examples of anticipated events and their realized market-moving news}
    \label{tab:anticipated_realized_examples}
    \vspace{-6pt}
    \begin{tabular}{>{\raggedright\setlength{\parindent}{0pt}\arraybackslash}p{0.08\linewidth}
                      >{\justifying\setlength{\parindent}{0pt}\arraybackslash}p{0.38\linewidth}
                      >{\raggedright\setlength{\parindent}{0pt}\arraybackslash}p{0.08\linewidth}
                      >{\justifying\setlength{\parindent}{0pt}\arraybackslash}p{0.38\linewidth}}
\toprule
        \multicolumn{2}{c}{Anticipation of future market moving uncertain events} &
        \multicolumn{2}{c}{Realization of anticipated market moving events} \\
        \midrule
        2003/1/24 Fri &
        The U.S. showdown with Iraq remained the biggest drag on stocks ahead of two key events early next week, analysts said. On Monday, the United Nations Security Council, whose members are coming out publicly in bulk against a rush to war, will receive a report on the U.N. inspectors' search for weapons. &
        2003/1/27 Mon &
        Article Title: Fears of War With Iraq Send Dow Below 8000
        \vspace{4pt}\newline Hopes that a report from chief United Nations weapons inspector Hans Blix might somehow point to a peaceful solution were dashed as Mr. Blix issued a tough statement noting that Iraq wasn't showing a willingness to disarm. \\
        \addlinespace[6pt]
        2019/10/1 Tue &
        Investors, who have been particularly sensitive to trade-related headlines in recent weeks, will be closely watching as the countries engage in planned high-level trade talks in Washington later this month. &
        2019/10/11 Fri &
        President Trump said just before the closing bell that the two countries reached a 'very substantial Phase One deal' and agreed not to implement tariffs set to go into effect next week
        \vspace{4pt}\newline China, meanwhile, said it would increase purchases of U.S. agricultural products \\
        \addlinespace[6pt]
        2007/4/10 Tue &
        ... said movement was limited because investors are awaiting the Federal Open Market Committee meeting minutes due for release Wednesday, which could shed more light on the Fed's thinking about the chances for the direction of interest rates. &
        2007/4/11 Wed &
        traders reacted to hawkish Fed minutes and a poor housing forecast
        \vspace{4pt}\newline minutes of the latest Federal Reserve policy meeting, which showed continued concern about inflation \\
        \addlinespace[6pt]
        2013/5/10 Fri &
        a number of key economic data are scheduled for release. On the week's docket are retail sales, wholesale and consumer inflation, industrial production and housing starts &
        2013/5/16 Thu &
        April's consumer-price index, a reading on inflation, fell 0.4\% on the month, more than forecast%
        \vspace{4pt}\newline Housing starts for April declined far worse than expected, and the Federal Reserve Bank of Philadelphia's May index of business activity unexpectedly fell \\
        \addlinespace[6pt]
        2008/7/28 Mon &
        People figure they should book some profits from the recent rally and wait to see how the numbers come out in two highly anticipated economic reports due out later in the week. &
        2008/8/1 Fri &
        Labor Department reported a decline of 51,000 jobs on nonfarm payrolls in July, better than anticipated, even though the decline represented the seventh straight month of job losses. \\
        \addlinespace[6pt]
        2024/10/29 Tue &
        Other Magnificent Seven members: Meta, Microsoft, Amazon and Apple, are set to report later this week. &
        2024/10/31 Thu &
        Microsoft forecast slower growth in its cloud business
        \vspace{4pt}\newline Meta said it expects significant growth in capital expenditures next year \\
        \addlinespace[6pt]
        \bottomrule
    \end{tabular}
    \footnotesize\justifying\noindent
    Note: The table reports examples of matched anticipation and realization text. The left columns show article text that describes a market-relevant event before realization. The right columns show later article text that attributes market movement to the realized event.
\end{table}

The examples in Table \ref{tab:anticipated_realized_examples} illustrate the matching logic.
In each case, the first article describes a specific event investors were waiting for before it was realized, and the later article attributes market movement to the realization of the same event.
The anticipated event is concrete enough to define a timing window, and the later reason describes the resolution of that uncertainty.
The examples cover several types of news.
The first concerns uncertainty around a potential war in Iraq: the market anticipated the U.N. weapons-inspection report, which was understood ex ante as relevant for war expectations and therefore for markets.
The second example, about U.S.-China trade negotiations, explicitly describes the market's prior view that trade-related developments would be the main market-moving factor at that time; the right column then shows the realized trade-negotiation news that our extraction method captures as the reason for the market rally.
The remaining examples include macroeconomic data releases, monetary-policy announcements, and corporate earnings announcements, all commonly understood as market-moving events.
This pattern suggests that the newspaper identifies the type of uncertainty that matters to markets and that our extraction method captures the realization of that uncertainty.

Using this conservative procedure, we find that 1,250 reason snippets, or 4.1\%, are anticipated, meaning they are matched with at least one anticipation text.
We also find that 710 trading days, or 12.5\%, have at least one anticipated reason snippet.
These rates should be interpreted as lower bounds because the anticipation source is limited to ``Today's Markets'' articles, a small subset of the textual information that reflects market expectations about future events.
Moreover, these articles are primarily backward-looking market recaps, so a broader search across news sources would likely identify more anticipated reasons.

Although anticipated reasons are only a small fraction of all snippets, they carry stronger explanatory power than average.
This implies that the explanatory power of news-based reasons remains strong, and may even be stronger, when we focus on the subset for which we have more confidence in ex-ante market relevance.

We evaluate the explanatory power of the anticipated reasons by treating them as a group, analogous to a cluster in Section \ref{subsec:2r2}.
The anticipated-reason group has active $R^2$ of 11.47\%, conditioning on days when at least one anticipated reason is active.
This is comparable to the Monetary policy cluster's active $R^2$ of 12.09\%, even though monetary policy is widely understood as an important market-moving category.
The comparison is notable because the Monetary policy cluster contains 4,952 snippets, or 16\% of all snippets, whereas anticipated reasons account for only 4.1\%.
The small anticipated subset therefore delivers explanatory power comparable to a major market-moving cluster, which is difficult to reconcile with a pure ex-post attribution story.

To account more rigorously for the mechanical effect of subset size on $R^2$, we bootstrap a baseline distribution using random subsets of reason snippets with the same size as the anticipated-reason group.
The anticipated-reason active $R^2$ of 11.47\% lies in the upper 1.2\% tail of this baseline distribution.
Thus we reject the null that anticipated reasons have the same explanatory power as a random subset of snippets of the same size, with a $p$-value of 1.2\%.
The detailed bootstrap procedure and results are reported in Appendix \ref{appsubsec:anticipated_reason_bootstrap}.

\subsection{News-based reasons align with structured real-world event data}

Next, we show that reason text aligns with structured real-world events.
Around scheduled events such as FOMC meetings and major macroeconomic releases, the corresponding reason categories become more likely to appear on event dates and around them in sensible lead-lag patterns.

We compile a list of scheduled market-moving events that appear in structured datasets and are commonly studied in the literature on financial market outcomes.
We pair each scheduled event type with a conceptually related reason cluster.
For example, the Monetary Policy cluster is paired with FOMC meetings, and the ``Price Level'' cluster is paired with BLS CPI and PPI releases.\footnote{
Appendix Table \ref{tab:event_matching_table} provides the full crosswalk between reason clusters, scheduled events, and their recurrence patterns.}

\captionsetup[subfigure]{labelformat=empty}
\begin{figure}[!t]
    \centering
    \begin{subfigure}{0.32\textwidth}
        \includegraphics[width=\textwidth]{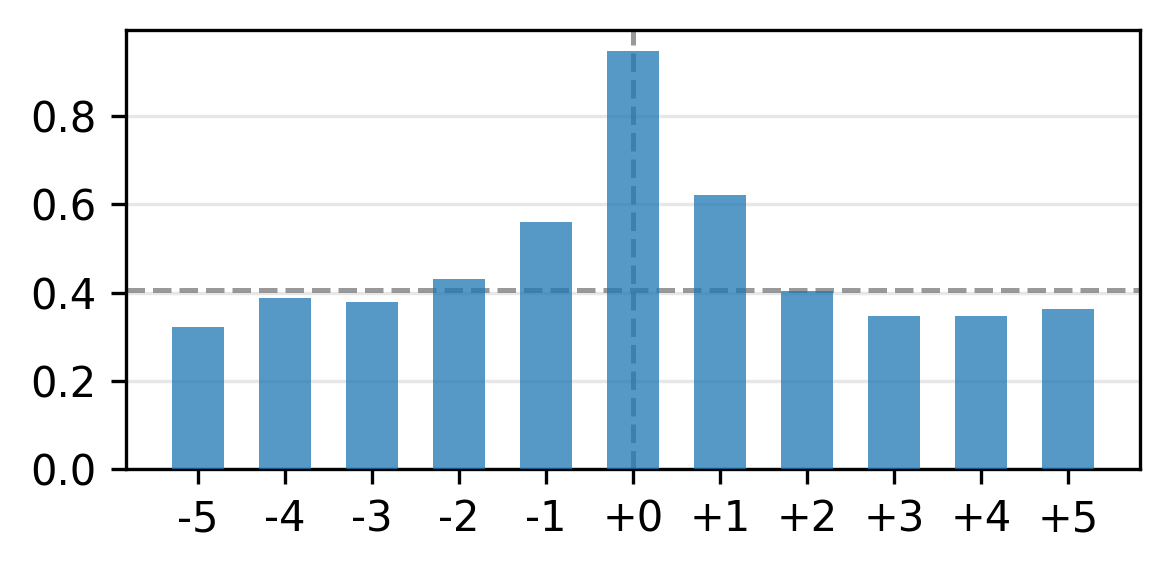}
        \caption{\shortstack{A. ``Monetary Policy'' around\\FOMC meetings}}
        \label{fig:all_hit_rates_mp_fomc}
    \end{subfigure}\hfill
    \begin{subfigure}{0.32\textwidth}
        \includegraphics[width=\textwidth]{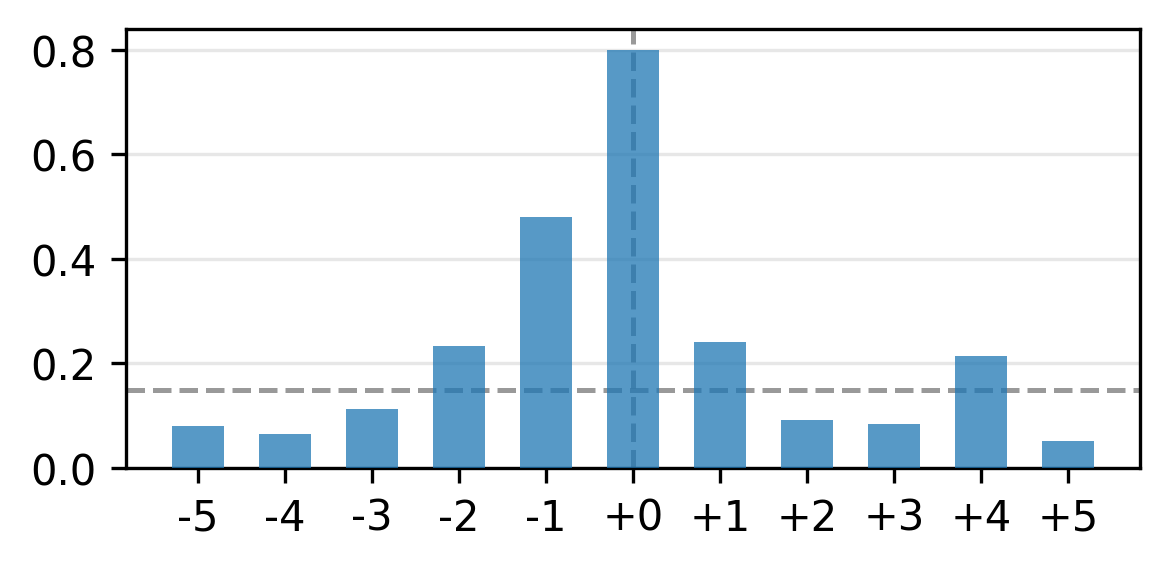}
        \caption{\shortstack{B. ``Labor Market'' around\\ BLS employment situation}}
    \end{subfigure}\hfill
    \begin{subfigure}{0.32\textwidth}
        \includegraphics[width=\textwidth]{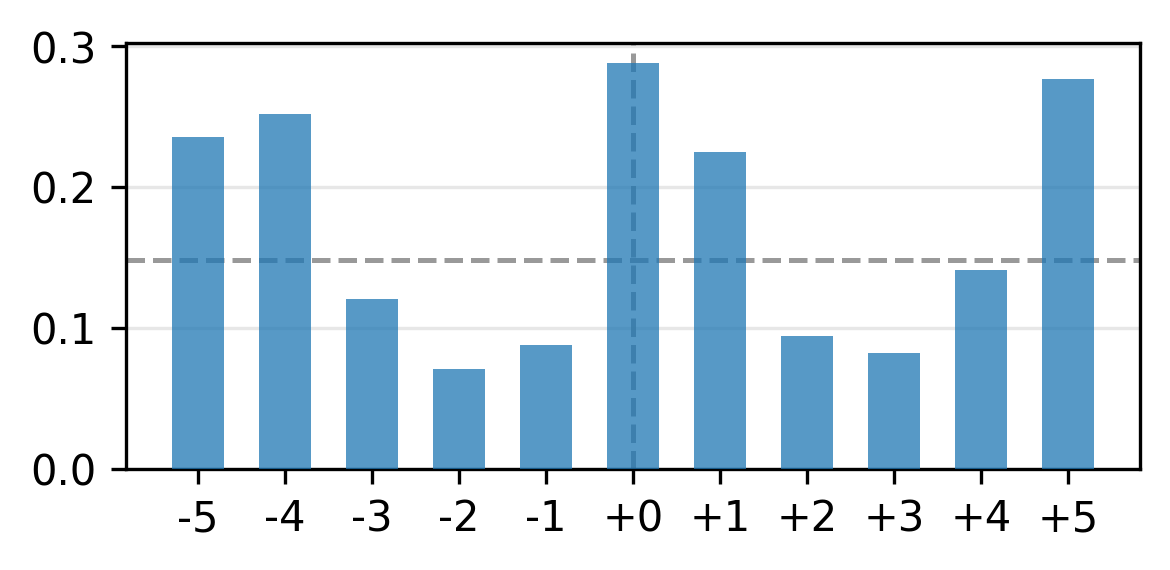}
        \caption{\shortstack{C. ``Labor Market'' around\\ Labor Dept.\ UI claims}}
        \label{fig:labor_unemployment_claims}
    \end{subfigure}

    \begin{subfigure}{0.32\textwidth}
        \includegraphics[width=\textwidth]{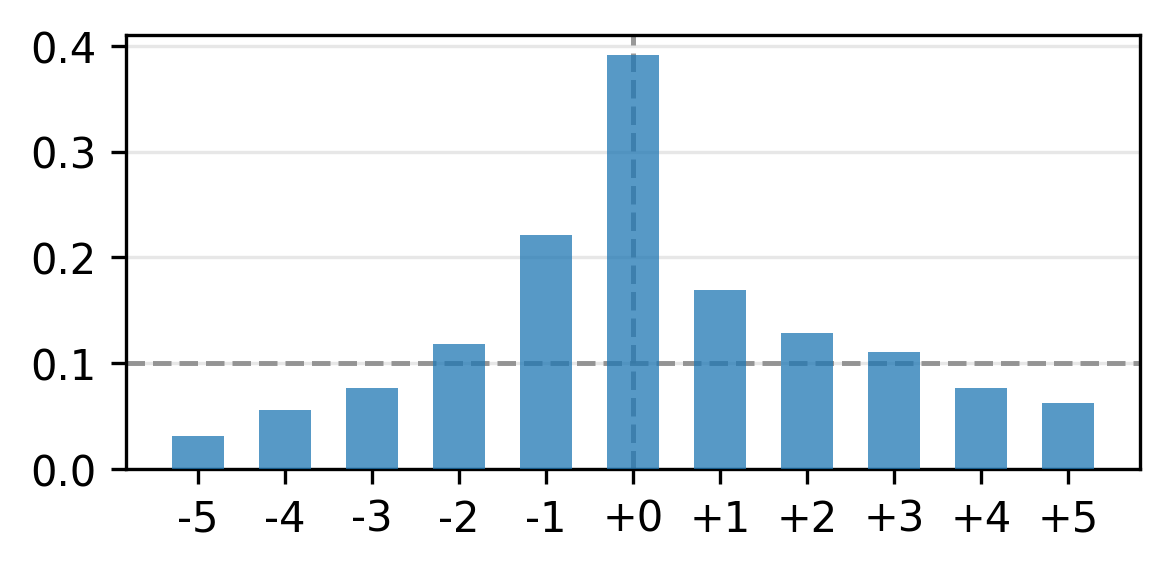}
        \caption{\shortstack{D. ``Price Level'' around\\ BLS CPI release}}
    \end{subfigure}\hfill
    \begin{subfigure}{0.32\textwidth}
        \includegraphics[width=\textwidth]{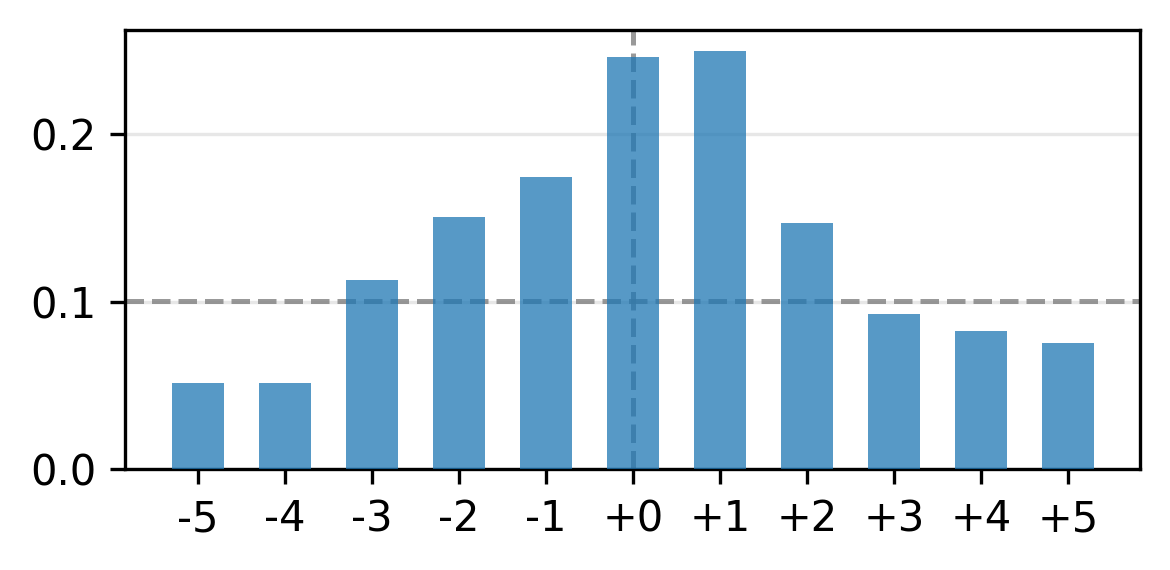}
        \caption{\shortstack{E. ``Price Level'' around\\ BLS PPI release}}
        \label{fig:inflation_ppi}
    \end{subfigure}\hfill
    \begin{subfigure}{0.32\textwidth}
        \includegraphics[width=\textwidth]{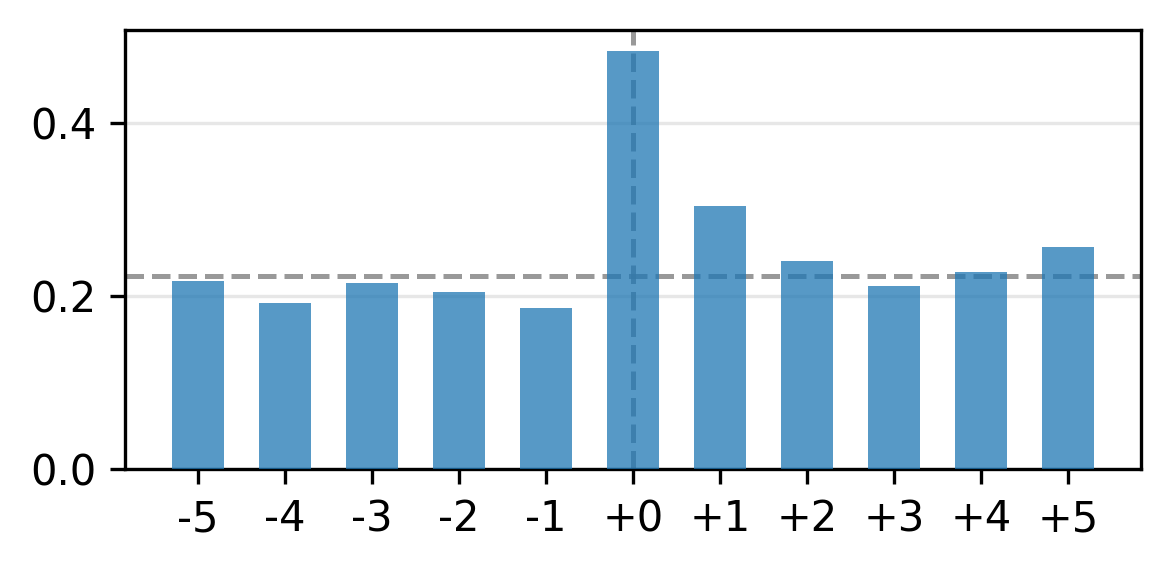}
        \caption{\shortstack{F. ``Economic Growth'' around\\BEA GDP advance estimates}}
        \label{fig:gdp}
    \end{subfigure}

    \begin{subfigure}{0.32\textwidth}
        \includegraphics[width=\textwidth]{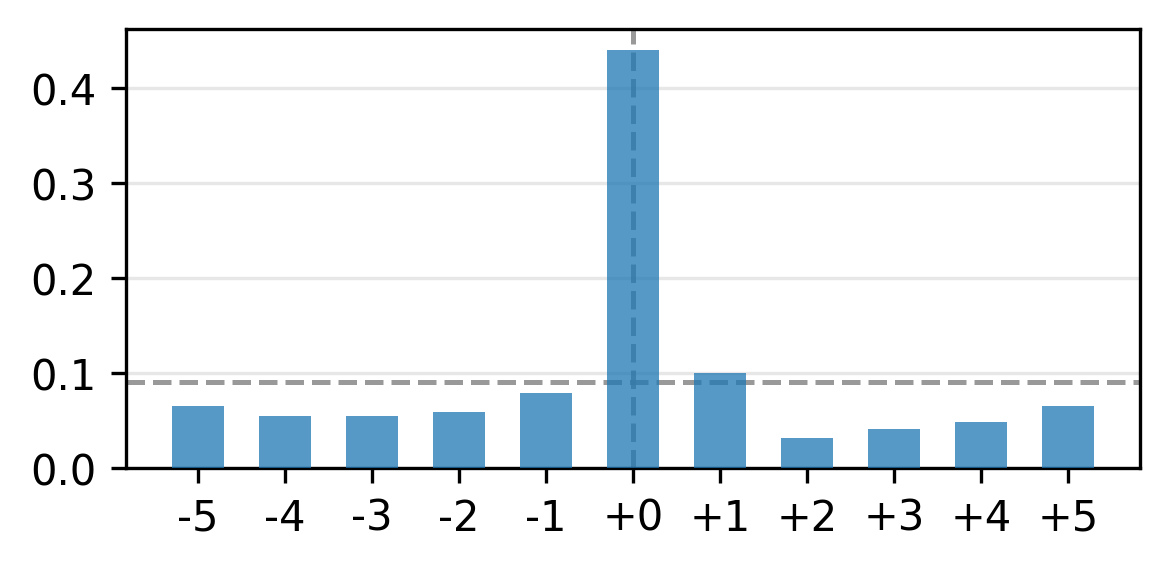}
        \caption{\shortstack{G. ``Consumer Activity'' around\\Census retail \& food sales}}
        \label{fig:consumer_retail_advance}
    \end{subfigure}\hfill
    \begin{subfigure}{0.32\textwidth}
        \includegraphics[width=\textwidth]{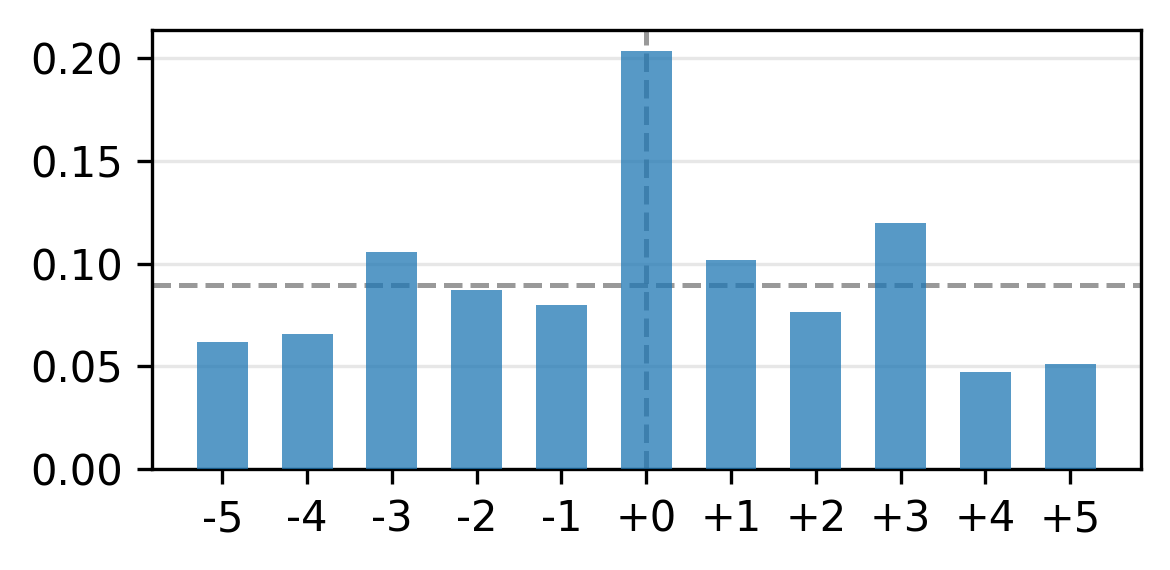}
        \caption{\shortstack{H. ``Consumer Activity'' around\\ BEA Personal Income}}
        \label{fig:consumer_personal_income}
    \end{subfigure}\hfill
    \begin{subfigure}{0.32\textwidth}
        \includegraphics[width=\textwidth]{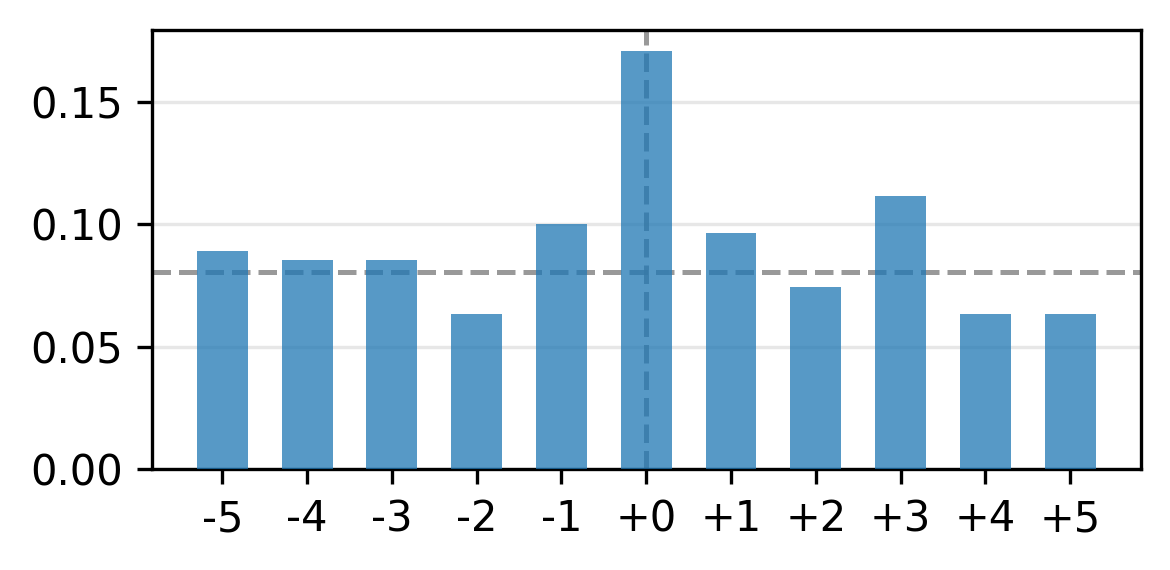}
        \caption{\shortstack{I. ``Global Economic Conditions''\\around U.S.\ intnl.\ trade release}}
        \label{fig:global_trade}
    \end{subfigure}

    \begin{subfigure}{0.32\textwidth}
        \includegraphics[width=\textwidth]{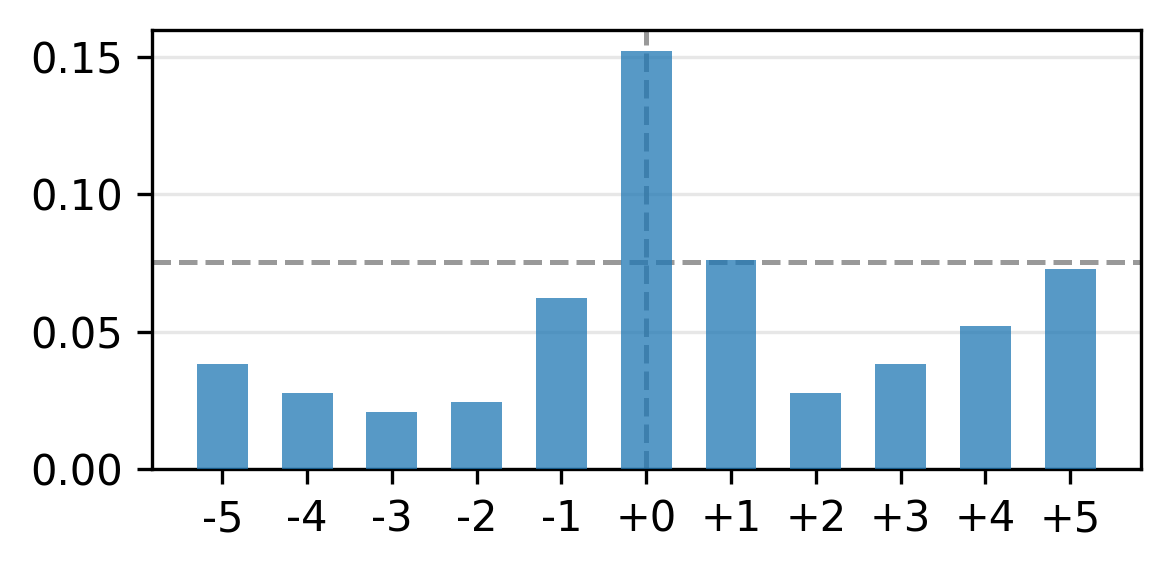}
        \caption{\shortstack{J. ``Manufacturing Activity''\\around Industrial Production}}
        \label{fig:manufacturing_industrial_production}
    \end{subfigure}\hfill
    \begin{subfigure}{0.32\textwidth}
        \includegraphics[width=\textwidth]{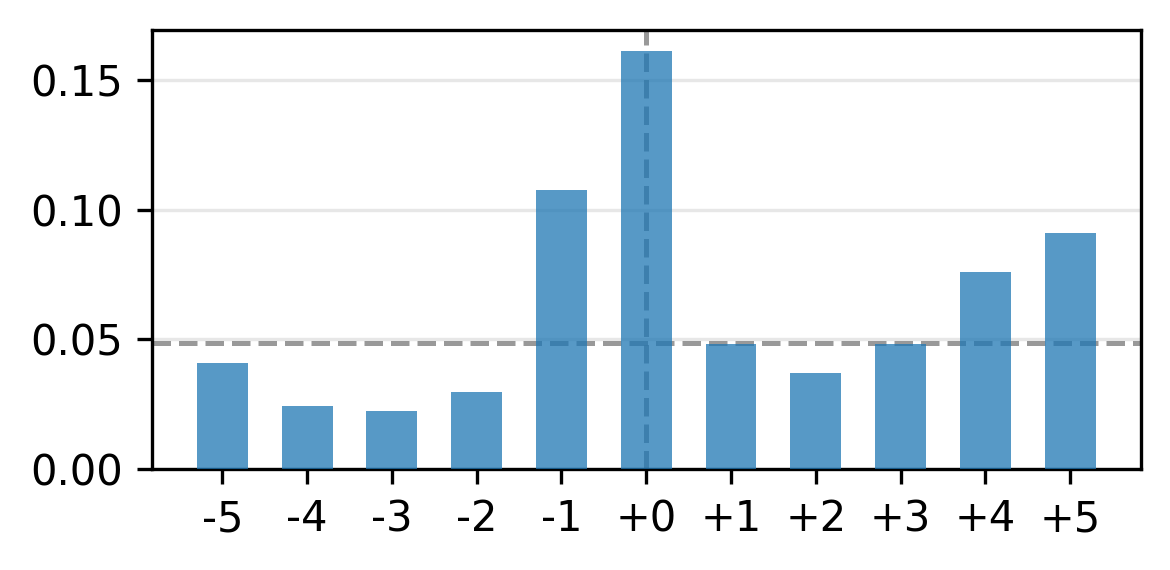}
        \caption{\shortstack{K. ``Housing Market'' around\\ Census/HUD housing starts}}
        \label{fig:housing_starts}
    \end{subfigure}\hfill
    \begin{subfigure}{0.32\textwidth}
        \includegraphics[width=\textwidth]{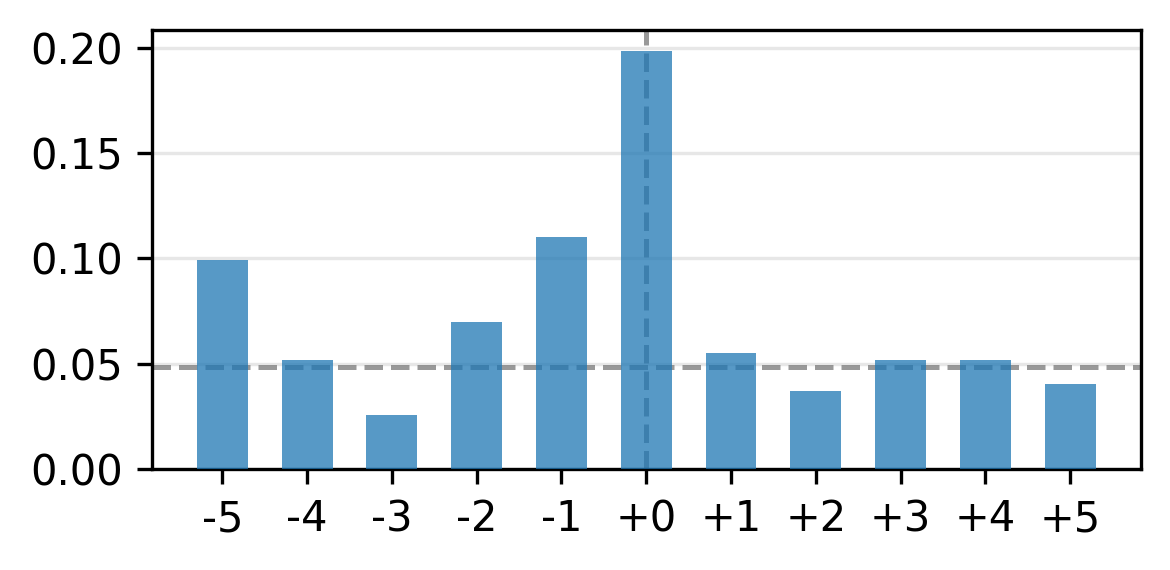}
        \caption{\shortstack{L. ``Housing Market'' around\\ Census/HUD new home sales}}
        \label{fig:housing_new_sales}
    \end{subfigure}

    \caption{Probability of reason-cluster occurrence around scheduled event dates}
    \label{fig:all_hit_rates}
    \justifying\footnotesize
    \noindent
    Note: 
    Each panel shows the occurrence of a reason cluster around the corresponding scheduled events. 
    Day 0 is the scheduled event day, with plus denoting the trading days after the event day and minus for before the event day. 
    The vertical axis is the sub-sample frequency with which the reason cluster is active on those days.
    The dashed horizontal line is the unconditional  full-sample frequency. 
    Details on the paired reason clusters and scheduled events, including the data sources and recurrence patterns, are provided in    
    Appendix Table \ref{tab:event_matching_table}.
\end{figure}

Figure \ref{fig:all_hit_rates} shows the occurrence of matched reason clusters around scheduled event dates.
Panel A shows that the Monetary Policy cluster is much more likely to be active on FOMC announcement days than on an average trading day, and its occurrence is also elevated in the days immediately before and after the announcement.
This pattern is consistent with market anticipation before the meeting and follow-up interpretation of policy news after the announcement.
The other panels follow the same format for macroeconomic releases and show similar event-centered peaks.
The elevated occurrence of the matched reason clusters around scheduled events supports the view that news-based reasons are grounded in real-world events and reflect the market's anticipation of, and reaction to, those events.

A few exceptions to a simple event-day peak reflect the timing structure of the scheduled events.
In Panel C, the Labor Department's weekly unemployment insurance claims report is the only weekly event in this exercise, so the window can overlap with the previous or next weekly release and generate an additional peak.
In Panels K and L, housing starts and new home sales are typically released only five trading days apart, so the non-monotonic patterns reflect interference between related housing-market releases.

We provide additional evidence that the reasons we extract---and the implied returns constructed from them---are aligned with economically meaningful market-moving forces. We focus on monetary-policy events, whose effects on market prices have been studied in the literature using high-frequency intraday returns. 
We use an existing measure developed by \citet{acosta2025financial} of the financial market effects of FOMC communications, based on high-frequency intraday responses of the S\&P 500 in windows of 30 to 100 minutes around scheduled monetary-policy announcements.
The intraday construction is intended to isolate the market response to monetary-policy news from other events that may affect the market on the same day.%
\footnote{Appendix~\ref{appsec:monetary_policy_validation} shows that the high-frequency market response a) appears on only a small fraction of days, b) has only moderate explanatory power for market returns on those days, and c) is constructed from realized market returns and therefore does not satisfy Criterion 1.}

We run the following regression across the monetary policy event days (i.e. FOMC meetings and/or minutes release days):
\begin{align*}
    \text{Intraday Monetary Policy Shock}_t = a + b \rhat_{\text{Monetary Policy}, t} + \epsilon_t,
\end{align*}
where $\rhat_{\text{Monetary Policy}, t}$ is the reason-based implied return from the Monetary Policy cluster on day $t$, as calculated in Section \ref{subsec:2r2}.\footnote{On monetary-policy event days when the Monetary Policy cluster is inactive, $\rhat_{\text{Monetary Policy}, t}$ is set to zero. Such inactive cases are rare for FOMC announcement days, as shown in Figure \ref{fig:all_hit_rates} Panel A central bar.}

\begin{table}[!t]
\centering
\begin{tabular}{lccc}
\toprule
 & FOMC meetings & Minutes releases & Combined \\
\midrule
Slope & $0.9393^{***}$ & $0.3046^{***}$ & $0.6719^{***}$ \\
      & $(4.24)$ & $(2.87)$ & $(5.16)$ \\[4pt]
Intercept & $-0.1019^{*}$ & $-0.0101$ & $-0.0467$ \\
          & $(-1.81)$ & $(-0.52)$ & $(-1.62)$ \\[4pt]
$N$ & $193$ & $182$ & $375$ \\[4pt]
$R^2$ & $0.0860$ & $0.0438$ & $0.0667$ \\
\bottomrule
\multicolumn{4}{@{}p{\linewidth}@{}}{\footnotesize Note: The dependent variable is the high-frequency S\&P 500 response around the monetary-policy event. The independent variable is the Monetary Policy cluster's reason-implied return, constructed from the pooled embedding-regression model. On event days when the Monetary Policy cluster is inactive, the reason-implied return is set to zero. $t$-statistics are in parentheses. $^{*}p<10\%$, $^{**}p<5\%$, $^{***}p<1\%$.} \\
\end{tabular}
\caption{Alignment of news-based monetary-policy reasons with high-frequency market responses}
\end{table}

The slope coefficient is positive and significant.
In the FOMC announcement sample, the estimate is close to one, suggesting that the reason-based implied return from the Monetary Policy cluster aligns with high-frequency market responses to FOMC meetings.
However, the regression $R^2$ is low, likely because both the news-based measure and the high-frequency market response are noisy proxies for the underlying monetary-policy shock.

The Monetary Policy reason is present on most FOMC announcement and minutes release days, consistent with the event-centered occurrence patterns in Figure \ref{fig:all_hit_rates}.
More importantly, conditional on a scheduled policy event, the high-frequency market response is more variable and larger in absolute magnitude when a Monetary Policy reason is present than when it is absent.\footnote{Appendix Figure \ref{fig:usmpd_sp500_dist_hit_missed_raw} reports the distributional comparison and its construction details.}
This pattern is consistent with selective journalistic reporting of economically important, market-moving news. For example, even though the FOMC meeting indeed happens, if the information content of the announcement is weak and the market response is small, the news source may simply not write about monetary policy as a reason for the market movement. This interpretation also suggests the flexibility of the news-based method in forming a set of explanations of market movements compared to following the scheduled events mechanically.

\section{Conclusion}

Roll's 100\% $R^2$ benchmark is an upper bound on how well researchers can explain market movements.
It makes minimal assumptions about the economics of price formation, except that a relationship between real-world events and price fluctuations exists, is sufficiently stable, and can be recovered after the fact.
Our contribution is to reinterpret this basic question into an empirical discipline: how close can finance research come to explaining observed market variation?

This paper shows that significant progress comes with constructing explanations using news text and AI tools.
We first show that conventional non-news explanations have little to no explanatory power in a standard OOS evaluation framework.
The central empirical result is that news-based reasons, extracted from Wall Street Journal market coverage, substantially improve explanatory power.
Large language models and generative AI allow us to extract, structure, and quantify the stated reasons embedded in unstructured news text.
The resulting reason text explains a significant share of daily and monthly market returns.
It also provides an interpretable taxonomy of market-moving narratives whose relative importance evolves over time, shedding light on what moves the market at a given point in history.

These results make substantial progress toward understanding the basic question of why markets fluctuate, while also clarifying what can and cannot be explained by the current measurement framework.
The remaining unexplained variation suggests that richer data and methods could push explanatory power further.
We deliberately use only a single article per trading day, which provides a conservative and transparent proof of concept.
Future research can use attention-based methods to extract and aggregate information from much broader news corpora, including the market-commentary documents produced by investment-bank research departments.
The criteria and baseline evaluation metric developed here also provide a common framework for that line of research \citep{hellum2025power}.

\clearpage
\begin{singlespace}

\end{singlespace}

\clearpage

\doublespacing

\begin{appendices}

	\renewcommand{\theequation}{A.\arabic{equation}}
	\renewcommand{\thefigure}{A.\arabic{figure}}
	\renewcommand{\thetable}{A.\arabic{table}}
	\renewcommand{\thepage}{A.\arabic{page}}

	\setcounter{equation}{0}
	\setcounter{figure}{0}
	\setcounter{table}{0}
	\setcounter{page}{1}
	\setcounter{footnote}{0}

	\begin{center}
		{\LARGE
		Online Appendix\\[6pt]
		Explaining Realized Returns: Roll's $R^2$ 40 Years Later}
	\end{center}

\section{Construction, definitions, and implementation details}

\subsection{Construction of non-news baseline variables}
\label{appsubsec:non_news}

This subsection provides construction details for the four sets of non-news baseline variables: macroeconomic statistics, cash flows, subjective expectations, and trading flows.

{\bf Quarterly frequency transformation.}
According to the \href{https://shillerdata.com/}{Shiller Data webpage}: ``Monthly dividend and earnings data are computed from the S\&P four-quarter totals for the quarters since 1926, with linear interpolation to monthly figures.''
Both survey expectations and flow data are also at the quarterly frequency in their original form. We apply the same monthly frequency procedure following Shiller. 

In particular, the following simple example illustrates the linear transformation: 
$Y_{\text{Mar}} = Y^{\text{(4Q)}}_{\text{Q1}}$,
$Y_{\text{Apr}} = \tfrac{2}{3}\,Y^{\text{(4Q)}}_{\text{Q1}} + \tfrac{1}{3}\,Y^{\text{(4Q)}}_{\text{Q2}}$,
$Y_{\text{May}} = \tfrac{1}{3}\,Y^{\text{(4Q)}}_{\text{Q1}} + \tfrac{2}{3}\,Y^{\text{(4Q)}}_{\text{Q2}}$,
$Y_{\text{Jun}} = Y^{\text{(4Q)}}_{\text{Q2}}$.

{\bf Macro statistics.}
We use the monthly macroeconomic statistics as the explanations. Specifically, we use the FRED-MD dataset \citep{mccracken2016fred}, which contains a large number of macroeconomic statistics at the monthly frequency.
Among the 126 series in the raw dataset, we only include 99 that are directly related to the real economy and remove the 27 series that are directly about or derived from financial asset prices, such as stock market indices, exchange rates, and interest rates. 
Specifically, we exclude the 22 variables in Group 6: ``Interest and exchange rates'' and the 5 variables in Group 8: ``Stock market''.

{\bf Cash flow.}
We use aggregate S\&P 500 dividend and earnings growth as cash-flow explanations.
Specifically, we use S\&P 500 aggregate earnings and dividend data described in \citet{shiller2015irrational}, downloaded from \href{https://shillerdata.com/}{Robert Shiller's webpage}.
We include four variables:
    Nominal Earnings Growth, 
    Nominal Dividend Growth, 
    Real Dividend Growth, 
    and Real Earnings Growth.
We exclude variables that directly incorporate asset-price information, such as price-earnings ratios, to satisfy criterion 1.

Following the Shiller data construction, monthly dividend and earnings series are obtained by linearly interpolating quarterly four-quarter totals.
We apply the same frequency transformation to the quarterly subjective-expectations and trading-flow data described below.

{\bf Survey expectations.}
We use quarterly subjective cash-flow-growth expectations from \citet{delao2021subjective}, downloaded from \href{https://www.sean-myers.net/data}{Sean Myers' webpage}.
The data aggregate I/B/E/S analyst forecasts for S\&P 500 firms and cover 2003Q1 through 2021Q3 in our sample.
The corresponding out-of-sample $R^2$ is therefore calculated over the months in which the survey-expectations data overlap with our news sample.
We retain four one-year expectation measures: expected log dividend growth, expected log earnings growth $e_t$, expected log earnings growth $d_t$, and expected log earnings growth based on adjusted earnings.
We exclude realized future growth and contemporaneous price-ratio variables to satisfy criterion 1.
We transform the quarterly expectations to the monthly frequency using the linear interpolation procedure described above.

{\bf Trading flows.}
We use sector-level net equity flows constructed following \citet{gabaix2021search}.
Specifically, we use $\Delta q_{i,t}^{\mathcal{E}}$, the net equity flow of investor sector $i$ in quarter $t$, as defined in their paper.
The resulting data contain 12 series corresponding to 12 investor sectors.
We transform the quarterly series to monthly frequency using the linear interpolation procedure described above.
Because the flow data end in 2019Q1, the corresponding out-of-sample $R^2$ is calculated over the months in which the flow data overlap with our news sample.

We replicate the data sources and construction procedure described in Appendix C of \citet{gabaix2021search} and validate our construction by visually comparing the resulting series with those plotted in Figure F.3 of their paper.

\subsection{Generative AI prompts for extraction, leakage screening, and return inference}
\label{appsubsec:prompts}

This subsection reports the prompts used to extract reason--price-movement pairs, screen reason text for leakage, and infer market returns from the retained reasons.

\noindent\textbf{Prompt 1. Extracting reason and price movement pairs from news articles}

\displayprompt{tables/prompts/movement_reason_prompt.tex}

\noindent\textbf{Prompt 2. Flagging reason text that contains leaked information about market movements}

\displayprompt{tables/prompts/flag_reason.tex}

\newpage
\noindent\textbf{Prompt 3. Inferring price movements from reasons}

\displayprompt{tables/prompts/reason_gpt_predict_prompt_display.tex}

\subsection{Cluster-level $R^2$ definitions and decomposition}
\label{appsubsec:2r2}

This appendix subsection provides the formal definitions of the overall and active-day cluster-level $R^2$ measures used in Figure \ref{fig:cluster_radar}, derives the approximate decomposition connecting Panels A, B, and C, and validates the approximation empirically.

Define cluster $c$'s active-day set as $\mathcal{A}_c := \{t: \text{cluster } c \text{ is active on day } t\}$, and let $|\mathcal{A}_c|$ denote the number of active days. We set $\rhat_{c,t}=0$ for $t\notin\mathcal{A}_c$.

The overall $R^2$ of cluster $c$ is defined as
\begin{align}
    R^2_{\textrm{overall},c}\defeq 1 - \frac{\sum_{t=1}^T\Paren{r_t - \rhat_{c,t}}^2}{\sum_{t=1}^T r_t^2} 
\end{align}

The active-day $R^2$ of cluster $c$ is defined as
\begin{align}
    R^2_{\textrm{active},c}\defeq 1 - \frac{\sum_{t\in \mathcal{A}_c}\Paren{r_t - \rhat_{c,t}}^2}{\sum_{t\in \mathcal{A}_c} r_t^2}
    \label{eq:active_r2}
\end{align}
where the summation is over $t\in\mathcal{A}_c$, the days when cluster $c$ is active.

We have the following identities that connect the two types of $R^2$:
\begin{align}
    R^2_{\textrm{overall},c} &= R^2_{\textrm{active},c} \frac{\sum_{t\in \mathcal{A}_c} r_t^2}{\sum_{t=1}^T r_t^2}, %
    \\
    &= 
    R^2_{\textrm{active},c}
    \times
    \frac{|\mathcal{A}_c|}{T}
    \times
    \frac{\frac{1}{|\mathcal{A}_c|}\sum_{t\in \mathcal{A}_c} r_t^2}{\frac{1}{T}\sum_{t=1}^T r_t^2}.
\end{align}
That is, the overall cluster $R^2$ can be decomposed as the product of the active-day cluster $R^2$ times the active-day percentage, adjusted by the relative variance (or average squared return, to be accurate) of market returns on active days compared to the overall sample.
In other words, the approximated decomposition in Figure \ref{fig:cluster_radar} that ``A $\approx$ B $\times$ C'' is accurate when the variance of market returns on active days is similar to that of the overall sample.

Note 1: The above identities are exact relationships in sample, regardless of the explanatory power of the clusters or the estimation method, including regardless of whether the evaluation is in sample or out of sample.
The identities also hold in subsamples of days, allowing us to evaluate the decomposition for positive and negative returns separately (the red and green versions of the radar charts in Figure \ref{fig:cluster_radar}).

Note 2: The $R^2$ values above do not demean the actual returns in calculating the total sum of squares in the denominators. The difference is numerically negligible when explaining daily returns, as the means are close to zero with a minuscule effect on the total sum of squares.
In cases where demeaning is more relevant (not here), the same decomposition relationship continues to hold with the demeaned $R^2$ values given the adjustment that the inactive days' implied returns are set to the sample mean return rather than zero.

\begin{figure}[htbp]
    \centering
    \caption{Validation of the cluster-level $R^2$ decomposition}%
    \label{fig:scatter_r2_comparison}
    \vspace*{-6pt}
    \includegraphics[width=1\textwidth]{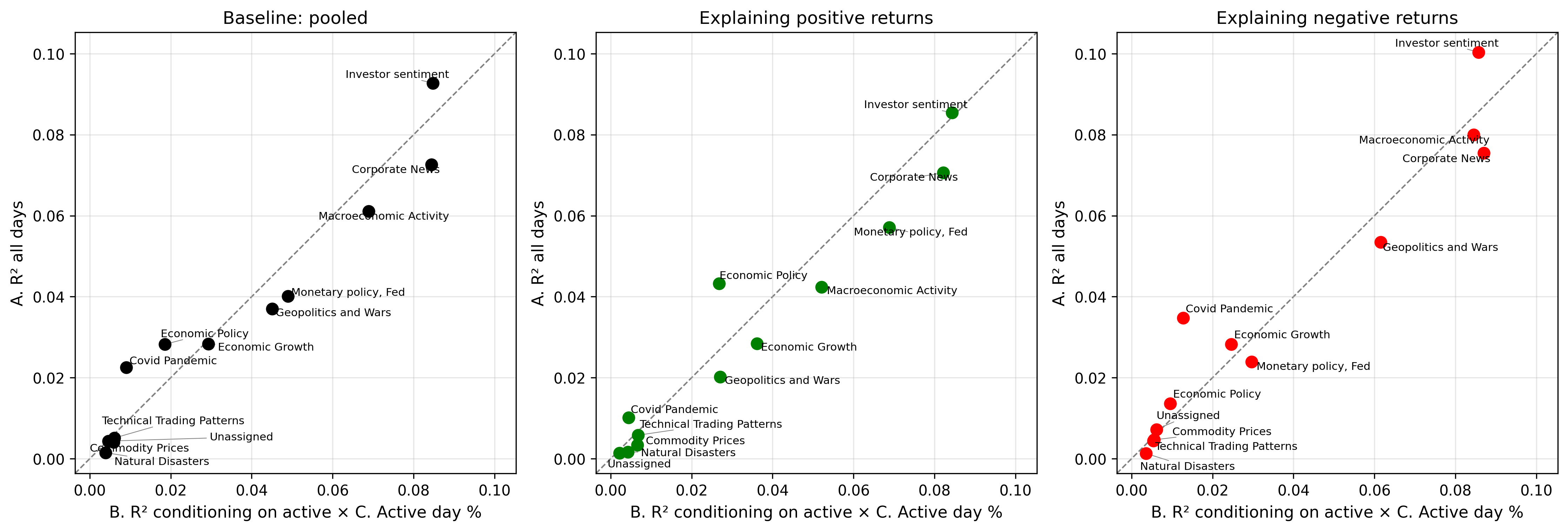}%
    \flushleft\justifying\footnotesize
    Note: The horizontal axis is the product of active-day $R^2$ and active-day share for each cluster, corresponding to Figure \ref{fig:cluster_radar} Panel B multiplied by Panel C. The vertical axis is overall cluster $R^2$, corresponding to Figure \ref{fig:cluster_radar} Panel A. Red and green points repeat the calculation for negative- and positive-return days, respectively. The 45-degree line indicates the approximation in which overall $R^2$ equals active-day $R^2$ multiplied by active-day share.
\end{figure}

Figure \ref{fig:scatter_r2_comparison} numerically validates the approximated decomposition relationship in Figure \ref{fig:cluster_radar} by plotting the overall $R^2$ against the product of active-day $R^2$ and active-day percentage for each cluster. The points are close to the 45-degree line, confirming the accuracy of the approximated decomposition relationship in our sample.
The deviations from the 45-degree line reflect the adjustment factor of relative average squared market returns on active days compared to the overall sample, which is close to one for most clusters, but can be higher for some clusters such as Investor sentiment and Covid pandemic.

\subsection{Scheduled-event matching details for Figure \ref{fig:all_hit_rates}}

This table provides the reason-cluster and scheduled-event crosswalk and timing details used in Figure \ref{fig:all_hit_rates}.

{
    \scriptsize\singlespacing
    \renewcommand{\baselinestretch}{0.9}\selectfont
    \setlength{\extrarowheight}{2pt}
    \begin{longtable}{
        >{\raggedright\arraybackslash}p{0.08\linewidth}
        >{\raggedright\arraybackslash}p{0.22\linewidth}
        >{\raggedright\arraybackslash}p{0.22\linewidth}
        >{\raggedright\arraybackslash}p{0.38\linewidth}}
    \caption{Reason-cluster and scheduled-event matches used in Figure \ref{fig:all_hit_rates}}
    \label{tab:event_matching_table}\\[-10pt]
    \toprule
     Figure \ref{fig:all_hit_rates} panel & Reason cluster & Scheduled event & Source and release cycle \\
    \midrule
    \endfirsthead
    \toprule
     Figure \ref{fig:all_hit_rates} panel & Reason cluster & Scheduled event & Source and release cycle \\
    \midrule
    \endhead
    \midrule
    \endfoot
    \bottomrule
    \multicolumn{4}{@{}p{\linewidth}@{}}{\footnotesize Note: Each row gives the reason cluster and scheduled event used for the corresponding panel of Figure \ref{fig:all_hit_rates}. Event date data from the Federal Reserve Bank of St.\ Louis, available at \url{https://alfred.stlouisfed.org}.}\\
    \endlastfoot
    A. & Monetary Policy cluster & FOMC meeting days & Fed board. Policy statement release and press conference on Wednesdays, eight times a year.\\
    \addlinespace[6pt]
    B. & Labor Market sub-cluster & BLS Employment Situation Summary & BLS. Includes total nonfarm payroll employment and the unemployment rate. Typically released on the third Friday after the conclusion of the reference week, i.e. the week that includes the 12th of the month. \\
    \addlinespace[6pt]
    C. & Labor Market sub-cluster & Unemployment Insurance Weekly Claims Report & Labor Department, ETA. Weekly. Typically Thursday. \\
    \addlinespace[6pt]
    D. & Price Level sub-cluster& Consumer Price Index & BLS. Monthly. Usually released between the 10th and 15th of each month. \\
    \addlinespace[6pt]
    E. & Price Level sub-cluster & Producer Price Index & BLS. Monthly. Typically a couple of days after the CPI release.\\
    \addlinespace[6pt]
    F. & Economic Growth cluster & GDP Advance Estimate & BEA. Quarterly GDP release with advance, second, and third estimates, typically near the end of months. \\
    \addlinespace[6pt]
    G. & Consumer Activity sub-cluster & Advance Monthly Sales for Retail and Food Services & Census Bureau.  Monthly. Usually around the middle of the month.\\
    \addlinespace[6pt]
    H. & Consumer Activity sub-cluster & Personal Income and Outlays & BEA. Including PCE (Personal Consumption Expenditures). Monthly. Typically near the end of the month. \\
    \addlinespace[6pt]
    I. & Global Economic Conditions sub-cluster & U.S. International Trade in Goods and Services & Census Bureau and BEA. Monthly. \\
    \addlinespace[6pt]
    J. & Manufacturing Activity sub-cluster & G.17 Industrial Production and Capacity Utilization & Fed board. Monthly. Typically mid-month.\\
    \addlinespace[6pt]
    K. & Housing Market sub-cluster & New Residential Construction & Census Bureau and HUD. Monthly. Usually on the 12th workday of the month.\\
    \addlinespace[6pt]
    L. & Housing Market sub-cluster & New Residential Sales & Census Bureau and HUD. Monthly. Usually on the 17th workday of the month.\\
    \end{longtable}
}

\subsection{Model-based variance attribution and explanatory power}
\label{appsubsec:variance_attribution}

Our framework asks how much realized return variation can be explained by observations that are neither about nor derived from asset prices (criterion 1) and that have a stable statistical relationship with returns (criterion 2).
This differs from allocating return variance across components of a decomposition or a calibrated model.
These exercises are useful and revealing in their respective settings, but their variance attributions do not establish explanatory power under our criteria.

\noindent\textbf{Campbell-Shiller decomposition.}
The Campbell-Shiller decomposition of returns into cash-flow news and discount-rate news delivers an identity with $R^2 = 100\%$ up to the linear approximation.
In canonical empirical implementations, discount-rate news is measured as the revision in expected future returns implied by a forecasting VAR that includes returns themselves, and cash-flow news is measured as the residual return component not explained by discount-rate news.
Both components are therefore derived from concurrent return data and do not satisfy criterion 1.
The resulting decomposition does not show how much return variation independently observed candidate explanations can account for out of sample.

\noindent\textbf{Gabaix-Koijen calibration.}
\citeauthor{gabaix2021search}'s \citeyearpar{gabaix2021search} \textit{In Search of the Origins of Financial Fluctuations: The Inelastic Markets Hypothesis} develops a model in which flows affect prices in an inelastic market.
In Section 5.4, they calibrate the general-equilibrium model and report that flow shocks account for approximately 90\% of stock-return variance.
Their model attributes unexpected return variation to dividend shocks and flow shocks, which are assumed to be uncorrelated.
The calibration matches annualized stock-return volatility of 15\%, with dividend-growth volatility set to 5\%.
With dividend shocks accounting for approximately 11\% of return variance, the calibrated flow-driven price component accounts for the remaining approximately 89\%.
Tables 4--5 report the calibration parameters and implied moments, and Appendix H.2 describes the calculation of the variance shares.
The flow contribution is therefore a model-based allocation of return variation.

\section{Additional empirical results}

\subsection{Monthly zero-shot aggregation and HAR filtering diagnostics}
\label{appsec:zero_shot_har}

\newcommand{\ma}{\textrm{ma}}

This appendix reports diagnostics that explain the poor performance of raw monthly aggregation and describes HAR filtering of reason-implied returns.
These diagnostics characterize the time-series properties of the implied-return series, while the main text uses zero-shot inference primarily as evidence that a daily reason-return mapping can be recovered without task-specific training on realized returns.

Table \ref{tab:GenAI_R2} Panel A, Column ``Raw'' reports the zero-shot results.
The daily $R^2$ is 25.58\%, comparable to the embedding-regression $R^2$ of 29.82\% reported in Panel B.
At the same time, aggregating raw daily implied returns $\widehat{r}_{t}^\text{zero-shot}$ to the monthly frequency performs poorly, producing a monthly $R^2$ of $-115.03\%$.
This failure reflects systematic biases in the reason-implied return series.

The daily implied return series $\rhat_t^\text{zero-shot}$ has a very negative mean and strong positive serial autocorrelation, both inconsistent with core time-series properties of market returns.
The mean of $\rhat_t^\text{zero-shot}$ is $-0.16\%$, which is sharply inconsistent with a positive long-run mean market return.
This likely reflects a combination of pessimistic bias in journalistic reporting and pessimistic bias in the LLM's interpretation of event implications.
We do not provide a definitive decomposition of these two sources.
Pessimism in the LLM output may also arise from subtle prompting choices.
For example, the word ``impact'' in the prompt ``predict the potential impact on the overall stock market'' may carry a negative connotation and push $\rhat_t^\text{zero-shot}$ downward.
The mean bias is not central for explaining price variation and can be removed by demeaning.

The more salient bias is serial persistence.
The daily first-order autoregressive coefficient of $\rhat_t^\text{zero-shot}$ is 16.19\% and highly statistically significant. 
The embedding-regression implied return series also exhibits positive persistence, though the bias is less severe than in the zero-shot series.
In contrast, market returns have very little serial correlation, and if anything, the AR(1) coefficient of daily S\&P 500 returns is negative (a well-known pattern dating back at least to \citealt*{roll1984simple}).
Moreover, in a heterogeneous autoregressive (HAR) setting, $\rhat_t^\text{zero-shot}$ is also significantly positively related to its lagged weekly and monthly moving averages.
\begin{table}[htbp]
\centering
\begin{tabular}{lcccccc}
\toprule
 & \multicolumn{3}{c}{AR(1)} & \multicolumn{3}{c}{HAR} \\
\cmidrule(lr){2-4}\cmidrule(lr){5-7}
 & $r_t$ & $\rhat_t^\text{emb-reg}$ & $\rhat_t^\text{zero-shot}$ & $r_t$ & $\rhat_t^\text{emb-reg}$ & $\rhat_t^\text{zero-shot}$ \\ \hline
const & $0.0397^{***}$ & $0.0429^{***}$ & $-0.1380^{***}$ & $0.0523^{**}$ & $0.0203^{**}$ & $-0.0471^{***}$ \\
       & $(2.60)$ & $(4.60)$ & $(-10.58)$ & $(2.45)$ & $(2.31)$ & $(-3.63)$ \\[4pt]
lag1 & $-0.1067^{***}$ & $0.1274^{***}$ & $0.1619^{***}$ & $-0.0899^{**}$ & $0.0722^{***}$ & $0.0800^{***}$ \\
       & $(-3.85)$ & $(8.54)$ & $(9.91)$ & $(-2.47)$ & $(4.38)$ & $(4.45)$ \\[4pt]
ma5 &  &  &  & $-0.0814$ & $0.1151^{***}$ & $0.1193^{***}$ \\
       &  &  &  & $(-0.97)$ & $(3.36)$ & $(3.41)$ \\[4pt]
ma22 &  &  &  & $-0.0061$ & $0.1388^{*}$ & $0.2357^{***}$ \\
       &  &  &  & $(-0.04)$ & $(1.95)$ & $(3.22)$ \\[4pt]
ma63 &  &  &  & $-0.2531$ & $0.2687^{***}$ & $0.2754^{***}$ \\
       &  &  &  & $(-1.14)$ & $(3.08)$ & $(3.39)$ \\
\midrule
$R^2$ & 0.0114 & 0.0162 & 0.0262 & 0.0134 & 0.0331 & 0.0595 \\
\bottomrule
\multicolumn{7}{l}{Note: \footnotesize Newey-West $t$-statistics in parentheses. $^{*}p<10\%$, $^{**}p<5\%$, $^{***}p<1\%$.} \\
\end{tabular}
\caption{Auto-regressions on daily returns and fitted returns}
\end{table}

This conditional bias is unlikely to arise solely from the LLM's interpretation because inference is performed independently day by day.
It appears to be rooted partly in persistence in the underlying reason text itself.
Journalistic reporting may remain focused on past events and adjust slowly to new information.
It is also possible that some articles review returns over a longer horizon and provide reasons that correspond to that broader period, which we then treat as reasons for that day's market movement.

These biases explain why raw monthly aggregation of generated daily implied returns fails.
The mean bias is modest at the daily frequency but becomes large when accumulated over a month.
Similarly, serial persistence accumulates at the monthly frequency and produces exaggerated implied monthly returns that overshoot realized returns.
We use HAR filtering to remove the mean and predictable persistence from the implied return series.
We interpret the residual from the HAR regression as the shock component of implied returns after filtering predictable persistence and evaluate the $R^2$ of this residual against market returns.
In particular, we run a full-sample heterogeneous autoregressive regression (HAR, \citealp*{corsi2009simple}) of daily implied returns from either embedding regression or zero-shot learning ($\rhat_{t}^\text{zero-shot}$ or $\rhat_{t}^\text{emb-reg}$)
on its lagged values and moving averages, and obtain the residual series $\ehat_{t}^\text{zero-shot}$ or $\ehat_{t}^\text{emb-reg}$:
\begin{align*}
    \rhat_{t} &= \beta_{0} + 
    \beta_{1} \rhat_{t-1} +
    \beta_{5} \ma_{5} \rhat_{t-1} + \beta_{22} \ma_{22} \rhat_{t-1} + \beta_{63} \ma_{63} \rhat_{t-1} + \ehat_{t}
\end{align*}
where $\ma_{5} \rhat_{t-1} \defeq \frac{1}{5} \Paren{\rhat_{t-1} + \cdots + \rhat_{t-5}}$ is the weekly moving average of $\rhat_{t}$, and the monthly and quarterly moving averages are defined similarly.
We treat the residual daily series $\ehat_{t}$ as the filtered implied return series and apply the same monthly aggregation and $R^2$ evaluation procedure used for $\rhat_{t}$.
Specifically, we aggregate daily residuals to the monthly frequency by summing $\ehat_{t}$ within each month, then evaluate monthly $R^2$ against realized monthly market returns.

The HAR filtering is applied only to the implied return series $\rhat_{t}$ and does not use realized market returns $r_t$.
Therefore, filtering does not introduce realized-return leakage: the filtered series remains consistent with Criterion 1 because it is not derived from asset returns.

Table \ref{tab:GenAI_R2}, Column ``HAR filtered'', reports the results.
Panels A and B report results for zero-shot inference and embedding regression, respectively; the corresponding HAR-filtered series are $\ehat_{t}^{\text{zero-shot}}$ and $\ehat_{t}^\text{emb-reg}$.
Filtering consistently improves $R^2$ for both methods and at both frequencies, indicating both the prevalence of persistence bias and the effectiveness of HAR filtering in removing it.
The zero-shot method benefits more from filtering, consistent with the more severe persistence bias in zero-shot implied returns.
At the daily frequency, zero-shot $R^2$ improves from 25.58\% to 31.88\%, exceeding the embedding-regression benchmark of 29.82\%.
The largest improvement is at the monthly frequency, where zero-shot $R^2$ rises from $-115.03\%$ to 49.84\% after filtering.
Filtering is especially effective at the monthly frequency because persistence bias accumulates when daily implied returns are summed to monthly returns.

For embedding regression, daily $R^2$ improves from an already high 29.82\% to 33.88\% after filtering, and monthly $R^2$ improves from 33.06\% to 44.42\%.
These monthly $R^2$ values are obtained by summing daily $\ehat_{t}^{\text{zero-shot}}$ or $\ehat_{t}^{\text{emb-reg}}$ to the monthly frequency.
This differs from the out-of-sample monthly $R^2$ of 40.30\% reported in Table \ref{tab:R2}, which is obtained from monthly ridge regression on monthly aggregated embeddings.
Monthly ridge regression already mitigates daily persistence bias to some extent, but HAR filtering with aggregation still produces further gains.

\subsection{Bootstrap significance of anticipated reasons}
\label{appsubsec:anticipated_reason_bootstrap}

\begin{figure}[h]
    \centering
    \caption{Active $R^2$ of anticipated reasons relative to random-subset benchmarks}
    \label{fig:anticipated_reason_bootstrap_active_days}
    \includegraphics[width=0.45\textwidth]{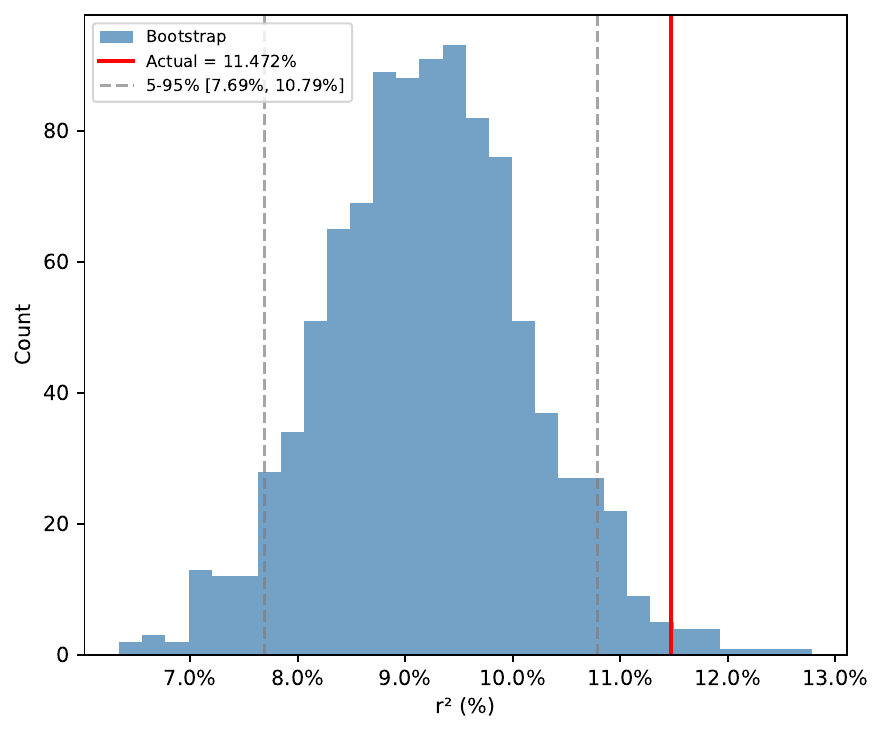}\par
    \footnotesize\justifying\noindent
    Note: The figure compares the active-day $R^2$ of the anticipated-reason group with a bootstrap distribution generated from 1,000 random subsets of reason snippets. Each random subset matches the anticipated-reason group in size, equal to 4.1\% of all snippets, and is evaluated using the same active-day $R^2$ formula. The red vertical marker indicates the observed anticipated-reason $R^2$ at 11.47\%.
\end{figure}

To evaluate the explanatory power of anticipated reasons, we treat the anticipated-reason snippets as a single group and calculate active-day $R^2$ on days when at least one anticipated reason is active.
This follows the cluster-level procedure in Section \ref{subsec:2r2}, except that the group is defined by anticipation--reason matching rather than by the clustering procedure.
Using the active-day $R^2$ defined in Equation \eqref{eq:active_r2}, with $c$ denoting the anticipated-reason group, the group has an $R^2$ of 11.47\%.

To assess whether this explanatory power is unusually high relative to that of an equally sized group, we construct a bootstrap reference distribution by randomly sampling 4.1\% of all reason snippets, matching the size of the anticipated-reason group.
For each draw, we treat the sampled subset as a group and calculate its active-day $R^2$ using the same procedure.
Repeating this exercise 1,000 times produces the bootstrap reference distribution shown in Figure \ref{fig:anticipated_reason_bootstrap_active_days}.
The observed $R^2$ of the anticipated-reason group lies in the upper 1.2\% tail of this distribution.

\subsection{High-frequency monetary-policy responses and its explanatory power for daily market returns}
\label{appsec:monetary_policy_validation}

Table \ref{tab:usmpd_hf_r2} reports a simple benchmark based on the U.S. Monetary Policy Event-Study Database (USMPD). For each event type, we use the corresponding USMPD high-frequency S\&P 500 return as a direct measure of the implied return associated with the event and compare it to the full daily S\&P 500 return on the same trading date. No parameters are fitted in this exercise. The reported statistic is the demeaned $R^2$ used throughout the paper,
\begin{align}
    R^2 = 1 - \frac{\sum_t (r_t - \widehat{r}_t)^2}{\sum_t (r_t - \bar{r})^2},
\end{align}
where $r_t$ is the daily return and $\widehat{r}_t$ is the corresponding USMPD high-frequency return. The four variables follow the sheet definitions on the first page of the USMPD workbook: \textit{Statements}, \textit{Press Conferences}, \textit{Monetary Events}, and \textit{Minutes}. The final column reports the event-window length as a fraction of the regular cash equity trading session of 390 minutes from 9:30 a.m.\ to 4:00 p.m.\ Eastern time.

\begin{table}[h]
    \caption{Explanatory power of high-frequency monetary-policy responses for daily market returns}
    \label{tab:usmpd_hf_r2}
    \begin{center}
    \begin{tabular}{lcccc}
        \toprule
        Event type & $N$ & $R^2$ (\%) & HF window len.\ (mins) & \% of day session\\
        \midrule
        Combined               & 375 & 14.40 & 100 or 30 &~ \\[6pt]
        ~~- FOMC announcements      & 193 & 19.31 & 100 (or 30 if no press conf.) & 25.6\% \\
        ~~~~- statements       & 193 & 1.87  & 30  & 7.7\% \\
        ~~~~- press conferences& 80  & 48.87 & 70  & 18.0\% \\[6pt]
        ~~- Minutes (3 weeks later)               & 182 & 5.36  & 30  & 7.7\% \\
        \bottomrule
    \end{tabular}
    \end{center}
    \footnotesize\justifying\noindent
    Note: The table reports $R^2$ values from using high-frequency intraday S\&P 500 responses around monetary-policy events directly as $\widehat{r}_t$ to explain full-day S\&P 500 returns on the corresponding event-day subsample. No regression parameters are estimated. $N$ is the number of event days from the USMPD dataset matched to our sample period of 2002--2024. The last column reports the event-window length as a percentage of the regular trading session, defined as 6.5 hours or 390 minutes.
\end{table}

Table \ref{tab:usmpd_hf_r2} first shows that identified monetary-policy events in structured datasets are infrequent.
The USMPD dataset contains 193 FOMC meetings and 182 Fed minutes release days during our sample period of 2002--2024.
Together, these events account for only a small share of trading days.
This is much narrower than the set of days on which monetary-policy reasons appear in the news, consistent with the view that structured event datasets capture an important but restricted subset of monetary-policy information.

Second, the high-frequency market response to monetary-policy events has only moderate explanatory power for the full daily market movement on those days.
This is notable because using intraday S\&P 500 returns to explain daily S\&P 500 returns violates Criterion 1: the explanatory variable is itself a market outcome.
Under a simple random-walk benchmark with independent, homoskedastic intraday increments, a subwindow's $R^2$ for the full-day return equals its share of the trading session.
The reported $R^2$ values generally do not materially exceed this duration-based benchmark, except for press conferences: the 70-minute window represents 18.0\% of the trading session but has an $R^2$ of 48.87\%.
The combined $R^2$ for the 375 event days is 14.40\%. The $R^2$ for FOMC announcements is higher at 19.31\%. The 80 press-conference events have a much higher $R^2$ of 48.87\%. The FOMC statement and minutes responses have much lower $R^2$ values of 1.87\% and 5.36\%, respectively.
These numbers are comparable to the active explanatory power of the news-based Monetary Policy cluster.
But the news-based reasons satisfy Criterion 1 and appear on a broader set of monetary-policy-relevant days.
This comparison highlights both the explanatory value of monetary-policy reason text and the limits of structured event datasets.

\subsection{News-reason coverage and the magnitude of monetary-policy responses}
\label{appsec:monetary_policy_news_coverage}

\begin{figure}[htbp]
    \centering
    \caption{Distribution of high-frequency S\&P 500 responses on monetary-policy event days by news-reason coverage}
    \label{fig:usmpd_sp500_dist_hit_missed_raw}
    \includegraphics[width=\textwidth]{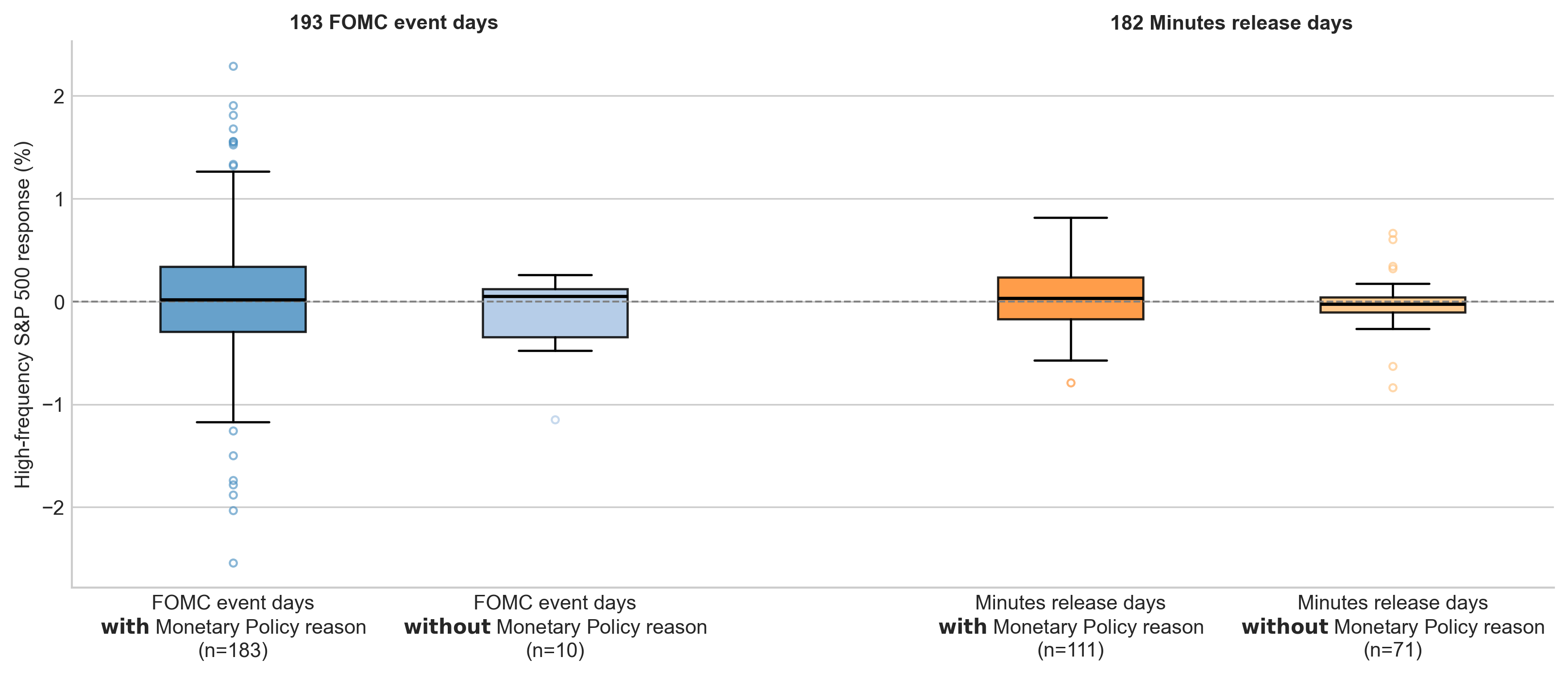}
    \footnotesize\justifying\noindent
    Note: The figure reports box-and-whisker distributions of high-frequency S\&P 500 responses on FOMC announcement and Fed minutes release dates. Event days are split by whether at least one extracted reason snippet is assigned to the Monetary Policy cluster. Sample sizes are reported below each group.
\end{figure}

We show that the presence of monetary-policy reasons in news coverage is associated with larger high-frequency market responses to scheduled monetary-policy events.
We divide monetary event days (FOMC announcement and Fed minutes, respectively) according to whether at least one Monetary Policy cluster snippet appeared on that day. 
Figure \ref{fig:usmpd_sp500_dist_hit_missed_raw} compares the distributions of high-frequency market responses across these groups.

First, this comparison shows that the Monetary Policy reason is present in news coverage on most FOMC announcement and minutes release days, as expected and shown in Figure \ref{fig:all_hit_rates}.
More importantly, high-frequency responses are more dispersed when the Monetary Policy cluster is active than when it is inactive. This pattern is consistent with selective journalistic reporting: when a scheduled communication contains little market-moving information and generates limited price variation, the market recap is less likely to identify monetary policy as a reason for the day's market movement. The comparison is descriptive, particularly because the cluster is absent on only 10 of the 193 FOMC announcement days.

\clearpage
\end{appendices}

\end{document}